\documentclass[%
 reprint,
 superscriptaddress,
 amsmath,amssymb,
 prx,
 floatfix,
]{revtex4-2}

\usepackage{graphicx}
\usepackage{dcolumn}
\usepackage{bm}
\usepackage{microtype}
\usepackage{dsfont}
\usepackage{mathtools}
\usepackage{comment}
\usepackage{hyperref}
\usepackage{enumitem}
\usepackage{braket}
\usepackage{enumitem}
\usepackage[english]{babel}

\usepackage[normalem]{ulem}

\hypersetup{colorlinks=true}

\usepackage{color} 
\usepackage[dvipsnames]{xcolor}

\usepackage{tikz}
\DeclareRobustCommand{\ringfouralpha}[1][1.0]{%
  \tikz[baseline=-0.6ex, scale=#1, x=1em, y=1em]{
    \node (NW) at (-1,  1) [circle, fill=black, inner sep=0pt, minimum size=0.6ex] {};
    \node (NE) at ( 1,  1) [circle, fill=black, inner sep=0pt, minimum size=0.6ex] {};
    \node (SE) at ( 1, -1) [circle, fill=black, inner sep=0pt, minimum size=0.6ex] {};
    \node (SW) at (-1, -1) [circle, fill=black, inner sep=0pt, minimum size=0.6ex] {};
    \draw (NW) -- node[above] {$\alpha$} (NE);
    \draw (NE) -- node[right] {$\alpha$} (SE);
    \draw (SE) -- node[below] {$\alpha$} (SW);
    \draw (SW) -- node[left]  {$\alpha$} (NW);
  }%
}
\DeclareRobustCommand{\linethreeTwoAlpha}[1][1.0]{%
  \tikz[baseline=-0.6ex, scale=#1, x=1em, y=1em]{
    \node (L) at ( 2, 0) [circle, fill=black, inner sep=0pt, minimum size=0.6ex] {};
    \node (C) at ( 0, 0) [circle, fill=black, inner sep=0pt, minimum size=0.6ex] {};
    \node (R) at ( -2, 0) [circle, fill=black, inner sep=0pt, minimum size=0.6ex] {};
    \draw (L) -- node[above] {$2\alpha$} (C);
    \draw (C) -- node[above] {$2\alpha$} (R);
  }%
}
\DeclareRobustCommand{\linetwoFourAlpha}[1][1.0]{%
  \tikz[baseline=-0.6ex, scale=#1, x=1em, y=1em]{
    \node (L) at (-1, 0) [circle, fill=black, inner sep=0pt, minimum size=0.6ex] {};
    \node (R) at ( 1, 0) [circle, fill=black, inner sep=0pt, minimum size=0.6ex] {};
    \draw (L) -- node[above] {$4\alpha$} (R);
  }%
}

\def\XXint#1#2#3{{\setbox0=\hbox{$#1{#2#3}{\int}$}
     \vcenter{\hbox{$#2#3$}}\kern-.5\wd0}}

\newcommand{\Hsum}[1]{%
  \sideset{}{^{(\mathrm{H})}}\sum_{#1}%
}
\newcommand{\hsum}[1]{%
  \scriptsize\sideset{}{^{(\mathrm{H})}}\sum_{#1}%
}
\newcommand{\hhsum}[1]{%
  \tiny\sideset{}{^{(\mathrm{H})}}\sum_{#1}%
}

\newcommand{\primesum}[0]{%
  \,\sideset{}{'}\sum%
}

\newcommand{\ie}{i.\,e.\,}
\newcommand{\eg}{e.\,g.\,}

\usepackage[dvipsnames]{xcolor}

\definecolor{myblue}{rgb}{0.3, 0.3, 4.0}

\definecolor{myred}{rgb}{0.82, 0.1, 0.26}

\begin{document}

\preprint{APS/123-QED}

\title{Exact and fast series expansions for quantum models with long-range interactions}

\author{Antonia Duft}%
\affiliation{%
 Department of Physics, Friedrich-Alexander-Universit\"at Erlangen-N\"urnberg, Staudtstrasse 7, D-91058 Erlangen, Germany
}
\author{Patrick Adelhardt} 
\affiliation{%
 Department of Physics, Friedrich-Alexander-Universit\"at Erlangen-N\"urnberg, Staudtstrasse 7, D-91058 Erlangen, Germany
}
\author{Jan Alexander Koziol}
\affiliation{%
 Department of Physics, Friedrich-Alexander-Universit\"at Erlangen-N\"urnberg, Staudtstrasse 7, D-91058 Erlangen, Germany
}
\affiliation{%
 Faculty of Physics, University of Vienna, Boltzmanngasse 5, AT-1090 Vienna, Austria
}
\author{Andreas A. Buchheit}
 %\email{andreas.buchheit@uni-saarland.de}
  \affiliation{%
 Department of Mathematics, Saarland University, D-66123 Saarbr\"ucken, Germany
}%
\affiliation{%
 Department of Mathematics, ETH Z\"urich, R\"amistrasse 101, CH-8092 Z\"urich, Switzerland
}%
\author{Kai Phillip Schmidt}%
\affiliation{%
 Department of Physics, Friedrich-Alexander-Universit\"at Erlangen-N\"urnberg, Staudtstrasse 7, D-91058 Erlangen, Germany
}

\date{\today}
             
\begin{abstract}
Over the past decade, high-order series expansions based on linked-cluster methods have become an important tool for studying low-energy properties of gapped quantum systems with long-range interactions.
We introduce a deterministic framework that removes a central computational bottleneck of this method. 
Our graph zeta method replaces the costly and statistically noisy Monte Carlo evaluation of high-dimensional lattice sums by a systematic, high-precision computation that delivers series coefficients within minutes on standard desktop hardware. 
The full momentum-dependent series is obtained in a single calculation, enabling high-resolution excitation spectra throughout the Brillouin zone without separate momentum-by-momentum evaluations.
Building on the mathematical developments of the companion paper~\cite{buchheit2026}, the method reformulates graph-embedding sums as graph zeta functions and decomposes them into blocks classified by their treewidth $\mathrm{tw}$. 
Low-treewidth blocks ($\mathrm{tw}\leq2$) admit closed expressions based on Epstein zeta functions, while higher-treewidth blocks ($\mathrm{tw}>2$) are evaluated using tensor-network bucket elimination. 
We benchmark the approach for transverse-field Ising models with power-law interactions in one, two, and three dimensions, reproducing previous Monte Carlo results at a fraction of the computational cost while enabling substantially denser parameter sampling. 
An open-source implementation makes the method directly applicable to general interactions and large parameter scans. As a further application, we compare microscopic interaction models for the stacked quasi-two dimensional transverse-field Ising triangular-lattice antiferromagnet $\mathrm{KTmSe}_2$ and find that a model including dipolar interactions best describes existing experimental data. 
The graph zeta method thus turns high-order linked-cluster expansions into a practical and deterministic tool for fast quantitative momentum-resolved modeling of short- and long-range quantum matter.
\end{abstract}

\maketitle

\section{Introduction}

Perturbative series expansions are a well-established method in quantum many-body and statistical physics \cite{Oitmaa2006,Gelfand2000,Mila2011,Guttmann1989}.
Their central aim is to determine power series of physical quantities around a well-controlled unperturbed limit.
A key advantage of perturbative approaches is that the linked-cluster theorem allows these series to be obtained directly in the thermodynamic limit, without finite-size extrapolation \cite{Oitmaa2006,Gelfand2000,Mila2011,Guttmann1989}.
This distinguishes linked-cluster expansions from many numerical approaches that are intrinsically formulated on finite systems or require controlled approximations, including exact diagonalization \cite{Sandvik2010,Mila2011}, tensor-network methods \cite{Schollwck2011,Orus2014,Paeckel2019,Cirac2021,Bauls2023}, quantum Monte Carlo \cite{Prokofev1998,Sandvik2002,Sandvik2003,Sandvik2010,Carr2010}, and variational approaches \cite{Toulouse2016,Becca2017} e.\,g. based on neural-network quantum states \cite{Lange2024,Medvidovi2024}.

Series expansions provide an exact perturbative result in the thermodynamic limit and thus establish an exact ground truth where the maximal perturbative order limits the considered quantum fluctuations, and against which other approximate numerical and analytical approaches can be assessed.
The unperturbed limit further provides a reference point from which perturbative processes can be understood, and thus provides a natural language for interpreting the results. 
In particular, the method of perturbative continuous unitary transformations (pCUTs) \cite{Knetter2000,Knetter2003} allows to determine coefficients as exact integer fractions, while other perturbative methods may at least provide coefficients with arbitrary numerical floating point precision \cite{Cloizeaux1960,Lwdin1962,Takahashi1977,Shavitt1980,Cederbaum1989,Yao2000,Oitmaa2006,Hormann2023}.

Perturbative series expansions become particularly powerful when high orders are reached.
In addition to providing high-precision estimates of observables, high orders give access to the quantum-critical regime through extrapolation techniques \cite{Guttmann1989}.
They can be reached by exploiting the linked-cluster theorem, which states that only linked processes contribute in the thermodynamic limit \cite{Coester2015}. 
Consequently, physical quantities in the thermodynamic limit can be computed on finite clusters, which is achieved most efficiently in practice by setting up a full graph decomposition \cite{Oitmaa2006}.
Such linked-cluster expansions constitute two steps. 
First, the perturbative contributions of all relevant graphs are calculated. 
Second, the graphs are embedded onto the lattice to obtain the bulk result. 

For models with finite-range interactions (including nearest-neighbor interactions), the embedding is simply a combinatorial problem, i.\,e.\,, the number of possible graph embeddings on the lattice yields an embedding factor which serves as a finite weight when summing up all graph contributions to determine the series in the thermodynamic limit. 
This structure changes qualitatively for long-range interacting systems, where the embedding problem becomes intrinsically nonlocal and cannot be formulated in the same combinatorial way \cite{Fey2016,Adelhardt2024}.
Nevertheless, linked-cluster expansions were generalized \cite{Coester2015} to quantum lattice models with arbitrary long-range interactions already a decade ago \cite{Fey2016}.
The key challenge in determining high-order perturbative series for long-range interacting models lies in the evaluation of nested high-dimensional infinite sums arising from the embedding of graph contributions \cite{Fey2016,Fey2019,Adelhardt2024}. 
Owing to the long-range nature of the interactions, arbitrary embeddings are allowed, since interactions can occur between any pair of lattice sites.

The prototypical example of a long-range interaction is a power-law interaction.
For such interactions, the direct summation of the high-dimensional sums is only feasible for one-dimensional systems and sufficiently fast decay.
Similar direct approaches become insufficient in higher spatial dimensions \cite{Fey2016,Fey2019}. 
An alternative approach was introduced in Ref.~\cite{Fey2019}, where Monte Carlo (MC) techniques were used to calculate these sums in higher dimensions \cite{Fey2019, Fey2020Diss, Adelhardt2020, Adelhardt2024, Adelhardt2025}.
This development laid the foundation for subsequent series-expansion studies on systems with long-range interactions, extending the approach to observables beyond the Hamiltonian \cite{Langheld2022,Adelhardt2023}, multi-flavored interactions \cite{Adelhardt2020,Adelhardt2023,Adelhardt2025}, and higher-spin systems \cite{Adelhardt2026}.

In parallel, in recent decades, long-range interacting quantum systems have attracted increasing attention due to significant progress in the control of atomic and molecular platforms \cite{Defenu2023}. 
In particular, trapped ion platforms \cite{Friedenauer2008,Kim2010,Islam2011,Britton2012,Schneider2012,Islam2013,Jurcevic2014,Richerme2014,Bohnet2016,Kiesenhofer2023,Guo2024}, Rydberg atom arrays \cite{Barredo2015,Labuhn2016,Bernien2017,Schauss2018,Leseleuc2019,Scholl2021,Semeghini2021,Ebadi2021,Scholl2022,Chen2023}, dipolar atoms \cite{Bloch2008,Baier2016,Chomaz2022,Su2023}, and polar molecules \cite{Carr2009,Moses2015,Moses2016,Bohn2017,Christakis2023} have enabled the realization of quantum many-body systems that provide direct experimental access to many-body physics in the presence of long-range interactions.
Long-range interactions also play an important role in condensed-matter systems, where three-dimensional dipolar transverse-field Ising materials constitute a natural setting in which their inclusion is essential for an accurate description \cite{Hansen1975,Bitko1996,Chakraborty2004,Tabei2008,Gingras2011,MartnezHidalgo2001,Buruzi2011,Li2010,Millis2010,Subedi2012,Gingras2011REOH}.

Although versatile, the MC-based embedding approach suffers from intrinsic statistical limitations.
In particular, the convergence of the sums is slow in the number of MC steps and requires averaging over multiple runs with independent random seeds, resulting in significant computational overhead \cite{KrauthBook}.
Consequently, each individual data point is computationally expensive to obtain, and the sampling density reasonably achievable in the parameter space is limited.
The resulting series coefficients carry an inherent error originating from the MC summation, in stark contrast to the exact coefficients obtained for nearest-neighbor models.
Especially for physically challenging problems, the accuracy of the series coefficients is essential to accurately determine the critical properties of the underlying physical system.

Overcoming these limitations requires moving from stochastic to deterministic evaluation of the embedding sums. 
This conceptual shift is enabled by recognizing that the high-dimensional embedding sums can be mapped exactly onto graph lattice sums for which efficient evaluation methods via generalized Epstein zeta functions have recently been developed.  
In particular, a solution for the embedding problem of simple chain and circle graphs can be based on the efficient numerical summation tools in Refs.~\cite{buchheit2024epstein,buchheit2025}, establishing the connection to the evaluation of generalized zeta functions. 
The general mathematical problem of evaluating high-dimensional lattice sums for arbitrary graph topologies is solved in the companion paper~\cite{buchheit2026}, published in parallel with this work.

In this paper, we build on these mathematical foundations to develop the graph zeta method for the deterministic evaluation of graph lattice sums arising in linked-cluster expansions for long-range interacting models. 
This solves the embedding problem for models with long-range interactions on Bravais lattices. 
The framework proceeds in three steps. 
First, a preparatory softcore mapping reformulates the original lattice sums such that all explicit summation constraints are removed, making the problem accessible to the proposed numerical treatment. 
Second, the contributing graphs are decomposed into elementary blocks and subsequently classified by topology.
Third, each block is evaluated analytically, semi-analytically, or numerically with controlled error, yielding deterministic series coefficients free of stochastic errors.
The resulting coefficients can be obtained on a personal computer in minutes on a single core, compared to the cluster-scale computations required by the MC approach, using our newly developed open-source Graph Zeta Library \cite{gzl2026}.

We demonstrate and benchmark the graph zeta method on the long-range transverse-field Ising model (LRTFIM) on chain, square, triangular, and cubic lattices in one to three dimensions.
The graph-zeta coefficients are benchmarked against existing MC series, confirming agreement within MC uncertainties while establishing the graph-zeta result as the more accurate reference. 
The substantially reduced computational cost over several orders of magnitude enables a high-resolution scan of the decay exponent, yielding a detailed characterization of the quantum phase diagram and the crossover between universality regimes. 
Additionally, the graph zeta method evaluates the full momentum-dependent dispersion across an entire Brillouin zone grid at the cost of a single momentum evaluation, compared with the expensive single-point sampling required by MC.
We demonstrate this feature by presenting finely resolved dispersion relations on the entire Brillouin zone for the dipolar LRTFIM on the triangular lattice.

Beyond these methodological advances and numerical benchmarks, the graph zeta method is directly relevant to experimental platforms that realize long-range interactions, including atomic and molecular quantum simulation platforms \cite{Britton2012,Browaeys2020,Monroe2021,Scholl2021,Leclerc2026,Sun2026,Chomaz2022,Cornish2024} and dipolar transverse-field Ising materials \cite{Hansen1975,Bitko1996,Chakraborty2004,Tabei2008,Gingras2011,MartnezHidalgo2001,Buruzi2011,Li2010,Millis2010,Subedi2012,Gingras2011REOH}.
It provides quantitative access to excitation spectra while naturally incorporating both long-range dipolar couplings and material-specific short-range exchange interactions. 
As a concrete application, we use the graph zeta method to compare competing transverse-field Ising descriptions of the quasi-two-dimensional triangular-lattice antiferromagnet $\mathrm{KTmSe}_2$ against existing inelastic neutron-scattering data \cite{Zheng2023KTmSe2}, finding that a nearest-neighbor model supplemented by the full long-range dipolar interaction gives a very accurate description of the measured dispersion. 

The remainder of this paper is structured as follows. 
Section~\ref{sec:setting} provides a brief introduction to the long-range series expansion approach, deriving the embedding lattice sums.
Section~\ref{Sec:softcore_mapping} develops the preparatory transformation of these sums. 
Section~\ref{Sec:evaluation} presents the graph zeta method for the efficient evaluation of lattice sums, with detailed algorithms for individual graph categories provided in Section~\ref{sec:evaluation_details}. 
Section~\ref{sec:benchmark} applies the graph zeta method to the \mbox{LRTFIM} and presents physical results and benchmarks. 
Section~\ref{sec:KTmSe2} applies the graph zeta method to the quantum Ising magnet $\mathrm{KTmSe}_2$ and compares candidate microscopic models directly with the measured inelastic neutron-scattering spectrum. 
Finally, Section~\ref{sec:OutlookAndConclusion} summarizes our main findings and outlines promising extensions and applications of the graph zeta method.

\section{From linked-cluster expansions to the hardcore embedding problem}
\label{sec:setting}

Linked-cluster expansions for quantum lattice models generically give rise to high-dimensional lattice sums that must be evaluated to obtain series coefficients in the thermodynamic limit. 
In this section we derive the explicit form of these sums for the ground-state energy and one-quasiparticle dispersion of quantum lattice models with algebraically decaying long-range interactions.

We begin by introducing perturbative continuous unitary transformations (pCUTs) \cite{Knetter2000,Knetter2003} as a cluster-additive series expansion method formulated as a linked-cluster expansion based on a full graph decomposition. 
We then perform a white-graph expansion \cite{Coester2015}. 
Matrix elements are computed on finite topologically distinct graphs with abstract perturbation parameters assigned to each graph edge.
These parameters remain abstract during the computation, yielding prefactors for each graph without yet specifying the physical interaction strengths. 
Embedding the graphs into the physical lattice by replacing the abstract parameters with the corresponding long-range interaction strengths gives rise to high-dimensional lattice sums subject to hardcore constraints.
It is this embedding step, common to any linked-cluster expansion with long-range interactions, that constitutes the central computational challenge which we call the {\it hardcore problem}. Due to the infinite range of long-range interactions the evaluation of such sums is demanding and numerical computations rely on Monte Carlo (MC) techniques~\cite{Fey2019}.

The hardcore lattice sums derived in this section are not specific to pCUTs but arise in any linked-cluster approach where graph contributions are embedded onto a lattice with long-range interactions. 
Other cluster-additive perturbative methods like multi-block orthogonal transformations \cite{Oitmaa2006}, projective cluster-additive transformations \cite{Hormann2023}, or matrix perturbation theory for appropriate physical quantities could serve the same role and the resulting lattice sums would take the identical structure.
The role of pCUTs in this work is to provide one way to access matrix elements and, in the end, the weights of the hardcore lattice sums.

This section provides a self-contained formulation of pCUTs and the graph-embedding which yields the starting point for the improved evaluation scheme developed in the following sections. 
For a more detailed review of pCUT, white-graph expansions, and the MC evaluation we refer to Ref.~\cite{Adelhardt2024}.

\subsection{Series expansions with perturbative continuous unitary transformations (pCUTs)}
The starting point of perturbation theory is a Hamiltonian of the form
\begin{align}
    H = H_0 + \lambda V\,,
\end{align}
where $H_0$ is a local unperturbed Hamiltonian and $V$ denotes the perturbation, with $\lambda$ controlling the strength of the perturbation. 
We assume that the unperturbed part $H_0$ is bounded from below and has an equidistant spectrum such that it can be expressed in terms of a quasiparticle (qp) counting operator, $Q = \sum_i t_i^\dagger t^{\phantom{\dagger}}_i$, where $t_i^{(\dagger)}$ are hardcore bosonic annihilation (creation) operators on a lattice site $i$. 
The unperturbed part can then be written as $H_0 = \epsilon_0 N + Q$, where $\epsilon_0$ denotes the ground-state energy density per site and $N$ the number of lattice sites.
The perturbation $V$ encodes interactions between lattice sites and can be decomposed as
\begin{align}
    V = \sum_{n=-N_{\rm max}}^{N_{\rm max}} T_n\ ,
    \label{eq:pert_structure}
\end{align}
with operators $T_n$ changing the number of qps by $n$ and $N_{\rm max}$ is the maximal number of qps that can be created or annihilated. 

We take the transverse-field Ising model (TFIM) on lattices with a single-site unit cell as a paradigmatic example and benchmark for the method we develop in this work.
Although we focus on the TFIM, the formalism readily extends to other quantum lattice models that admit a suitable perturbative limit \cite{Adelhardt2024}.
The Hamiltonian for the TFIM reads
\begin{align}\label{eq:H_TFIM}
    H = h\sum_{\bm{i}} \sigma_{\bm{i}}^z - \frac{J}{2} \sum_{\bm{i} \neq \bm{j}} K(\bm i - \bm j) \sigma_{\bm{i}}^x \sigma_{\bm{j}}^x \ ,
\end{align}
where $\sigma_{\bm{i}}^\gamma$ ($\gamma\in{x,z}$) denote Pauli matrices acting on spin-$1/2$ degrees of freedom, $h>0$ is the transverse magnetic field and $J>0$ ($J<0$) corresponds to \mbox{(anti-)}ferromagnetic interactions. 
The strength of the Ising interactions between lattice sites $\bm i$ and $\bm j$ depends on their distance as described by the translationally invariant kernel $K(\bm i-\bm j)$, which we assume to be regularized at zero distance and absolutely summable over the lattice, and which we leave unspecified at this point.

To set up a perturbative expansion, we choose the transverse-field term as the unperturbed Hamiltonian $H_0$ and the Ising interaction as the perturbation $V$, and introduce the associated dimensionless perturbation parameter $\lambda=J/(2h)$.
Employing the Matsubara-Matsuda transformation \cite{Matsubara1956}, the spin operators can be mapped to hardcore bosonic operators $t_i^{(\dagger)}$ that annihilate (create) local spin flip excitations on top of the unperturbed ground state $\ket{0}=\ket{\downarrow \dots \downarrow}$. 
In this representation, the perturbation then decomposes as $V = T_{-2} + T_0 + T_2$.

In general, the unperturbed Hamiltonian $H_0$ and the perturbation $V$ cannot be diagonalized simultaneously. 
We resort to perturbation theory and derive Taylor series expansions for physical quantities in the expansion parameter $\lambda$ directly in the thermodynamic limit.
We employ the method of perturbative continuous unitary transformations (pCUTs) \cite{Knetter2000,Knetter2003}, which enables the efficient calculation of high-order series.
In pCUTs, the Hamiltonian is mapped to an effective, qp-conserving Hamiltonian
\begin{align}
	H_{\text{eff}} = H_0 + \sum_{k=1}^{\infty} \lambda^{k} \sum_{\substack{\vert\bm{m}\vert=k, \\ M(\bm{m})=0}} C(\bm{m})T(\bm{m})\,,
\end{align}
where $C(\bm m)$ are model-independent, rational coefficients and $M(\bm m) = \sum_i m_i = 0$ ensures qp-conserving processes in the operator products $T(\bm m) = T_{m_1} \cdots T_{m_k}$ with $m_i \in [-N_{\rm max}, N_{\rm max}]$. 
As a consequence of qp conservation, the original complicated many-body problem reduces to an effective few-body problem, allowing one to treat different qp-conserving sectors separately.
The operator sequences $T(\bm m)$ are universal in the sense that they depend only on the decomposition of the perturbation $V$ into the operators $T_n$ (c.\,f.~Eq.~\eqref{eq:pert_structure}) and not on the details of the microscopic model.
Note that it is straightforward to generalize pCUTs to multiple perturbation parameters $\lambda_1\dots \lambda_n$ and multiple quasiparticle flavors $t_i^{\alpha}$ \cite{Knetter2003,Coester2015}.

The model-independent structure of the effective Hamiltonian in pCUTs comes at a cost. 
The effective Hamiltonian $H_{\text{eff}}$ is not yet normal ordered, and the model-specific information enters only upon normal ordering. 
This step can be carried out either at the operator level, by exploiting the hardcore-bosonic commutation relations, or by explicitly evaluating matrix elements on finite clusters. 
In practice, the latter approach is typically more efficient in terms of computational effort.

In the following, we are interested in the 0qp sector which gives the ground-state energy density $\epsilon_0$ of the system, and the 1qp sector from which we can obtain the 1qp dispersion $\omega(\bm k)$ and the associated excitation gap ${\Delta=\min_{\bm{k}} \omega(\bm k )= \omega(\bm{k}_c)}$ with the critical gap momentum $\bm{k}_c$. 
The relevant matrix element for the ground-state energy is $E_0 = \epsilon_0 N = \bra{0}H_{\text{eff}}\ket{0}$, where $\ket{0}$ is the unperturbed ground state. 
The 1qp-conserving sector of the Hamiltonian on lattices with a trivial unit cell can be written as the hopping problem 
\begin{equation}
        H_\text{eff}^\text{1qp} = E_0 + \sum_{\bm i , \bm \delta} a(\bm \delta) t^{\dagger}_{\bm i} t^{\phantom{\dagger}}_{\bm i+\bm\delta}\,,
        \label{eq:1qp_hopping_ham}
\end{equation}
where the amplitude of a qp hopping process from site $\bm i$ to $\bm{i} + \bm{\delta}$ is given by the matrix element \mbox{$a(\bm{\delta}) =  \bra{1; \bm{i}+\bm{\delta}} \mathcal{H}_{\text{eff}} \ket{1; \bm{i}}-E_0\delta_{\bm{\delta},\bm{0}}$}. 
Here $\ket{1; \bm i}$ denotes a 1qp spin-flip excitation in the basis of $H_0$ at site $\bm i$. 
The ground-state energy needs to be subtracted from diagonal matrix elements to obtain the irreducible contribution of the hopping amplitude.
Due to the translational invariance of the lattice, a Fourier transformation diagonalizes the 1qp effective Hamiltonian
\begin{equation}
        H_\text{eff}^\text{1qp} = E_0 + \sum_{\bm k} \omega(\bm k) t^{\dagger}_{\bm k} t^{\phantom{\dagger}}_{\bm k}\,,
        \label{eq:1qp_ham}
\end{equation}
and we obtain the 1qp dispersion $\omega(\bm k)$ as desired.
The goal and main effort of the series expansion approach is to determine these matrix elements and perform the normal ordering to obtain a power series in the perturbation parameter $\lambda$ for $E_0$ and $\omega(\bm k)$.
The pCUT formalism can be extended to multi-site unit cells and higher qp sectors as well \cite{Knetter2003}.

\subsection{White-graph expansion}

As a consequence of the linked-cluster theorem, which states that only linked processes contribute to the overall contribution in the thermodynamic limit, we can carry out the normal ordering of the effective Hamiltonian via a linked-cluster expansion, which is most efficiently set up as a full graph decomposition. 

\begin{figure}[t]
    \centering
    \includegraphics[width=\linewidth]{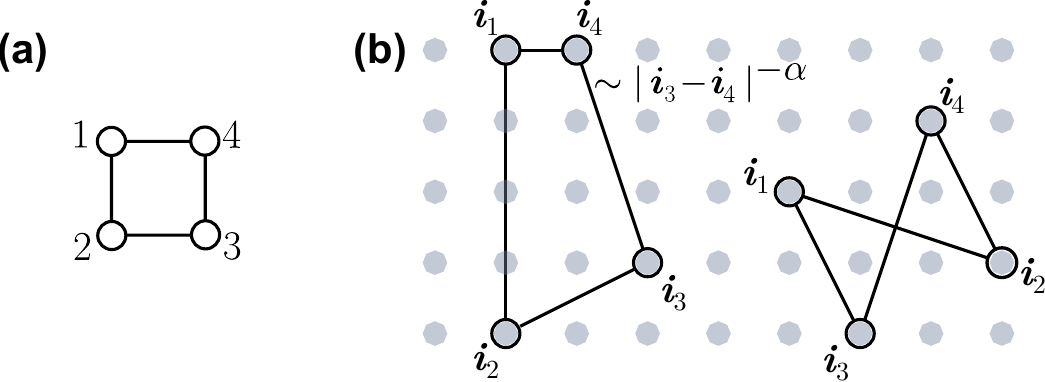}
    \caption{Simple connected example graph with possible embeddings on the square lattice. (1) Example graph $\mathcal{G}$ with vertex set $\mathcal{V}=\{1,2,3,4\}$ and edge set \mbox{$\mathcal{E}=\{(1,2),(2,3),(3,4),(4,1)\}$}. (2) Possible embeddings of the graph on the square lattice, where the perturbation parameters associated with the edges are replaced with the physical algebraically decaying long-range interactions depending on the current embedding. Configurations with overlapping vertex positions are not allowed.}
    \label{fig:graph_embedding}
\end{figure}

We compute the relevant matrix elements by applying $H_\text{eff}$ to all topologically distinct finite clusters, i.\,e.\,, different graphs, up to a given number of edges and sum up their contributions appropriately. 
A graph $\mathcal{G}$ is given by a set of edges $\mathcal{E}$ and vertices $\mathcal{V}$ where an edge $e\in\mathcal{E}$ is a pair of vertices $e=(u,v)$ with $u,v\in\mathcal{V}$. 
In Fig.~\ref{fig:graph_embedding}\,(a), we show an example graph with vertex set $\mathcal{V}=\{1,2,3,4\}$ and edge set $\mathcal{E}=\{(1,2),(2,3),(3,4),(4,1)\}$.

In the following, we denote the ground-state energy on a graph $\mathcal{G}$ as 
\begin{equation}
    E_0(\mathcal{G}) = \bra{0}H_{\text{eff}}\ket{0}_{\mathcal{G}}
\end{equation}
and hopping contributions from vertex $\mu$ to $\nu$ as 
\begin{equation}
    t_{\mu;\nu}(\mathcal{G}) = \bra{1; \nu} H_{\text{eff}} \ket{1;\mu}_{\mathcal{G}}\ - E_0(\mathcal{G})\delta_{\mu,\nu}.
\end{equation}
In an intuitive picture, each order of perturbation corresponds to acting on an edge of a graph. 
As a consequence, in order $\mathfrak{o}$ perturbation theory, only graphs with $\mathfrak{o}$ or fewer edges can contribute. 
It is important to note that we implicitly consider only reduced graph contribution, i.\,e.\,, only contributions in which the perturbation has acted on every edge of the graph at least once. Contributions from proper subgraphs are discarded.

In the conventional graph decomposition for short-range models, each edge is assigned a specific interaction type or strength, which serves as a topological attribute of the graph and is referred to as the color of the edge. 
In the simplest case of a single interaction type or strength, each edge has the same color, but for different types of interactions, a linked-cluster expansion using graphs with multiple edge colors must be carried out, which exponentially increases the number of possible graphs at a given order. 
After computing the perturbative contributions at the level of individual graphs, these graphs must be embedded onto the lattice. 
Since different embeddings of the same graph yield identical contributions, it suffices to count the total number of embeddings for each graph contribution. The embedding problem therefore reduces to a pure combinatorial problem \cite{Gelfand2000, Oitmaa2006,Muehlhauser2024PhD}.
The resulting embedding factor serves as a weight by which the graph contribution is multiplied.

For long-range interactions, however, a linked-cluster expansion based on graphs with colored edges is not possible since the long-range interactions would lead to infinitely many graphs already in first order of perturbation theory and infinitely many edge colors. 
This fundamental issue can be circumvented by a white graph expansion~\cite{Coester2015}. 
In this approach, the colors of edges are discarded as explicit topological attributes, thus the name \textit{white graphs}. 
Instead, each edge $e$ is assigned an abstract perturbation parameter $\lambda_e$, such that the contributions are of the form
\begin{equation}
   v_\mathcal{G}\prod_{e\in\mathcal{E}} \lambda_e^{m_e} 
   \label{eq:pcut_white_graph}
\end{equation}
in order $\mathfrak{o}=\sum_e m_e$, where $v_\mathcal{G}$ is a contribution-specific prefactor, and we introduce $m_e$ as the multiplicity of an edge $e$. 
The multiplicity $m_e$ tells us how often the perturbation acted on edge $e$. 

At a given perturbative order $\mathfrak{o}$, we generate all simple connected graphs with $\mathfrak{o}$ edges and determine all contributions from calculating the ground-state energies $E_0(\mathcal{G})$ and hopping amplitudes $t_{\mu;\nu}(\mathcal{G})$ in terms of the abstract perturbation parameters $\lambda_e$. 
For these contributions, the physical coupling strengths and thus the original edge colors are reintroduced only during the final embedding step, where the abstract parameters $\lambda_e$ are replaced by the physical interaction kernel for the given lattice embedding.
Up to this point, the graph expansion is independent of the specific lattice but depends only on the given model. The lattice information enters in the embedding procedure which we describe next.

\subsection{Graph embedding for long-range interactions}

After generating all simple connected graphs up to a desired perturbative order $\mathfrak{o}_{\text{max}}$ and evaluating the corresponding matrix elements as white-graph contributions of the form~\eqref{eq:pcut_white_graph}, one obtains a large set of terms that must be embedded onto the lattice. 
Each abstract perturbation parameter $\lambda_e$ in the product is associated with an edge $e=(u,v)$  connecting two vertices $u$ and $v$. 
Due to the long-range nature of interactions, which allow couplings between arbitrary pairs of lattice sites, there are infinitely many embeddings of the graph vertices onto the lattice. 
In the thermodynamic limit, the correct contribution is therefore obtained by summing over all such embeddings. 
Each vertex position gives rise to an infinite sum, subject to additional hardcore constraints imposed by the hardcore bosonic operators $t_i^{(\dagger)}$, which prohibit multiple vertices from occupying the same lattice site. 
See Fig.~\ref{fig:graph_embedding}\,(b) for two valid embeddings of the previous example graph on a square lattice. 
 
We can write physical quantities such as the ground-state energy density $\epsilon_0$ as a sum over graphs $\mathcal{G}$, where each contribution is weighted by its embedding factor $w(\mathcal{G},\Lambda)$ on a given lattice $\Lambda$,
\begin{equation}
		\epsilon_0 = - \frac{1}{2} + \sum_{\mathfrak{o}=2}^{\mathfrak{o}_{\text{max}}} \sum_{\mathcal{G}} w(\mathcal{G},\Lambda)E_{0}^{(\mathfrak{o})}(\mathcal{G})\ . 
	\label{eq:lr_embedding_ansatz}
\end{equation}
Here, the summand 1/2 is the contribution of the unperturbed part $H_0$, $w(\mathcal{G},\Lambda)$ denotes the embedding factor of a graph $\mathcal{G}$ on the lattice $\Lambda$, and $E_0^{(\mathfrak{o})}(\mathcal{G})$ denotes the ground-state energy contribution in order $\mathfrak{o}$ associated with the graph $\mathcal{G}$.
Note that evaluating the matrix elements on the graph level yields a sum of terms of the form~\eqref{eq:pcut_white_graph}. 
In contrast to Ref.~\cite{Adelhardt2024}, we treat the edge multiplicities $m_e$ as an intrinsic property of a graph. 
Graphs that share the same vertex and edge sets but differ in their multiplicities correspond to distinct perturbative processes. 
In previous work, in particular in Ref.~\cite{Adelhardt2024}, such contributions were grouped as different contributions on the same graph. 
For the approach developed here, it is necessary to distinguish them explicitly, and therefore we regard them as separate graphs from the outset.
The difference, however, is only conceptual and both formulations are equivalent.

For long-range interactions, the embedding factor reads
\begin{equation}
    w(\mathcal{G},\Lambda) = \sum_{c\in \mathcal{C}\vert_{\bm{i}_{n_s}=\bm{0}}} \frac{1}{s_{\mathcal{G}}} \,.
\end{equation}
We must divide by the symmetry factor $s_{\mathcal{G}}$ to avoid over-counting due to graph symmetries 
\footnote{
In Ref.~\cite{Adelhardt2024}, the embedding factor was incorrectly defined as a sum over all configurations $\mathcal{C}$, rather than over configurations restricted to $\bm{i}_{n_s}=\bm{0}$. 
Although this expression is mathematically incorrect, it corresponds to the computation in the Monte Carlo (MC) simulations. 
The reported MC results nevertheless remain valid, since the resulting extensive factor appears identically in both the target sum and the corresponding reference sum and therefore effectively cancels in their ratio.
}. 
We introduce the configuration space $\mathcal{C}$ of a graph $\mathcal{G}$ which are all possible graph embeddings with a given configuration $c$ specified by the set of vertex positions $\{\bm{i}_1, \bm{i}_2, \dots, \bm{i}_{n_s} \}$.
Here, $n_s$ denotes the number of vertices of a graph.
If we calculated the extensive ground-state energy $E_0$, we would just take the sum over the entire configuration space, involving all graph indices. 
However, for the ground-state energy density, we need to restrict the configuration space by fixing an arbitrary site $\mathcal{C}\vert_{\bm{i}_{n_s}=\bm{0}}$ to account for translational invariance.
The ground-state energy density then reads
\begin{equation}
	\begin{split}
		\epsilon_0 
        &= - \frac{1}{2} + \sum_{\mathfrak{o}=2}^{\mathfrak{o}_{\text{max}}} \sum_{\mathcal{G}} \sum_{c\in \mathcal{C}\vert_{\bm{i}_{n_s}=\bm{0}}} \frac{1}{s_\mathcal{G}} E_{0}^{(\mathfrak{o})}(\mathcal{G},c) \lambda^\mathfrak{o} \ . \\
	\end{split}
	\label{eq:lr_embedding_gs}
\end{equation}
At this stage, the abstract perturbation parameters $\lambda_e$ are replaced by the interaction kernel, making the dependence on the physical perturbation parameter $\lambda$ explicit and the graph contribution $E_{0}^{(\mathfrak{o})}(\mathcal{G},c)$ now directly depends on the current configuration $c$. For illustration purposes, we now explicitly consider an algebraically decaying power-law interaction for which the kernel in Eq.~\eqref{eq:H_TFIM} is given by 
\begin{equation}\label{eq:power-law-kernel}
     K(\bm r\neq \bm 0)= \frac{1}{|\bm{r}|^{\alpha}} \,, \quad K(\bm0) = 0\,.
\end{equation}
The decay exponent $\alpha$ tunes from nearest-neighbor interactions for $\alpha=\infty$ to uniform all-to-all coupling for $\alpha=0$. Often, the decay exponent $\sigma = \alpha - d$ is used in literature \cite{Dutta2001,Defenu2017,Koziol2021,Langheld2022}, where $d$ is the dimension. 
The contribution of a graph in a given configuration then takes the form
\begin{align}
	\frac{1}{s_{\mathcal{G}}}E_0^{(\mathfrak{o})}(\mathcal{G}, c) = v_{\mathcal{G}} \prod_{e=(u,v)\in \mathcal{E}} \frac{1}{|\bm i_v - \bm i_u|^{m_{e}\alpha}}\ ,
\end{align}
where $m_{e}$ again denotes the multiplicity of edge $e$ and $v_\mathcal{G}$ is a prefactor specific to the perturbative process on the graph.

We can write sums over the configuration space $\mathcal{C}$ more explicitly in terms of the vertex positions. 
Since each configuration corresponds to a choice of lattice sites for vertices $\bm i_1,\dots,\bm i_{n_s}$, the sum over $c\in \mathcal{C}$ is replaced by an independent sum over these vertex positions, subject to the hardcore constraint that no two vertices occupy the same site
\begin{align}
    \sum_{c\in \mathcal{C}}  \to \Hsum{\bm{i}_1, \dots, \bm{i}_{n_s}} \equiv\sum_{\bm{i}_1, \dots, \bm{i}_{n_s}}\prod_{u<v}(1-\delta_{\bm{i}_u,\bm{i}_{v}}) \ ,
    \label{eq:hardcore_sums}
\end{align}
where $({\rm H})$ indicates the exclusion of overlapping site positions, which is enforced by the Kronecker deltas in the product over all vertex pairs $u,v\in \mathcal{V}$ \footnote{Note that in previous publications \cite{Fey2019,Adelhardt2024} the notation $\sideset{}{'}{\sum}$ was used to denote the hardcore sum, but to avoid confusion with the mathematical literature we use the $({\rm H})$ superscript instead.}.
Restricting the configuration space $\mathcal{C}$ by fixing certain vertex positions (for example to $\mathcal{C}\vert_{\bm{i}_{n_s}=\bm{0}}$) excludes these from the summation while the product over vertex pairs remains unchanged.
For each graph, this yields a contribution of the form
\begin{align}
    v_\mathcal{G}\, {\rm H}^{\rm 0qp}(\mathcal{G})=  v_{\mathcal{G}} \hspace{1.5em} \Hsum{\bm{i}_1, \dots, \bm{i}_{n_s-1}} \hspace{.5em}\prod_{e\in\mathcal{E}} \frac{1}{\vert \bm{i}_v - \bm{i}_u \vert ^{m_{e}\alpha}} 
    \label{Eq:hardcore_sum_final_0qp}
\end{align}
that we need to compute to solve the embedding problem. 

To address the 1qp problem, we must encode the relevant information of hopping processes by explicitly specifying the initial and final vertices $\mu$ and $\nu$. 
This is achieved by introducing \textit{colored graphs} $\mathcal{G}_C$, in which these two vertices are distinguished by additional graph attributes (colors). 
Our white-graph approach essentially remains unchanged.
We still consider white-graph contributions with abstract perturbation parameters assigned to each edge.
The two vertices associated with the hopping process are now marked by the color attributes.

We want to calculate the 1qp dispersion defined in Eq.~\eqref{eq:1qp_ham}, which takes the form
\begin{align}
	\omega(\bm{k}) &= a(\bm{0}) + \sum_{\bm{\delta}\neq \bm{0}} a(\bm{\delta}) \mathrm{e}^{-\mathrm{i}\bm{k\delta}}\  \\
    &= a(\bm{0}) + 2\tilde{\sum_{\bm{\delta}\neq\bm{0}}} a(\bm{\delta}) \cos(\bm{k\delta})\ ,
	\label{eq:dispersion}
\end{align}
where the restricted sum $\tilde{\sum}$ takes into account the hermiticity of hopping processes implying $a(\bm\delta)=a(-\bm\delta)$.

Analogously to the ground-state energy, we rewrite all desired hoppings $a(\bm\delta)$  in terms of an embedding problem
\begin{align}
     a(\bm\delta)  &= \delta_{\bm{0},\bm{\delta}} + \sum_{\mathfrak{o}=1}^{\mathfrak{o}_{\text{max}}} \sum_{\mathcal{G}_C } w(\mathcal{G}_C, \Lambda_C)\,t^{(\mathfrak{o})}_{\mu;\nu}(\mathcal{G}) \ ,
\end{align}
where $t_{\mu ; \nu}^{(\mathfrak{o})}(\mathcal{G})$ denotes hopping processes at a given perturbative order $\mathfrak{o}$ and $\delta_{\bm{0},\bm{\delta}}$ arises from the unperturbed part similar to the constant term in the 0qp case. 
The vertices $\mu$ and $\nu$ are mapped to physical lattice sites $\bm{i}_\mu$ and $\bm{i}_\nu$, so that $\bm{\delta} = \bm{i}_\nu - \bm{i}_\mu$. The embedding thus corresponds to a embedding problem with color-matching between the (vertex-)colored graph $\mathcal{G}_C$ and a site-colored lattice $\Lambda_C$. 

For long-range interactions, we again need to replace the embedding factor $w(\mathcal{G}_C, \Lambda_C)$ by a sum over all configurations. 
This time, we replace
\begin{equation}
\begin{split}
\sum_{\mathcal{G}_C}w(\mathcal{G}_C, \Lambda_C) &= \sum_{\mathcal{G}_C}\sum_{c\in \mathcal{C}\vert_{\bm{i}_\mu=0,\bm{i}_\nu=\bm{\delta}}} \hspace{.2em}\frac{1}{s_{\mathcal{G}}} \\
&= \sum_{\bar{\mathcal{G}}_C} \sum_{c\in \mathcal{C}\vert_{\bm{i}_\mu=0,\bm{i}_\nu=\bm{\delta}}} \hspace{.2em}\frac{s_{\mathcal{G}_C}}{s_{\mathcal{G}}} \,, 
\end{split}
\label{eq:hopping_contribution}
\end{equation}
where we need to restrict the configuration space $\mathcal{C}$ by fixing the hopping sites to $\bm{i}_\mu=\bm{0}$ and $\bm{i}_\nu=\bm{\delta}$ accounting for the hopping constraint and translational invariance. 
From the first to the second line in Eq.~\eqref{eq:hopping_contribution}, we exploit graph symmetries to reduce the number of of colored graphs from $\mathcal{G}_C$ to a symmetry-reduced subset $\bar{\mathcal{G}}_C$, while compensating for it by multiplying with the symmetry number of the colored graph $s_{\mathcal{G_C}}$.
Consider, for instance, a star graph with $n$ legs and a single central vertex with a hopping process from the central vertex to one of the legs. Due to symmetry the perturbative processes to any of the legs are identical, so it suffices to calculate the contribution once and multiply with the symmetry number of this specific colored graph.

In contrast to the short-range case, the sum over hopping distances $\bm{\delta}$ does not need to be treated separately. Instead, it is implicitly generated by the sum over all configurations, since long-range interactions allow for hopping processes between arbitrary lattice sites at any perturbative order. Thus, we can write
\begin{equation}
    \sum_{\bm{\delta}\neq \bm 0} \hspace{0.5em}\sum_{c\in \mathcal{C}\vert_{\bm{i}_\mu=0,\bm{i}_\nu=\bm{\delta}}} =  \sum_{c\in \mathcal{C}\vert_{\bm{i}_\mu=0}} 
\end{equation}
and the desired 1qp dispersion as
\begin{equation}
	\omega(\bm{k}) = 1 +  \sum_{\mathfrak{o}=1}^{\mathfrak{o}_{\text{max}}}  \sum_{ \bar{\mathcal{G}}_C } \sum_{c\in \mathcal{C}\vert_{\bm{i}_\mu=0}} \frac{s_{\mathcal{G}_C}}{s_{\mathcal{G}}}\cos(\bm k\bm\delta) t_{\mu;\nu}^{(\mathfrak{o})}(\mathcal{G},c) \lambda^{\mathfrak{o}} \, .
	\label{eq:lr_omega_intermediate}
\end{equation}

As before, considering power-law interactions and substituting the abstract white-graph contribution with their algebraic long-range form, we obtain
\begin{equation}
	\frac{s_{\mathcal{G}_C}}{s_{\mathcal{G}}}t_{\mu;\nu}^{(\mathfrak{o})}(\mathcal{G}, c) =v_{\mathcal{G}_C}\prod_{e\in \mathcal{E}} \frac{1}{|\bm i_v - \bm i_u|^{m_{e}\alpha}}\ ,
	\label{eq:parameter_sub}
\end{equation}
where $v_{\mathcal{G}_C}$ is the process-specific prefactor of the hopping. Finally, rewriting the sum over all configurations \mbox{$c\in \mathcal{C}\vert_{\bm{i}_\mu=0}$} as a sum over vertex positions under hardcore constraints as in Eq.~\eqref{eq:hardcore_sums} and relabeling the vertices such that $\mu$ is assigned index $n_s$, i.\,e.\,, $\bm i_\mu = \bm i_{n_s} = \bm 0$, we are left with the hardcore sum
\begin{align}
    v_{\mathcal{G}_C}\,{\rm H}^{\rm 1qp}(\mathcal{G}_C,\bm k)=v_{\mathcal{G}_C}\hspace{1.5em}\Hsum{\bm{i}_1, \dots, \bm{i}_{n_s-1}}  \mathrm{e}^{-\mathrm{i}\bm{k}\bm{\delta}}\prod_{e\in\mathcal{E}} \frac{1}{\vert \bm{i}_v - \bm{i}_u\vert ^{m_{e}\alpha}}\ ,
    \label{Eq:hardcore_sum_final_1qp}
\end{align}
that we again need to evaluate to solve the 1qp long-range embedding problem. The resulting hardcore sums $\rm{H}^{\rm 0qp}$ and $\rm{H}^{\rm 1qp}$ in Eqs.~\eqref{Eq:hardcore_sum_final_0qp} and ~\eqref{Eq:hardcore_sum_final_1qp} share the same structure and differ only in the presence of the momentum-dependent phase $\mathrm{e}^{-i\bm{k} \bm{\delta}}$. 
Indeed, the 0qp case is recovered from the 1qp expression by setting $\bm k = \bm 0$ and $\mu=\nu$ which renders the phase trivial and the coloring redundant. 
We treat both cases on the same footing in the following, writing $\mathcal G$ for graphs that carry either no (0qp) or two (1qp) colored vertices and considering the coloring $\mu,\nu$ as graph property. 
The central computational task is then to evaluate the unified hardcore sums 
\begin{align}
    \mathrm H(\mathcal G,\bm k) &= \hspace{1.5em}\Hsum{\bm{i}_1, \dots, \bm{i}_{n_s-1}}   \mathrm{e}^{-\mathrm{i}\bm{k}(\bm{i}_\nu - \bm{i}_\mu)} \prod_{e\in\mathcal{E}} \hspace{.2em}\frac{1}{\vert \bm{i}_v - \bm{i}_u\vert ^{m_{e}\alpha}}\ 
    \label{Eq:hardcore_sum_final_unified}
\end{align}
for all relevant graphs, adding them up with the perturbative prefactors $v_\mathcal{G}$ to obtain the final series coefficients. 

Although we explicitly considered power-law long-range interactions in the last steps, we can easily write down the graph hardcore sums for general regularized, absolutely summable interaction kernels $K(\bm r)$ depending only on relative distances as
\begin{align}
    \mathrm H(\mathcal G,\bm k) = \hspace{1.5em}\Hsum{\bm{i}_1, \dots, \bm{i}_{n_s-1}}  \mathrm{e}^{-\mathrm{i}\bm{k}(\bm{i}_\nu - \bm{i}_\mu)} \prod_{e\in\mathcal{E}} \hspace{.2em} K_e(\bm i_v - \bm i_u)\,.
    \label{Eq:hardcore_sum_general}
\end{align}
Apart from the important instance of the power-law kernel $K_e(\bm r) = |\bm r|^{-m_e\alpha}$ (with ${K_e(\bm 0)=0}$), other examples include finite-range interactions, screened interactions with ${K_e\sim \mathrm{e}^{-\gamma m_e|{\bm r}|} |{\bm r}|^{-m_e\alpha}} $, and anisotropic dipolar interactions ${K_e\sim (1-3\cos^2\theta)^{m_e} |{\bm r}|^{-3m_e}} $ relevant to experimental platforms based on ultracold dipolar atoms (e.\,g. Erbium) in optical lattices \cite{Chomaz2022,Su2023}, dipolar molecules \cite{Carr2009,Moses2015,Moses2016,Bohn2017,Christakis2023}, or Rydberg atom experiments \cite{Browaeys2020}.

The calculation of such high-dimensional sums subject to hardcore constraints is a non-trivial task and conventional, brute force numerical summation techniques eventually fail 
\cite{Fey2016,Fey2019}. 
Monte Carlo (MC) methods developed for long-range power-law interactions provide a viable alternative, allowing the computation of high-dimensional sums with sufficient precision on one- and two-dimensional lattices~\cite{Adelhardt2024}. This approach, referred to as pCUT+MC~\cite{Adelhardt2024}, has been successfully applied to a range of long-range interacting models, including quantum spin models with Ising, XY, and XXZ interactions~\cite{Fey2019,Koziol2019,Adelhardt2020,Langheld2022,Adelhardt2023,Adelhardt2024,Adelhardt2025}, as well as spin-one models~\cite{Adelhardt2026}.

The advantage of a MC-based approach is that the structure imposed by the hardcore constraints can be exploited.
All contributions arising from graphs with the same number of vertices $n_s$ can be combined into a single integrand and sampled collectively. Consequently, it is sufficient to perform MC simulations for each perturbative order and fixed number of sites $n_s$. However, this approach has significant drawbacks as well. MC simulations converge slowly, with statistical errors scaling as $\sim N_{\text{steps}}^{-1/2}$, and reliable results require averaging over multiple runs with independent random seeds \cite{KrauthBook}. Moreover, since higher perturbative orders involve graphs with more sites, the dimension of each MC summation grows with order. This leads to substantial computational cost, requiring resource-expensive HPC-scale runs. Simulations must be repeated for each set of parameters, including the decay exponent $\alpha$, the momentum $\bm{k}$, and additional model parameters, such as interaction anisotropies. In practice, this restricts the method to a relatively limited set of accessible data points \cite{Adelhardt2025}.

These limitations motivate the development of alternative approaches that allow for a more efficient evaluation of the hardcore lattice sums in Eqs.~\eqref{Eq:hardcore_sum_final_unified} and \eqref{Eq:hardcore_sum_general}, ideally in analytically closed form.
This will allow for improved accuracy and a significantly denser sampling of physical observables at substantially reduced computational cost.

\section{Softcore Mapping: from hardcore to softcore sums}\label{Sec:softcore_mapping}

In this section we develop an exact reformulation of the hardcore lattice sums in Eq.~\eqref{Eq:hardcore_sum_general}.
The hardcore constraint in these sums explicitly prohibits vertices from occupying the same lattice site, so that no pair of summation indices can be associated with the same lattice site. 
However, this constraint is incompatible with the central mathematical objects used for the precise evaluation in this work, which require unconstrained lattice sums.
Therefore, the goal is to remove the explicit constraints between vertices such that the hardcore sums are reformulated into unconstrained \textit{softcore sums} of the form
\begin{align}
    {\rm S}(\mathcal{G},\bm k)&= \sum_{\bm{i}_1,\dots,\bm{i}_{n_s-1}}e^{-\mathrm{i}\bm{k}(\bm i_\nu - \bm i_\mu)} \prod_{e\in\mathcal{E}}  K_e({\bm i_v-\bm i_u}) \, .
	\label{eq:softcore_sum_gen} 
\end{align}
The \textit{softcore mapping} we develop for this purpose is purely combinatorial and independent of the choice of the interaction kernel $K_e(\bm r)$.
It rewrites the series coefficients of physical quantities, which are given by a weighted sum over hardcore sums $\mathrm H(\mathcal G, \bm k)$ with prefactors $ v_\mathcal{G}$ for all contributing graphs at a given perturbative order, as a weighted sum over softcore sums $\mathrm S(\mathcal G, \bm k)$ with prefactors $ v_{\mathcal{G}}^{(s)}$. 
Concretely, the mapping provides a linear map between the set of hardcore prefactors $v_\mathcal{G}$ and softcore prefactors $v_{\mathcal{G}}^{(s)}$ at each perturbative order. 

To gain an intuition for the mapping, we consider the sum associated with the (uncolored) circle graph with four sites with indices for the vertex set $\mathcal{V}=\{1,2,3,4\}$ with edge set \mbox{$\mathcal{E}=\{(1,2),  (2,3), (3,4), (4,1)\}$} depicted in Fig.~\ref{fig:graph_embedding}.
We assume each edge has multiplicity $m_e=1$ for simplicity, in general multiplicities differ.
For power-law long-range interactions, for instance, each edge contributes a factor $|\bm i_v - \bm i_u|^{-\alpha}$ to the summand.
Each hardcore sum can then be expressed in terms of a sum of softcore sums, 
\begin{widetext}
\begin{equation}
\begin{aligned}
\label{eq:examplehardcoresoftcore}\Hsum{}\left[\ringfouralpha \right]&=\primesum_{\bm{i}_1, \bm{i}_2, \bm{i}_3}{\vphantom{\sum}} (1-\delta_{\bm i_1, \bm i_3})(1-\delta_{\bm i_2, \bm i_4}) \frac{1-\delta_{\bm i_1, \bm i_2}}{|\bm i_1-\bm i_2|^\alpha} \frac{1-\delta_{\bm i_2, \bm i_3}}{|\bm i_2-\bm i_3|^\alpha} \frac{1-\delta_{\bm i_3, \bm i_4}}{|\bm i_3-\bm i_4|^\alpha} \frac{1-\delta_{\bm i_1, \bm i_4}}{|\bm i_4-\bm i_1|^\alpha}\\
     &= \sum{\vphantom{\sum}} \left[\ringfouralpha \right] -2	\sum{\vphantom{\sum}} \left[\linethreeTwoAlpha\right] + 	\sum{\vphantom{\sum}} \left[\linetwoFourAlpha\right]\,.
\end{aligned}
\end{equation}
\end{widetext}
The prime sum indicates that for power-law interactions we have $K_e(\bm 0)=0$ (c.\,f.\,Eq.~\eqref{eq:power-law-kernel}), \ie, we set the interaction kernel at $\bm{i}_u=\bm{i}_v$ to zero.
The hardcore sum $\sum{\vphantom{\sum}}^{\rm (H)}$ of a graph is expressed as a sum over softcore sums $\sum{\vphantom{\sum}}$ of graphs obtained by contractions of the original graph. 
Each softcore sum has a prefactor that is determined by the contraction procedure.

We define a valid contraction of a graph as the merging of two vertices into a single vertex, provided the two vertices are non-adjacent. This restriction ensures consistency with the exclusion of divergent summands arising from assigning identical positions to adjacent vertices in the lattice sum.
After performing a contraction, the resulting graph may contain multiple edges between the same pair of vertices. 
In such cases, these edges are merged into a single edge, and their multiplicities are summed.
We note that a representation in terms of multi-graphs (with multiple edges between two vertices) is always equivalent to a representation using the associated simple graphs with merged edge kernels.

When calculating contractions on a graph level, there will be multiple contraction paths (sequences of contractions) that lead to the same final contracted graph. 
To avoid overcounting, we consider only unique contraction results while keeping the vertex labeling of the original graph.
Representatives of the isomorphism equivalence classes of these unique contraction results determine the required softcore sums. 
The matrix elements of the linear map between hardcore and softcore sums are determined from the cardinality of the isomorphism equivalence classes.
Therefore, the central element of the mapping procedure is the calculation of unique contractions of graphs.

An efficient algorithm for calculating all unique contractions can be obtained by formulating the problem as a partitioning problem on the vertex set. 
The rule for this is that each subset of every resulting partition contains no two vertices connected by an edge of the graph.
We start with a graph $\mathcal{G}$ given by the set of vertices $\mathcal{V}$ with vertex colorings and the set of edges $\mathcal{E}$ with multiplicities.
We determine the set $\mathcal P$ of all partitions $P$ of $\mathcal V$ such that no subset $p\in P$ contains two distinct vertices $u,v$ connected by an edge, or equivalently, $(u,v)\notin\mathcal E$ for all distinct $u,v\in p$.
Each of the partitions $P\in \mathcal{P}$ is equivalent to a unique contraction of the original graph, which we reconstruct as follows.
For each subset $p\in P$, bijectively assign a unique index $\tilde{u}$ representing the vertices of the contracted graph.
This contracted graph contains an edge between vertices $\tilde{u}$ and $\tilde{v}$ if there is an edge in the original graph between indices in the subsets $p(\tilde{u})$ and $p(\tilde{v})$.
The multiplicity of an edge in the contracted graph is given by the sum of multiplicities of all edges between the subsets $p(\tilde{u})$ and $p(\tilde{v})$.
The color of a vertex $\tilde{u}$ in the contracted graph is given by the union of colors of all original vertices in the corresponding subset $p(\tilde{u})$.
This procedure yields a list of graphs $G_{\text{c}}(\mathcal{G})$ representing the unique contractions of $\mathcal G$. 
We show for the graph already used as an example in Eq.~\eqref{eq:examplehardcoresoftcore} the partitions with the resulting contractions in Eq.~\eqref{eq:allpartitionsexample}.

\begin{equation}
\begin{aligned}
		\footnotesize \mathcal{P}& \left[  \ringfouralpha \right] = \bigg\{ &&\\
				\footnotesize & \quad \big\{\{1\},\{2\},\{3\},\{4\}\big\}, \quad &\longrightarrow& \quad \ringfouralpha\\
				\footnotesize & \quad \big\{\{1\},\{3\},\{2,4\}\big\}, \quad &\longrightarrow& \quad \linethreeTwoAlpha\\
				\footnotesize & \quad \big\{\{1,3\},\{2\},\{4\}\big\},  \quad &\longrightarrow& \quad \linethreeTwoAlpha\\
				\footnotesize & \quad \big\{\{1,3\},\{2,4\}\big\}, \quad &\longrightarrow& \quad \linetwoFourAlpha \\
		\bigg\}& &&
\end{aligned}
\label{eq:allpartitionsexample}
\end{equation}

We can now further group these graphs into equivalence classes $x\in G_{\text{c}}(\mathcal{G})/\sim$, where isomorphic graphs, including their multiplicities and vertex colorings, are identified..
Isomorphic graphs represent the same softcore sum.
For the conversion from hardcore to softcore sums it is important to retain one representative $[x]$ of each equivalence class and the number of elements in each of them $|x|$.

Consider a \textit{mothergraph} $\mathcal{G}_0$ and representatives of all its contractions $\mathcal{G}_0,...,\mathcal{G}_N$. 
Note that we consider $\mathcal{G}_0$ also as a trivial contraction of itself.
We can now write a matrix that provides the hardcore to softcore sum mapping.
We write a matrix $\bm M$ where we arrange in each matrix element $\bm M_{nm}$ the number of graphs that are isomorphic to $\mathcal{G}_m$ when considering the unique contractions of $\mathcal{G}_n$.
If the graphs $\mathcal{G}_0,...,\mathcal{G}_N$ are ordered appropriately, this matrix $\bm M$ is an upper triangular matrix of integers with ones on the diagonal.
It is invertible using only integer operations.
The inverse matrix $\bm M^{-1}$ therefore expresses the the hardcore sums in terms of the softcore sums:
\begin{align}
\underbrace{\begin{pmatrix} \rm \Hsum{}\left[\mathcal{G}_0 \right]\\ \vdots\\ \hsum{}\left[\mathcal{G}_N \right]\end{pmatrix}}_{\hhsum{\vphantom{\sum}}[\bm{\mathcal{G}}]} = 
\bm{M}^{-1}\cdot
\underbrace{\begin{pmatrix} \rm \sum{\vphantom{\sum}}\left[\mathcal{G}_0 \right]\\ \vdots\\ \rm \sum{\vphantom{\sum}}\left[\mathcal{G}_N \right] \end{pmatrix}}_{\rm \sum{\vphantom{\sum}}[\bm{\mathcal{G}}]}\, .
\end{align} 
In this construction, Eq.~\eqref{eq:examplehardcoresoftcore} corresponds to multiplying the appropriate row in $\bm M^{-1}$ by $\sum{\vphantom{\sum}}[\bm{\mathcal{G}}]$ to obtain the desired $\rm \sum{\vphantom{\sum}}^{\rm (H)}[\mathcal{G}]$. 
Thus, determining the mapping for a single mothergraph also determines the mappings for all of its contractions.

In the perturbative series expansions considered in this work, the final results for physical quantities are obtained by taking a scalar product of the vector of hardcore coefficients $\bm{v}^{\text{T}}_{\bm{\mathcal{G}}}$ (obtained from perturbation theory) with the hardcore graph sums $\sum{\vphantom{\sum}}^{\rm (H)}[\bm{\mathcal{G}}]$.
We can now use the softcore mapping
\begin{align}
    \bm{v}^{\rm T}_{\bm{\mathcal{G}}} \cdot \Hsum{} [\bm{\mathcal{G}}]= \underbrace{\bm{v}^{\rm T}_{\bm{\mathcal G}}\bm M^{-1}}
    _{\bm{v}^{(s)\rm T}_{\bm{\mathcal G}}} \cdot\ \sum\vphantom{\sum}[\bm{\bm{\mathcal{G}}}]
\end{align}
to define softcore coefficients ${\bm{v}^{(s)\rm T}_{\bm{\mathcal{G}}}}$.
This allows the final calculation to be performed entirely in terms of softcore graph sums.
As an example, we provide the matrix $\bm M$ again for the four-site circle graph used as a mothergraph:
\begin{align}
\begin{pmatrix} \rm \tiny \sum\left[ \ringfouralpha\right]\\ \\\tiny \sum{\vphantom{\sum}}\left[ \linethreeTwoAlpha\right]\\ \\\tiny \sum{\vphantom{\sum}}\left[\linetwoFourAlpha\right]\end{pmatrix} &= 
\underbrace{\begin{pmatrix} 1 & 2 & 1 \\
0 & 1 & 1 \\ 0 & 0 & 1\end{pmatrix}}_{\bm M}
    \begin{pmatrix} \rm \hhsum{}\left[ \ringfouralpha\right]\\ \\\hhsum{}\left[ \linethreeTwoAlpha\right] \\ \\\hhsum{}\left[\linetwoFourAlpha\right]\end{pmatrix}
	\label{eq:mapplicationexample}
\end{align}

The softcore sums of all distinct valid contractions of a graph contribute to its hardcore sum. For example, a single 11-bond chain has 15\,541 distinct contractions. 
However, the complete 0qp and 1qp expansions require the evaluation of approximately as many nonzero softcore contributions as hardcore contributions, because many contractions coincide with graphs already present in the expansion \footnote{
This general property can be understood from the fact that a linked perturbative process in order $\mathfrak{o}$ is given by the action of $\mathfrak{o}$ bond operators, generating all contributing hardcore graphs.
Node contractions correspond to perturbative processes on smaller graphs (less number of vertices) with the same number of acting operators.
}.
The softcore mapping is an instance of what is known in combinatorics as Möbius inversion over set partitions \cite{rota1964}. The mapping is also independent of lattice and interaction kernel and can therefore be precomputed once and reused. While the mapping itself does not reduce the computational complexity of the resulting sums, it enables the block factorization introduced in the following section, thereby reducing the computational complexity.

So far, we have reformulated the graph embedding problem in terms of softcore sums free of explicit hardcore constraints.
We will now use recent mathematical developments in the evaluation of lattice sums and their meromorphic continuations \cite{buchheit2024epstein,buchheit2025,buchheit2026} to efficiently evaluate these softcore sums.

\section{Graph-zeta method for computing softcore sums: an overview}
\label{Sec:evaluation}
In this section we introduce the graph zeta method for the efficient and numerically exact evaluation of softcore graph sums. 
For a detailed description of the underlying mathematical framework, rigorous proofs, and the convergence analysis, see the companion paper \cite{buchheit2026}.
By this we enable the deterministic evaluation of graph embeddings for systems with long-range interactions.  
So far, brute force summation was only practical for the simplest possible embeddings in one dimension \cite{Fey2016,Fey2019}, while higher-dimensional problems required treatment by computationally expensive statistic Monte Carlo summations \cite{Fey2019,Adelhardt2024}.

With the graph zeta method, we create a paradigm shift in the study of general embedding problems, going well beyond a simple backend change from Monte Carlo to exact sums. 
It includes both quantitative and qualitative advancements.
On the quantitative side, computation times for state-of-the-art 0qp and 1qp series expansions for models such as the long-range transverse-field Ising model \cite{Adelhardt2024} are reduced from cluster-timescales \footnote{Typical calculations required on the order of $10^4-10^5$ core hours.} to minutes on a standard desktop hardware, even for three-dimensional lattices, while reproducing Monte-Carlo results at better precision. 
Block-factorizing the arising high-dimensional sums reduces scaling from exponential in the total number of nodes to exponential in the number of nodes in the largest block, while caching permits evaluating each block only once. 
Subsequently, each block is classified by its treewidth $\mathrm{tw}$ and evaluated using the cheapest applicable method, further reducing the scaling exponent substantially.
On the qualitative side, the method allows for an analytic study of the perturbative series, as the leading-order singularities in $\bm k$ are determined analytically, with important potential for extraction of quantum critical exponents. The study of these analytical properties is based on the analysis of what we call graph zeta functions, which encompass all softcore sums as special cases, and whose computation and assembly we discuss in this section.

The softcore lattice sum for a graph $\mathcal{G}$ reads
\begin{align}
	\label{eq:softcore_sum}
    {\rm S}(\mathcal{G},\bm k)&= \sum_{\bm{i}_1,\dots,\bm{i}_{n_s-1}}e^{-\mathrm{i}\bm{k}(\bm i_\nu - \bm i_\mu)} \prod_{e\in\mathcal{E}}  K_e({\bm i_v-\bm i_u}) \, .
\end{align}
In the following, we will put our focus on the evaluation of softcore sums for power-law long-range interactions. For distinction, we explicitly denote the regularized power-law kernel as $\mathcal{K}_{\gamma}$.
Most of the concepts discussed here, however, are interaction-agnostic and apply to any absolutely summable translationally invariant and even kernel $K_e$ and only the particular evaluation of individual blocks depends on the power-law form and must be adapted accordingly. 

The central idea of the graph zeta method is to exploit the graph structure underlying these lattice sums. 
In Sec.~\ref{sec:basic_idea}, we start by defining the central mathematical objects of the method, Epstein zeta functions and graph zeta functions, and discuss their relation to the graph lattice sums. 
We introduce a block decomposition to factorize graph lattice sums into elementary blocks in Sec.~\ref{sec:block_decomposition}. We classify the blocks according to their topology in Sec.~\ref{sec:block_categories}. Each block is evaluated using the most efficient available method depending on its category with most blocks computable analytically or semi-analytically in terms of generalized zeta functions. We discuss this in Secs.~\ref{sec:elementary_blocks} -- \ref{sec:tn_blocks}. Finally, we discuss further computational speedup via block caching in Sec.~\ref{sec:block_caching}. 
The purpose of this section is to establish an overview and the conceptual framework underlying the graph zeta method. We focus on the overall structure and recent methodological developments, while a detailed discussion of the evaluation of individual block categories is given in Sec.~\ref{sec:evaluation_details}.

Throughout this section, we use standard graph-theoretic notions such as cutvertices, block decompositions, series-parallel reduction, and treewidth. For background and an introduction to graph theory, see Refs.~\cite{diestel2025graph,bondy2008graph}.

\subsection{Epstein zeta functions and graph zeta functions}\label{sec:basic_idea}

The key idea of the graph zeta method is to exploit recent advancements in the efficient computation of Epstein zeta functions and their generalizations \cite{buchheit2024epstein,buchheit2025,buchheit2026}.
We therefore begin by introducing the central mathematical objects. First, we recall 
the definition of Epstein zeta functions \cite{epstein1903theorieI,epstein1903theorieII}
from Ref.~\cite{buchheit2024epstein},
\begin{align}\label{eq:zeta_k}
    Z_{\Lambda,\nu}({\bm k}) = \sum_{\bm x \in\Lambda} \mathrm{e}^{-\mathrm{i} \bm x\cdot \bm k}\, \mathcal{K}_\gamma(\bm x)\,,
\end{align}
where $\Lambda$ denotes a $d$-dimensional Bravais lattice, $\bm k \in \mathbb{R}^d$ the momentum and $\mathcal K_{\gamma}(\bm x \neq \bm 0)=|\bm x|^{-\gamma}$ with $\mathcal K_{\gamma}(\bm 0)=0$ is a regularized power-law kernel with $\gamma\in\mathbb{C}$. 

Second, we introduce the notion of graph zeta functions \cite{buchheit2026}.
For a graph $\mathcal{G}$ with two nodes $s,t\in\mathcal{V}$ specified as \textit{terminals}, the graph zeta function is given by
\begin{align}
    \zeta_{\mathcal G}(\bm k) = \sum_{\{\bm i_v\}_{v\neq p}}e^{-\mathrm{i} \bm x_{(s,t)}\cdot \bm k} \prod_{e\in \mathcal E} \mathcal K_{\gamma_e}(\bm x_e)\,.
\end{align}
Here, each edge $e=(u,v)$ is associated with an exponent $\gamma_e>d$ and a vector $\bm x_{e} = \bm i_v - \bm i_u$. 
Due to translational invariance one \textit{pinning vertex} $p$ is removed from the summation over the positions of all vertices, denoted by $\{\bm i_v\}_{v\neq p}$. 
The position of the pinning vertex is fixed to an arbitrary lattice point, e.\,g.\,, zero. 
The choice of pin does not alter the sum, which has large implications for the computational complexity, as we will see in the following.

Both Epstein zeta functions and graph zeta functions are closely related to the softcore sums in Eq.~\eqref{eq:softcore_sum}.
To make this relation more explicit, we perform a change of variables from vertex coordinates $\bm{i}_u, \bm i_v$ to edge-distance variables $\bm x_e = \bm i_v - \bm i_u$.
For a graph with $n_s$ nodes the number of independent vertex summation variables is ${n_s-1}$. 
We may therefore fix one vertex, say vertex $n_s$, at the origin, $\bm{i}_{n_s}=\bm{0}$. The remaining $n_s-1$ vertex coordinates can then be represented equivalently by $n_s-1$ independent distance vectors, $\bm{x}_1,\ldots,\bm{x}_{n_s-1}$. 
In this edge-distance representation, the softcore graph sum ${\rm S}(\mathcal{G},\bm{k})$ for power-law interactions in Eq.~\eqref{eq:softcore_sum} takes the form 
\begin{align}
    {\rm S}(\mathcal{G},\bm k) = \sum_{\{\bm{i}_v\}_{v\neq n_s}} \mathrm{e}^{-\mathrm{i} \bm x_{(\mu,\nu)}\cdot \bm k}\prod_{e\in\mathcal{E}} \mathcal{K}_{m_e\alpha}(\bm x_e)\,.
    \label{eq:softcore_sum_z}
\end{align}
Thus, the softcore lattice sum is a special graph zeta function with terminals given by the hopping nodes, $s\equiv\mu$ and $t\equiv\nu$ ($s=t$ for 0qp graphs) \footnote{We note that the 1qp hopping processes considered here have the symmetry $t_{\mu;\nu}=t_{\nu;\mu}$ due to the hermiticity of the Hamiltonian; for other observables like the spectral weight this may in general not be the case.}, pinning vertex $p=n_s$ and with per-edge exponent $\gamma_e=m_e\alpha$. In the following, we transition to using the more general terminology of graph zeta functions and terminals. Despite focusing on power-law edge kernels $K_e(\bm x) \equiv \mathcal{K}_{\gamma_e}(\bm x)$, we note that the techniques described in the following sections apply directly to edge kernels of the form
\begin{equation}
K_e(\bm x)=a(\bm x)+\sum_{j} b_j \mathcal K_{\gamma_j}(\bm x)\,,
\label{eq:gen_kernel}
\end{equation}
with $\gamma_j>d$ and $a(\bm x)$ short-ranged. 
Further, the techniques also apply to general summable interaction kernels $K_e$, where, however, individual evaluators require adaptation. 
 
If all edge-distances $\bm x_e$ were independent, the sum in Eq.~\eqref{eq:softcore_sum_z} would simply factorize into a product of Epstein zeta functions, which is true for $\mathcal G$ a tree, as we shall see in the next section. 
For more highly connected graphs, however, the edge-distance vectors are not all independent once $n_e>n_s-1$.
In this case, a subset of ${n_e-n_s+1}$ edge-distance vectors $\bm{x}_e$ can be expressed as linear combinations of the remaining ones.
These constraints are determined by the topology of the graph.
Consequently, the evaluation of the softcore sums is governed by the graph topology. 
This motivates the strategy of the graph zeta method: classify graphs according to their topology and evaluate each category by the most efficient available method, while caching reusable results.

\subsection{Block decomposition}\label{sec:block_decomposition}

\begin{figure}[t]
    \centering
    \includegraphics[width=\linewidth]{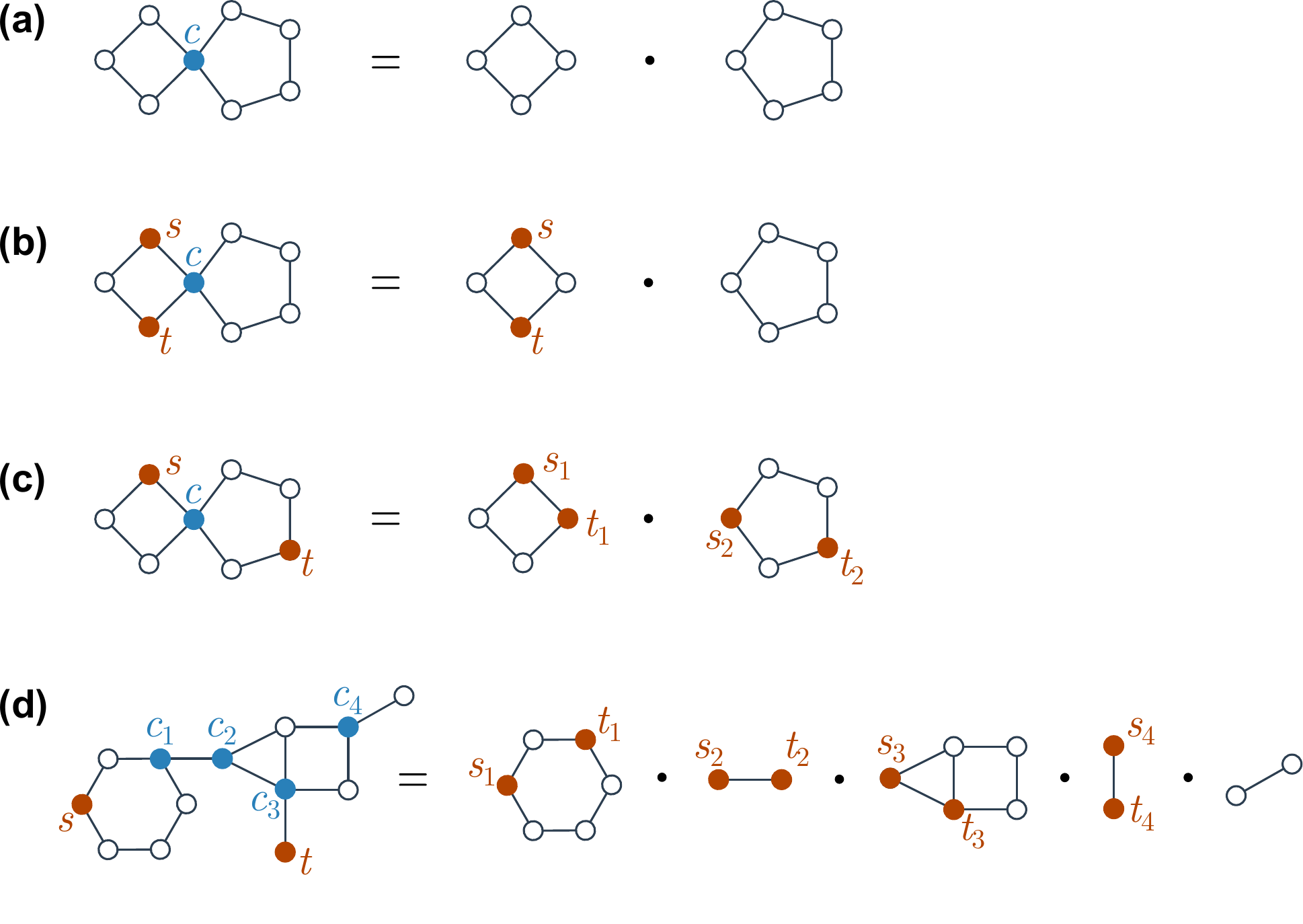}
    \caption{Block decomposition of a graph into (a)-(c) two and (d) multiple blocks. The respective graph zeta function factorizes into a product over the graph zeta functions for the individual blocks. 
    (a) For zero momentum or equivalently equal or no terminals, the graph zeta function factorizes at the cutvertex $c$ and both blocks contribute at zero momentum. (b) For both terminals within the same subgraph, that block contributes at the respective finite momentum while the other subgraph is treated as zero-momentum decoration. (c) For terminals in different subgraphs, the momentum is routed through the cutvertex and terminals $(s_1,t_1)$ and $(s_2,t_2)$ of the subgraphs are assigned accordingly. Both blocks contribute at finite momentum. (d) Graphs with more than one cutvertex, $c_1,\dots c_4$, are block-decomposed recursively. Blocks on the simple path between the terminals $s$ and $t$ constitute the spine $\mathcal{S}$ and are evaluated at finite momentum while remaining blocks constitute the off-path attachments $\mathcal{A}$ and are evaluated as zero-momentum decorations.}
    \label{fig:block_decomposition}
\end{figure}

A central simplification arises from the fact that, after the softcore mapping (see Sec.~\ref{Sec:softcore_mapping}), the lattice sums contain no explicit hardcore constraints. Together with the freedom to choose the pinning vertex $p$ due to translational invariance (see Sec.~\ref{sec:basic_idea}), this allows for a recursive factorization of the graph zeta functions into smaller substructures, based on the standard block-cut decomposition of graphs \cite{diestel2025graph}. 
For proofs of the following statements for general interaction kernels, see Ref.~\cite{buchheit2026}.
We call a vertex $c \in \mathcal{V}$ a \emph{cutvertex} (or articulation point) if $\mathcal{G}$ can be decomposed into two subgraphs $\mathcal{G}_1$ and $\mathcal{G}_2$ with disjoint edge sets that intersect only at $c$, see Fig.~\ref{fig:block_decomposition} (a)--(c).

At zero external momentum $\bm k = \bm 0$, choosing the pinning vertex at a cutvertex, $p=c$, immediately implies factorization of the graph zeta function,
\begin{equation}
    \zeta_\mathcal{G}(\bm 0) = \zeta_{\mathcal{G}_1}(\bm 0)\,\zeta_{\mathcal{G}_2}(\bm 0)\,.
\end{equation} 
This is visualized in Fig.~\ref{fig:block_decomposition} (a).
For finite momenta, terminals must be assigned consistently in the decomposition. 
We choose the subgraphs such that $s\in\mathcal{G}_1$. 
If both terminals $s,t$ lie in $\mathcal{G}_1$, the subgraph $\mathcal{G}_2$ serves as pure decoration and is evaluated at $\bm k = 0$ [see Fig.~\ref{fig:block_decomposition} (b)]. 
If $s$ and $t$ belong to different subgraphs, \ie, $s\in\mathcal{G}_1$ and $t\in\mathcal{G}_2$, the momentum is routed through the cutvertex [Fig.~\ref{fig:block_decomposition} (c)]. We obtain the terminals of the subgraphs as $(s_1,t_1) = (s,c)$ and $(s_2,t_2)=(c,t)$ and the factorization
\begin{equation}\label{eq:zeta_mult}
    \zeta_\mathcal{G}(\bm k) = \zeta_{\mathcal{G}_1}(\bm k)\,\zeta_{\mathcal{G}_2}(\bm k).
\end{equation}

For graphs with more than one cutvertex, $c_1,\dots,c_m$, this procedure can be repeated recursively, always moving the pinning vertex $p$ to the corresponding cutvertex. 
This procedure terminates at \emph{blocks}, \ie, graphs without cutvertices [see Fig.~\ref{fig:block_decomposition} (d)].
We call the set of blocks $B$ on the simple path between the terminals $s$ and $t$ ($s\to c_1\to \dots c_n\to t$) the \emph{spine} $\mathcal S$. 
Intuitively, if one thinks of $s$ and $t$ as the start and end point of a hopping process, this path can be interpreted as the hopping path the quasiparticle takes in this process and the spine as blocks or graph parts involved therein. 
Calling all remaining blocks of the graph the off-path \emph{attachments} $\mathcal A$, the graph zeta function factorizes as 
\begin{equation}
    \zeta_\mathcal{G}(\bm k) = \left( \prod_{B\in \mathcal S} \zeta_{B}(\bm k) \right) \left( \prod_{B\in \mathcal A} \zeta_{B}(\bm 0) \right).
\end{equation}

The block decomposition is computed with linear ${\mathcal{O}(|\mathcal V|+|\mathcal E|)}$ effort by Tarjan's depth-first search \cite{Tarjan1972}.
After the decomposition, the complexity of the evaluation is no longer determined by the total number of nodes $|\mathcal{V}|$, but only by the complexity of individual blocks.
This procedure applies to any interaction $K_e(\bm x)$ that only depends on relative distances, provided that we have access to an evaluator for its lattice Fourier transform.
Having reduced graph evaluations to the computation of their underlying blocks, we now proceed with a discussion of the individual block categories.

\subsection{Block categories}\label{sec:block_categories}

As elaborated on before, the evaluation of a graph depends on its topology, such that after block decomposition, the computational complexity is determined by the topology of the individual blocks rather than by the topology of the full graph. 
This suggests a classification of blocks according to the structure of their constraints which is naturally quantified by the \textit{treewidth} $\mathrm{tw}$ (see Ref.~\cite{diestel2025graph} for an introduction and Ref.~\cite{bodlaender1998} for its connection to computational complexity).
We now explicitly introduce the block categories relevant to the setting discussed in Sec.~\ref{sec:setting}. 
\begin{figure}[t]
    \centering
    \includegraphics[width=\linewidth]{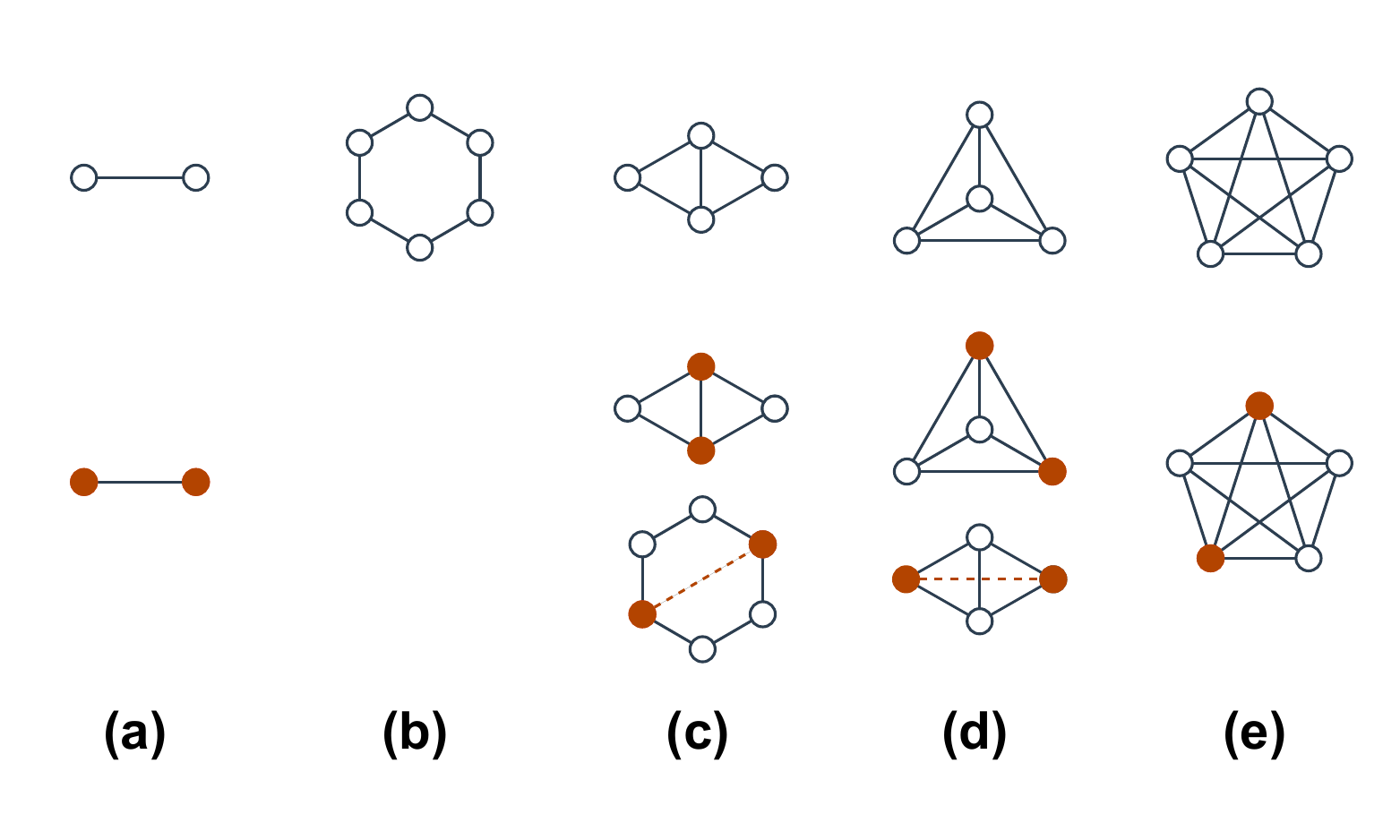}    
    \caption{Block categories for evaluation. Upper row: blocks without terminals. Lower row: blocks with terminals (colored nodes). The categories from (a)--(e) are: (A) bridges, (B) zero-momentum circle graphs, (C) series-parallel (SP) blocks, including finite-momentum circle graphs, and overlap graphs with (D) $\mathrm{tw}=3$ and (E) $\mathrm{tw}=4$. For blocks with terminals, we consider the augmented block with an additional edge between terminals (orange dashed line) for the classification.}
    \label{fig:block_categories}
\end{figure}

We first consider blocks without terminals. For representative examples for each category, (A)--(E) we refer to Fig.~\ref{fig:block_categories} (a)--(e). The simplest blocks are bridges, \ie, single edges, corresponding to $\mathrm{tw}=1$ and category (A). 
Blocks with $\mathrm{tw}=2$ contain loops. We distinguish between circle graphs, in which no edge belongs to more than one loop, and overlap graphs, in which at least one edge is shared by two or more loops. For vanishing external momentum (no terminals), circle graphs form category (B) while overlap graphs form category (C). The latter coincide precisely with series-parallel (SP) blocks and constitute the central class addressed by the zeta algebra below. 
The separation between categories (B) and (C) is motivated by implementation considerations, as the corresponding most efficient evaluation procedures differ substantially.
Blocks with higher treewidth exhibit increasingly nested loop overlap structures and blocks with $\mathrm{tw}=3$ and $\mathrm{tw}=4$ define categories (D) and (E), respectively.

For 1qp blocks, the positions of the terminals must additionally be taken into account. Here, complexity is determined by the treewidth of the augmented block $\mathcal {\tilde G}$ with an additional edge between the terminals
\begin{equation}
    \mathcal {\tilde G}=\mathcal G+(s,t), \quad s\neq t\,,
\end{equation}
and $\tilde {\mathcal G}=\mathcal G$ otherwise. In the following, we use the shortened notation
\begin{equation}
    \mathrm{tw}=\mathrm{tw}(\tilde{\mathcal{G}})\,.
\end{equation}
As a consequence, while a series-parallel block without momentum belongs to category (C), through an additional edge between the terminals it may either stay in category (C) or move to category (D) depending on the position of the terminals. Further, a circle graph without momentum lies in category (B), while a circle graph with momenta belongs to category (C). Representative examples are again illustrated in Fig.~\ref{fig:block_categories}.

The treewidth is bounded by the number of edges and therefore by the perturbative order, including the additional edge between terminals and counting edges with multiplicity $m_e$ as $m_e$-many edges.
Using perturbative continuous unitary transformations together with a white-graph expansion as described in Sec.~\ref{sec:setting}, the expansion for the TFIM is computationally feasible up to \mbox{$\mathfrak{o}_\text{max}\leq 13$} for the 0qp case and \mbox{$\mathfrak{o}_\text{max}\leq 11$} for the 1qp case, so that graphs have at most \mbox{$n_e=13$} edges. 
We find that at these orders, the maximally reached treewidth is $\mathrm{tw}=4$.
Regardless, the evaluation methods we employ for $\mathrm{tw}\geq3$ could in principle be applied to such blocks with higher treewidth $\mathrm{tw}\geq 5$. 

In the remainder of this section we briefly outline the evaluation strategy associated with each class. Detailed algorithms for the construction of the respective graph zeta functions are presented in Sec.~\ref{sec:evaluation_details}.

\subsection{Elementary blocks: analytical computation}\label{sec:elementary_blocks}
We start with elementary blocks that can be computed analytically. 
Bridges, category (A), are simply evaluated as Epstein zeta functions $Z_{\Lambda,\gamma}({\bm k})$ in Eq.~\eqref{eq:zeta_k}. The properties and efficient computation of the Epstein zeta function have previously been established in Ref.~\cite{buchheit2024epstein}, with a high-performance implementation available in EpsteinLib \cite{epsteinlib}. 

The second analytically computable block are circle graphs which evaluate at zero momentum in category (B). For such blocks the graph zeta function can be written as an efficiently computable integral over Epstein zeta functions \cite{buchheit2025,robles2025exact}, 
\begin{align}\label{eq:graph-zeta-circle}
    \zeta_\mathcal{G}(\bm 0) =\frac{1}{|\mathrm{BZ}|} \int_{\text{BZ}} \prod_{e\in\mathcal{E}} Z_{\Lambda,m_e\alpha}(\bm k)\,\mathrm{d}\bm k\,,
\end{align}
with $|\mathrm{BZ}|$ the volume of the Brillouin zone. For a derivation of this expression we refer to Sec.~\ref{sec:evaluation_details} and Ref.~\cite{buchheit2025}.

Both of these blocks evaluate analytically to machine precision even for exponents $\alpha$ close to the system dimension $d$.
Importantly, these blocks and their corresponding graphs form the objects that converge most slowly using direct methods for exponents close to the spatial dimension $d$. The graph zeta method thus transforms the hardest objects into the easiest ones to evaluate.

\subsection{Series-parallel blocks: semi-analytic zeta algebra}\label{sec:sp_blocks}
For series-parallel (SP) blocks, corresponding to treewidth $\mathrm{tw}=2$ and category (C), we develop an efficiently computable algebra based on products and convolutions of Epstein zeta functions. For details and proofs, see Ref.~\cite{buchheit2026}.

\begin{figure}[t]
    \centering
    \includegraphics[width=\linewidth]{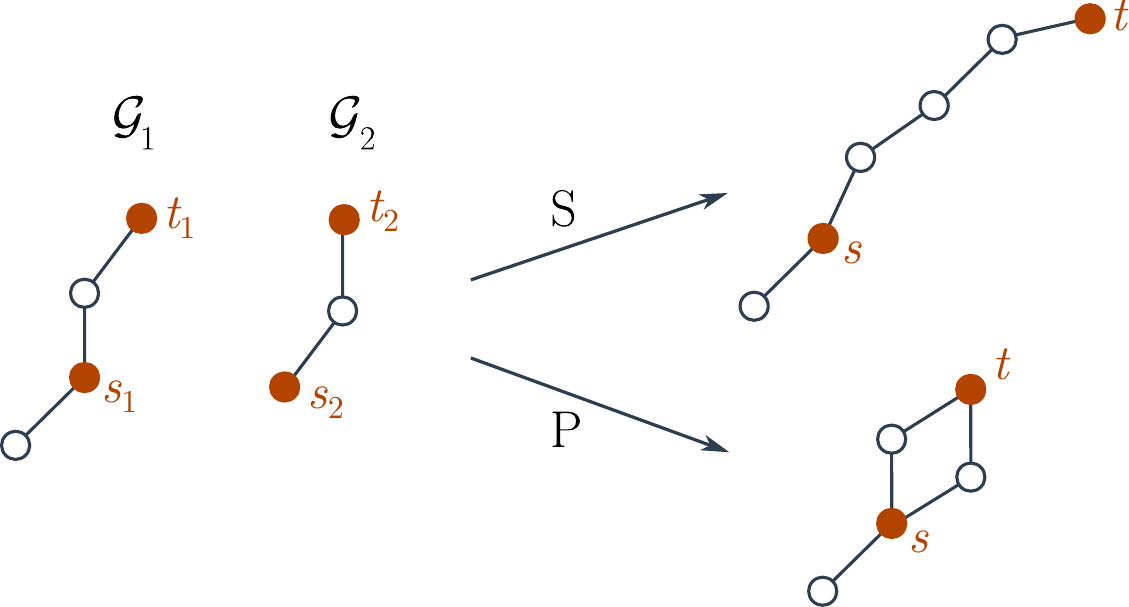}
    \caption{Serial (S) and parallel (P) composition of the two-terminal graphs $\mathcal{G}_1$ and $\mathcal{G}_2$ with terminals $(s_1,t_1)$ and $(s_2,t_2)$ depicted as colored nodes. Serial composition identifies the terminals $t_1=s_2$, yielding terminals $(s,t) = (s_1,t_2)$. Parallel composition identifies ${s=s_1=s_2}$ and $t=t_1=t_2$.} 
    \label{fig:sp_composition}
\end{figure}

A key result is that every SP block can be generated recursively from elementary bridge blocks using only the standard two-terminal graph compositions: serial composition and parallel composition. On the level of graph zeta functions, these operations induce a closed algebra generated by multiplication and convolution. For an illustration of serial and parallel composition see Fig.~\ref{fig:sp_composition}.

The serial composition of two graphs with terminals $(s_1,t_1)$ and $(s_2,t_2)$ identifies the vertices $t_1 = s_2$, yielding a graph with terminals $(s_1,t_2)$.
As discussed in the previous section, this corresponds to multiplication of the graphs [see Eq.~\eqref{eq:zeta_mult}].

The parallel composition identifies the $s$ and $t$ terminals respectively, such that $s\equiv s_1 = s_2$ and $t \equiv t_1 = t_2$. 
This corresponds to a convolution of the respective graph zeta functions over the Brillouin zone, 
\begin{equation}
    (\zeta_{\mathcal{G}_1} * \zeta_{\mathcal{G}_2})(\bm k)=\frac{1}{|\mathrm{BZ}|}\int_{\mathrm{BZ}}\zeta_{\mathcal{G}_1}(\bm k')\zeta_{\mathcal{G}_2}(\bm k-\bm k')\,\mathrm d\bm k'.
\end{equation}
Choosing the pinning vertex as $p=s_1=s_2$ ($\bm i_{s_1},\bm i_{s_2}=0$), this follows from the basic Fourier identity 
\begin{equation}
    \left(e^{- {\rm i} \bm k\cdot\bm i_{t_1}} *e^{-{\rm i} \bm k\cdot\bm i_{t_2}}\right)(\bm k)=e^{-{\rm i} \bm k\cdot\bm i_{t_1}}\delta_{\bm i_{t_1},\bm i_{t_2}}\,,
\end{equation}
where $\bm i_{t_1},\bm i_{t_2} \in \Lambda$. 

To compute the convolution efficiently, graph zeta functions are represented as
\begin{equation}\label{eq:graph_zeta_fourier}
    \zeta_\mathcal{G}(\bm k)=\sum_{\bm x\in\Lambda}e^{- {\rm i} \bm k\cdot\bm x}a(\bm x)+\sum_j b_j Z_{\Lambda,\gamma_j}(\bm k)\,,
\end{equation}
with coefficients $b_j\in \mathds R$, exponents $\gamma_j>d$ and Fourier coefficients $a(\bm x)\in \mathds R$. Note that this corresponds directly to the lattice Fourier transform of a general kernel of the form~\eqref{eq:gen_kernel}. Therefore, the generalization to such kernels is straightforward.
The Epstein zeta contributions capture the leading order singularities in $\bm k$, while the Fourier series represents the remaining regular part and decays rapidly. 
We note that for large exponents as well as finite-range interactions with $\alpha=\infty$ the convolution can already be computed precisely using the truncated Fourier series. In particular, exact integer evaluation is provided for the nearest neighbor case. For exponents close to the system dimension, however, the singularities in $\bm k$ need to be finely resolved. 
Importantly, this representation can be preserved under both multiplication and convolution, enabling an efficient semi-analytic evaluation of all SP blocks.

Any SP block can be constructed recursively from elementary bridges through recursive serial and parallel compositions, \ie, multiplications and convolutions of graph zeta functions. 
In Section \ref{sec:overlap_graphs} we develop an algorithmic language that encodes this recursive construction and directly translates the block into an evaluation sequence.
From a graph-theoretical perspective, the resulting procedure corresponds to a bottom-up traversal of the series-parallel reduction tree \cite{Duffin1965,ValdesTarjanLawler1982}.
This approach reduces the computational complexity of a SP block from exponential to linear in the number of nodes $|\mathcal{V}|$.

\subsection{Highly connected blocks}\label{sec:tn_blocks}

So far, we have computed simple bridges and cycles using analytic formulas, which have subsequently served as ingredients for the SP algebra, fully covering blocks with treewidth $\mathrm{tw}\leq2$. For larger treewidths, corresponding to categories (D) and (E) in Fig.~\ref{fig:block_categories} and beyond, the additional constraint structures cannot be generated solely by serial and parallel composition of two-terminal bridges anymore. Equivalently, such graphs contain $K_4$ minors (see the representative example for category (D) in Fig.~\ref{fig:block_categories}). Viewing the compositional algebra as a reduction process, we can first eliminate all 2-connected parts of the graph, yielding a highly-connected object. The following procedure then completes the method, providing access to any graph \cite{buchheit2026}. 

Consider the graph zeta function on a torus $\Lambda_n$ with $n$ grid points per dimension, 
\begin{equation}
    \zeta_\mathcal{G}^{(n)}(\bm k) = \sum_{\{\bm i_v \in \Lambda_n\}_{v\neq p}}e^{-i \bm x_{(s,t)}\cdot \bm k} \prod_{e\in \mathcal E} \mathcal K_{\gamma_e}( \bm x_e)\,.
\end{equation}
We use tensor network bucket elimination to reduce the scaling from exponential in the number of nodes $|\mathcal{V}|$ to exponential in the treewidth $\mathrm{tw}$ \cite{Dechter1999,MarkovShi2008}.

Each edge $e=(u,v)$ is represented by a two-index kernel tensor $T_{u,v}$, while each vertex coordinate corresponds to a tensor index. The graph zeta function is obtained by contracting the resulting tensor network over all internal (non-terminal) vertex indices, while pinning the terminal position $\bm i_s$ to zero. Eliminating the other terminal $t$ last through FFT yields the full momentum grid at once.

It is a standard result that an optimal contraction order evaluates the sum in $n^{(\mathrm{tw}+1)d}$ operations and $n^{\mathrm{tw}\,d}$ memory. Using that translational invariance of the initial tensors is preserved under repeated contractions, one vertex position can always be fixed to zero, reducing the exponent of work and memory by one unit.
Additionally, whenever the highest-width contraction is a convolution, using FFT saves a further order in work, reducing the scaling in work to $n^{(\mathrm{tw}-1)d}\log (n^d)$. We find that this optimal scaling is obtained for more than $99\,\%$ of graphs occurring in the TFIM in the orders considered here, in particular for all graphs with $\mathrm{tw}=4$. Since $\mathrm{tw}\leq 4$ for all these graphs, numerical work increases at worst with cubic law.
For an equivalent representation in momentum space, $\mathrm{tw}-1$ momentum channels are required. For $\mathrm{tw}=2$, the resulting operations are equivalent to the single-momentum multiplications and convolutions we derived in the previous section. 

It is important to note that the zeta algebra from Sec.~\ref{sec:sp_blocks}, providing an analytic treatment of long-range kernel tails, is not yet available for $\mathrm{tw}> 2$. This is due to a more complicated singularity structure arising for the tensor network. However, this is not problematic as series-parallel reduction, seen as a preprocessing step for $\mathrm{tw}>2$ blocks, results in highly connected (at least 3-connected) lattice sums which converge rapidly on their own. We further accelerate convergence using Richardson extrapolation \cite{sidi2003practical}. 

\subsection{Block caching}\label{sec:block_caching}
For a given perturbative order, the final series coefficient is obtained from a large number of graph contributions which are summed over after individual evaluation. 
A further substantial reduction of the computational cost is achieved by exploiting redundancy at the level of blocks.
While there is little redundancy between the graphs themselves, the redundancy between the blocks is high.
We therefore cache block evaluations and identify isomorphic blocks \cite{Shervashidze2011WL,Cordella2004VF2}. 
Since the number of graphs grows rapidly with perturbative order, this reuse of block evaluations significantly reduces the overall computational cost. We further cache Epstein zeta grid evaluations and series-parallel blocks occurring within tensor contractions.

\section{Explicit construction of graph zeta functions}\label{sec:evaluation_details}

In the previous section we established the conceptual framework of the graph zeta method by introducing the block decomposition, classifying blocks according to their topology and establishing the mathematical groundwork for the evaluation of the block categories.

The central task of this section is to turn these concepts into an explicit computational framework for systematically assembling graph zeta functions for increasingly complex graph topologies.
We develop their practical construction and evaluation from elementary bridges (single edges), providing explicit algorithmic recipes together with illustrative examples throughout. 

We begin in Sec.~\ref{sec:basic_evaluation_categories} with the simplest blocks---elementary bridges and zero-momentum circle graphs---and illustrate how these results suffice for the fully analytic evaluation of any tree graph. 
In Sec.~\ref{sec:overlap_graphs} we discuss the construction of the remaining, more difficult blocks---SP blocks and blocks of higher-treewidth---from elementary bridges. 
The resulting algorithms provide a systematic procedure for assembling graph zeta functions for SP blocks. 
The remainder of this section develops them in detail.
This framework is implemented in the open-source high-performance Graph Zeta Library \cite{gzl2026}, which computes the graph zeta function on the full momentum grid for arbitrary combinations of power-law and short-range edge kernels as in Eq.~\eqref{eq:gen_kernel}, arbitrary $d$-dimensional lattices, and an arbitrary input graph. 
Beyond single graphs, it also allows for the computation of full 0qp and 1qp series expansions for gapped quantum systems, starting from the associated graphs and combinatorial prefactors, while exploiting redundancies through caching. 
In the following, for simplicity of presentation, we focus on power-law-only kernels.

\subsection{Analytic evaluation of elementary bridges, tree graphs and zero-momentum loops}\label{sec:basic_evaluation_categories}

The elementary block category (A) depicted in Fig.~\ref{fig:block_categories} (a) consists of bridges, \ie, graphs given by a single edge.
The graph zeta function directly reduces to an Epstein zeta function, 
\begin{align}
    \zeta_\mathcal{G}(\bm k) = Z_{\Lambda,m_e\alpha}(\bm k)\,,
\end{align}
where $\bm k=0$ if the bridge has terminals $s=t$ (or no terminals). 
An analytic evaluation of bridges is enabled by the implementation of Epstein zeta functions in EpsteinLib \cite{epsteinlib}. 

Tree graphs represent the simplest nontrivial application of the block decomposition. Every tree decomposes into bridge blocks and its graph zeta function thus factorizes completely into bridge contributions.
For finite momenta $\bm k$, the bridges on the unique path connecting the terminals $s$ and $t$ carry the external momentum and constitute the spine $\mathcal{S}$, while remaining edges are attachments $\mathcal{A}$ that contribute at zero momentum (see Sec.~\ref{sec:block_decomposition}). 
The resulting graph zeta function is therefore
\begin{align}
    \zeta_\mathcal{G}(\bm k) = \left( \prod_{e\in\mathcal{S}} Z_{\Lambda,{m_e\alpha}}(\bm{k}) \right) \left(\prod_{e\in\mathcal{A}} Z_{\Lambda,{m_e\alpha}}(\bm 0)\right)\,.
\end{align}

\begin{figure}[t]
    \centering
    \includegraphics[width=.85\linewidth]{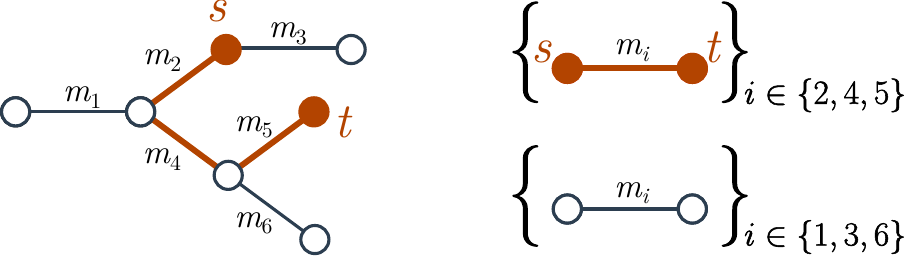}
    \caption{Full block-decomposition of a tree graph into elementary bridge blocks. Terminals $s,t$ are depicted as orange nodes with the simple path connecting them highlighted as orange edges. The multiplicities of the edges are denoted by $m_1,\ldots,m_6$. By separation at the cutvertices the graph decomposes into six bridges. Bridges along the path $s\to t$ form the spine $\mathcal{S}$ and carry momentum while remaining bridges are the attachment $\mathcal{A}$ and evaluated at zero momentum.}
    \label{fig:tree_graph}
\end{figure}

An example is sketched in Fig.~\ref{fig:tree_graph}. 
Hence, tree graphs evaluate fully analytical with computational effort linear in the number of nodes, $n_s-1$.

The simplest non-tree-like graphs are circle graphs in category (B) with $n_e=n_s$ edges and sites (see Fig.~\ref{fig:block_categories} (b)). We consider only the zero-momentum case, since circle graphs with a finite momentum admit a different evaluation and belong to category (C), and derive the graph zeta function given in Eq.~\eqref{eq:graph-zeta-circle}.

Circle graphs are subject to a single constraint such that only $n_e-1$ of the edge-distance variables $\bm x_e$ are independent.
Equivalently, a circle graph can be viewed as an open chain with the additional closure constraint $\sum_{e}\bm x_e = \bm 0$.
Expressing the constraint as a Kronecker delta, the graph zeta function takes the form 
\begin{align}
      \zeta_{\mathcal G}(\bm 0) = \sum_{\{\bm x_e\}_{e}}\prod_{e\in \mathcal E} \mathcal K_{\gamma_e}(\bm x_e) \,\delta_{\bm 0,\sum_{e}\bm x_e}\,.
\end{align}

As discussed in detail in Ref.~\cite{buchheit2025}, this constraint can be represented in Fourier space, restoring factorization at the cost of an additional integration over the Brillouin zone
\begin{align}
    \zeta_\mathcal{G}(\bm 0) =  \frac{1}{|\rm BZ|} \int_{\text{BZ}} \prod_{e\in\mathcal{E}} Z_{\Lambda,m_e\alpha}(\bm k)\,\mathrm{d}\bm k\,,
\end{align}
where $|\rm BZ|$ is the Brillouin-zone volume. 
This provides an analytic Fourier-space representation of the 0qp circle graph contribution in terms of products of single-edge Epstein zeta functions.

\subsection{Constructing series-parallel and higher-treewidth blocks}\label{sec:overlap_graphs}

In this subsection, we discuss the evaluation of SP blocks and higher-treewidth blocks from categories (C)--(E), \ie, graphs with overlapping loops and finite-momentum circle graphs.
For obtaining the graph zeta function for these block the structure of these overlaps is essential.
To expose this structure, we introduce a reduced representation, the \textit{construction graph}, which encodes how a given graph is assembled from elementary bridges and is thus agnostic of the interaction.

\subsubsection{Determination of the construction graph}\label{sec:construction_graph}

The \textit{construction graph} is a reduced representation of a $\mathrm{tw}\geq 2$ graph that encodes the structural information required for its evaluation. 
The aim is to separate the structure of overlaps between loops, and (in 1qp graphs), the location of external momentum, from geometric details such as loop size or edge multiplicities. 

\begin{figure}[t]
    \centering
    \includegraphics[width=.95\linewidth]{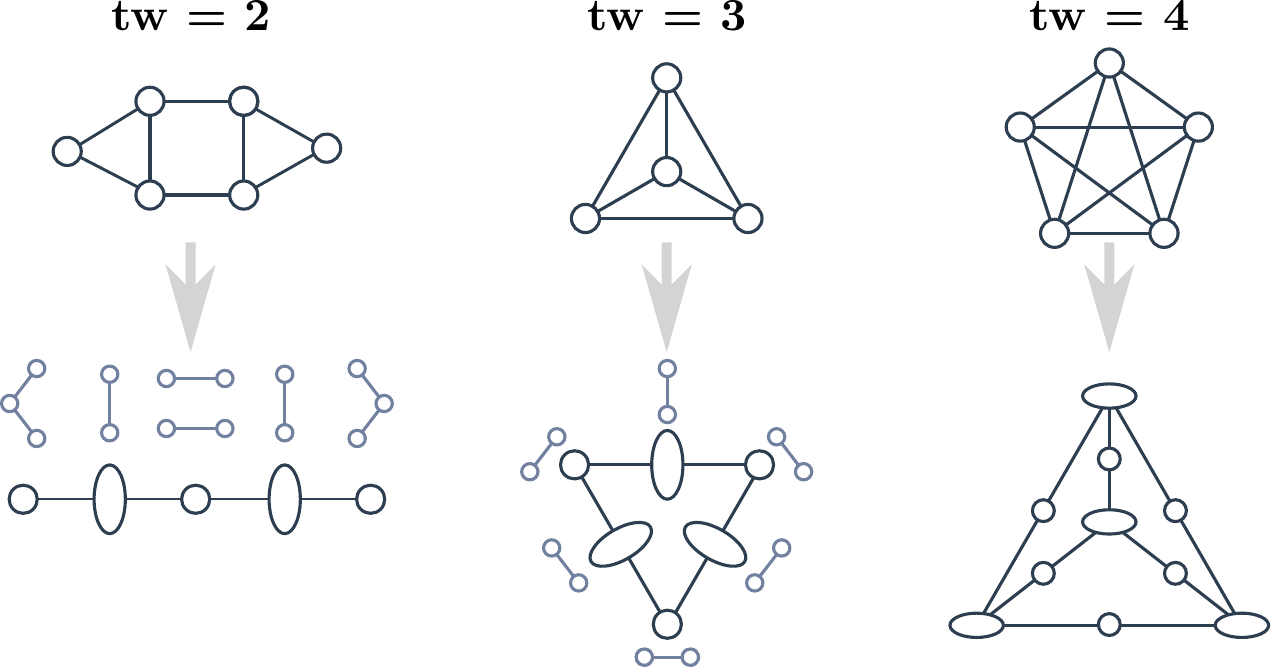}
    \caption{Representative graphs with $\mathrm{tw}\in\{2,3,4\}$ from categories (C), (D) and (E), respectively, from Fig.~\ref{fig:block_categories} (upper row) alongside the respective construction graphs determined by applying the algorithm described in the main text (lower row). Circular nodes in the construction graph are cycle nodes that represent cycles of the original graph while oval nodes are segment nodes which represent segments of overlapping edges between cycles in the original graph. The gray subgraphs at the nodes of the construction graph depict which part of the original graph is represented by the respective node. For category (E) with $\mathrm{tw}=4$, we drop the explicit depiction of the node-subgraphs for readability. Each node represents a single edge of the original graph like in category (D), $\mathrm{tw}=3$. Cycle nodes only represent the subgraph that is contained exclusively by that cycle such that each edge of the original graph is represented by exactly one edge of the construction graph.}
\label{fig:construction_graph_categories}
\end{figure}

\begin{figure}[t]
    \centering
    \includegraphics[width=\linewidth]{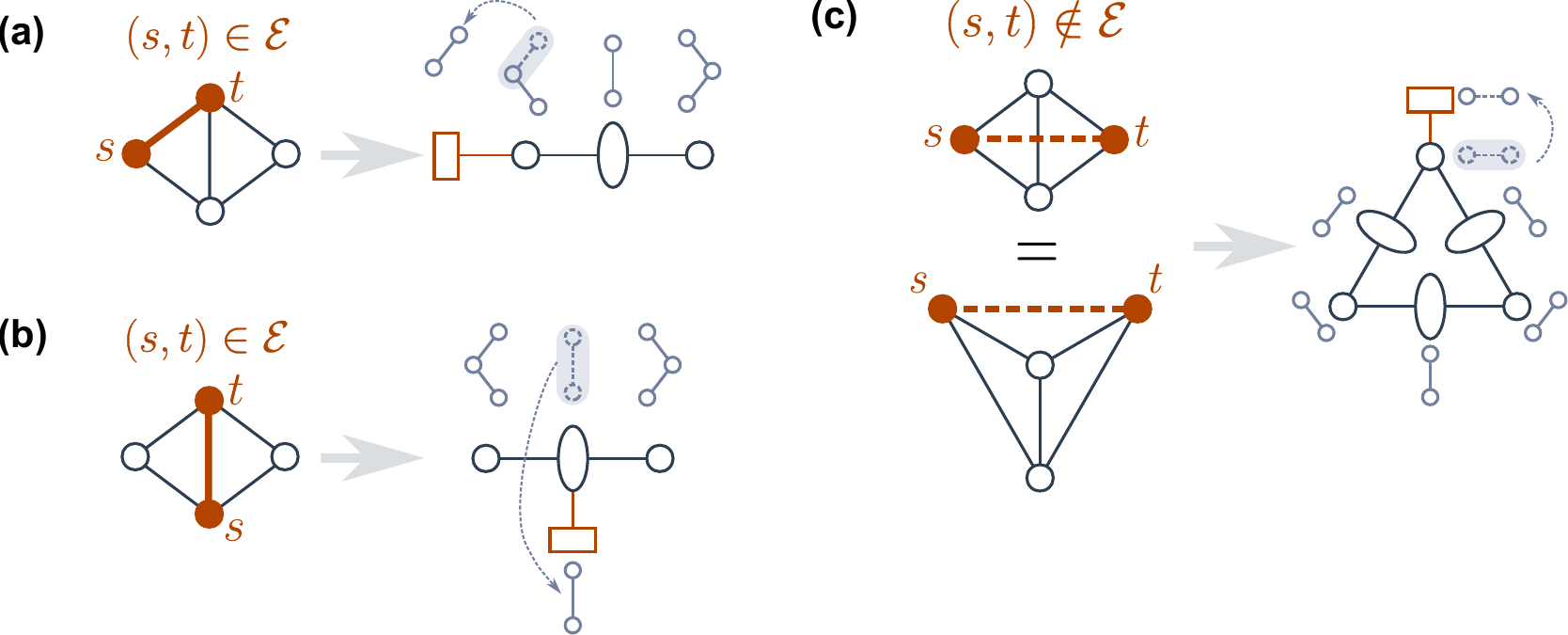}
    \caption{Dependence of the 1qp construction graph on the position of terminals $s,t$. (a)--(c) Three 1qp graphs with the same underlying 0qp graph corresponding to three distinct hopping processes from $s$ to $t$. For each case, the original graph is shown on the left and the construction graph on the right. The gray subgraphs at the construction graph nodes depict which part of the original graph is represented by the respective node. (a) and (b) The edge $(s,t)\in\mathcal{E}$ is part of the graph. First, the construction graph is determined according to the 0qp algorithm. 
    The 1qp construction graph is given by moving the edge $(s,t)$ in the construction graph from its original node to a new hopping-node (orange rectangle) as indicated by the arrow and the shaded, dashed graph part. The hopping node is connected to the original node. 
    (c) The edge $(s,t)\notin\mathcal{E}$ is not a graph edge and needs to be added as an additional edge with multiplicity $m_{(s,t)}=0$ prior to determining the construction graph, indicated as orange dashed line. The construction graph is determined as before.}
    \label{fig:SP_evaluation}
\end{figure}

We first define the construction graph for graphs without terminals. 
We then extend the algorithm to 1qp graphs and encode the terminals between which the external momentum is associated. 
In this sense, the 1qp construction graph is an extension of the 0qp case, using identical structural rules and adding a single element encoding the terminals.
For an illustration of representative construction graphs see Figs.~\ref{fig:construction_graph_categories} and \ref{fig:SP_evaluation}.

\paragraph{0qp graphs}
For zero-momentum graphs, the construction graph contains two types of nodes: cycle-nodes, which represent loops in the original graph, and segment-nodes, which represent segments of overlapping edges shared by multiple loops. 

The algorithm to determine the construction graph consists of the following two steps:
\begin{enumerate}
    \item Choose a cycle basis of the block. For each cycle $i$ in the basis, define one \textit{cycle node} $c_i$ representing the edges contained exclusively in that cycle (see circles in Fig.~\ref{fig:construction_graph_categories}).
    \item For each maximal segment of edges that is contained in two or more basis cycles $j_1,j_2,\ldots$, define a \textit{segment node} $s_{j_1,j_2,\ldots}$ representing the respective shared edges (see ovals in Fig.~\ref{fig:construction_graph_categories}). Connect the segment node to all corresponding cycle nodes $c_{j_1},c_{j_2},\ldots$.
\end{enumerate}

\paragraph{1qp graphs} 
For 1qp graphs, the construction graph contains an additional node type, the \textit{hopping node}, which encodes the position of the terminals $(s,t)$ (or hopping nodes $\mu,\nu$) and thus of the finite external momentum. 

The algorithm to determine the construction graph is an extension of the procedure for a 0qp graph, illustrated in Fig.~\ref{fig:SP_evaluation}:
\begin{enumerate}[start=1]
  \item If $(s,t)\notin\mathcal{E}$, augment the graph by adding an \textit{additional edge} $e_\text{add}=(s,t)$ with multiplicity $m_{e_\text{add}}=0$ (or, more generally, unit weight kernel). This ensures $s$ and $t$ are in the same cycle and guarantees a designated edge representing the terminal pair for step 3, to which the momentum can subsequently be assigned.
  \item Determine the construction graph according to the 0qp algorithm.
  \item Define a \textit{hopping node} $h$ (see rectangles in Fig.~\ref{fig:SP_evaluation}). Connect $h$ to the cycle or segment node containing the edge $(s,t)$ and move this edge from the respective cycle or segment node to $h$. By construction, $h$ has degree 1, it is connected to this single node.
\end{enumerate}
We note that the edge $e_\text{add}=(s,t)$ which is added in step 1 is already used for the categorization of blocks by the treewidth of the augmented graph as explained in Sec.~\ref{sec:block_categories}. In particular, this promotes the finite-momentum circle graph to a series-parallel block. We explicitly list the step here for completeness of the algorithm. 

\paragraph{Interpretation}
This algorithm yields a connected construction graph that captures the overlap structure relevant for the evaluation of the graph contribution. Evaluation algorithms can operate exclusively on the construction graph. We note that the construction graph is intended as a self-contained depiction and algorithmic language that provides a direct intuition for the structure of a graph and the sequence of operations required for its evaluation. The current implementation instead relies on existing libraries for SP decomposition for $\mathrm{tw}=2$ blocks and numerical approaches for $\mathrm{tw}>2$, see Sec.~\ref{sec:tn_blocks}. 

The construction graph is not unique but depends on the choice of cycle basis. While the final graph contribution is invariant under this choice, different bases may lead to construction graphs of varying complexity, which can affect the efficiency of the evaluation. Throughout this work, we therefore select a cycle basis that minimizes the complexity of the resulting construction graph, in practice corresponding to the minimal-treewidth construction graph.

The construction graph also provides a direct structural intuition on the graph categories (C)--(E), as illustrated in Fig.~\ref{fig:construction_graph_categories}. For category (C) with $\mathrm{tw}=2$, the construction graph is acyclic and thus a tree. For categories (D) and (E) with $\mathrm{tw}=3$ and $\mathrm{tw}=4$, the construction graphs contain loops. Category (D) corresponds to a single loop or a tree-like arrangement of overlapping loops, whereas category (E) exhibits nested loop structures with higher connectivity. We note the similarity between construction graph (E) and original graph (D). 

In the 1qp case, the topology of the construction graph depends explicitly on the location of the terminals. 
This is illustrated in Fig.~\ref{fig:SP_evaluation}, where the same underlying SP graph with $\mathrm{tw}=2$ from category (C) hosts three distinct placements of hopping processes. 
In cases (a) and (b), the terminals are already connected by an existing edge $(s,t)$ and no additional edge is required. The hopping node $h$ is attached either to a cycle node (a) or to a segment node (b), and the resulting construction graph remains a tree graph.
In case (c), the edge $(s,t)$ is not contained in the original graph. After adding this edge, the augmented graph is no longer a SP graph.
The construction proceeds as in the 0qp case and the hopping node is then attached to the cycle node containing the added edge $e_\text{add}=(s,t)$. The resulting construction graph contains a loop.
This shows that identical underlying 0qp graphs can lead to different construction graphs once hopping is included. Consequently, the complexity classification must be performed on the augmented graph.
Once the construction graph is defined, 0qp and 1qp graphs are treated on equal footing in all subsequent evaluation steps.

We briefly put the construction graph in connection with the existing literature. For series-parallel blocks, the construction graph is a circle-basis variant of the series-parallel decomposition tree of two-terminal networks \cite{Duffin1965,Tarjan1972}. For blocks of higher treewidth, cycle nodes, segment nodes, and loops of the construction graph correspond to the S-, P-, and R-nodes of the SPQR tree \cite{dibattista1996}. Meanwhile, only the 3-connected R-node skeletons require the numerical evaluation of Sec.~\ref{sec:tn_blocks}.

\subsubsection{Evaluation of tree-construction graphs for series-parallel blocks}\label{sec:tree_construction_graphs}
SP blocks yield tree-construction graphs, whose evaluation is based on the serial and parallel composition of elementary bridges---corresponding to multiplication and convolution of Epstein zeta functions, respectively, as discussed in Sec.~\ref{sec:sp_blocks}.
The sequence of operations is encoded in the construction graph itself, and the same evaluation algorithm applies to both 0qp and 1qp graphs, differing only in how the root of the graph traversal is chosen.

\begin{enumerate}
    \item \textbf{Per-node contribution.} For each node $n$ of the construction graph, compute its isolated contribution as the serial composition of the represented edges $\mathcal{E}_n$ by multiplication,
            \[ Z_{n}(\bm k) = \prod_{e\in\mathcal{E}_{n}} Z_{\Lambda,m_e\alpha} (\bm k)\]  
    with $Z_n(\bm k)=1$ if $\mathcal{E}_n=\emptyset$. 

    \item \textbf{Root choice.} For 0qp graphs, choose an arbitrary endpoint of the tree as root. This is always a cycle node, $c_\mathrm{root}$. 
    For 1qp graphs, the root is fixed to be the hopping node $h$. \\
    In both cases, orient all edges of the tree towards the root and evaluate by bottom-up traversal from the leaves.

    \item \textbf{Initialization.} For a leaf node $n$, initialize
            \[ F_n(\bm k) = Z_n(\bm k) \,.\]

    \item \textbf{Bottom-up evaluation.} For an internal cycle node $c$ with child contributions $F_1,\ldots ,F_r$, accumulate by serial composition
            \[ F_c(\bm k) = Z_c(\bm k) \cdot F_1(\bm k) \cdot \ldots \cdot F_r (\bm k)\,,\]
    and for an internal segment node $s$ by parallel composition, 
            \[ F_s(\bm k) = Z_s (\bm k)* F_1(\bm k) *\ldots * F_r(\bm k) \,.\]
            
    \item \textbf{Root termination.} The graph zeta function is obtained by a final convolution of the root with its child, 
            \[ \zeta_\mathcal{G}(\bm k) = Z_{\mathrm{root}} (\bm k) * F_\text{child} (\bm k)\,.\]
            Because the root has degree 1 by construction it has exactly one child contribution. This rule covers both the 0qp and the 1qp case: for 0qp, $Z_\mathrm{root} = Z_{c_\mathrm{root}}$ and for 1qp, $Z_\mathrm{root} = Z_h$.
\end{enumerate}
\begin{figure}[t]
    \centering
    \includegraphics[width=.9\linewidth]{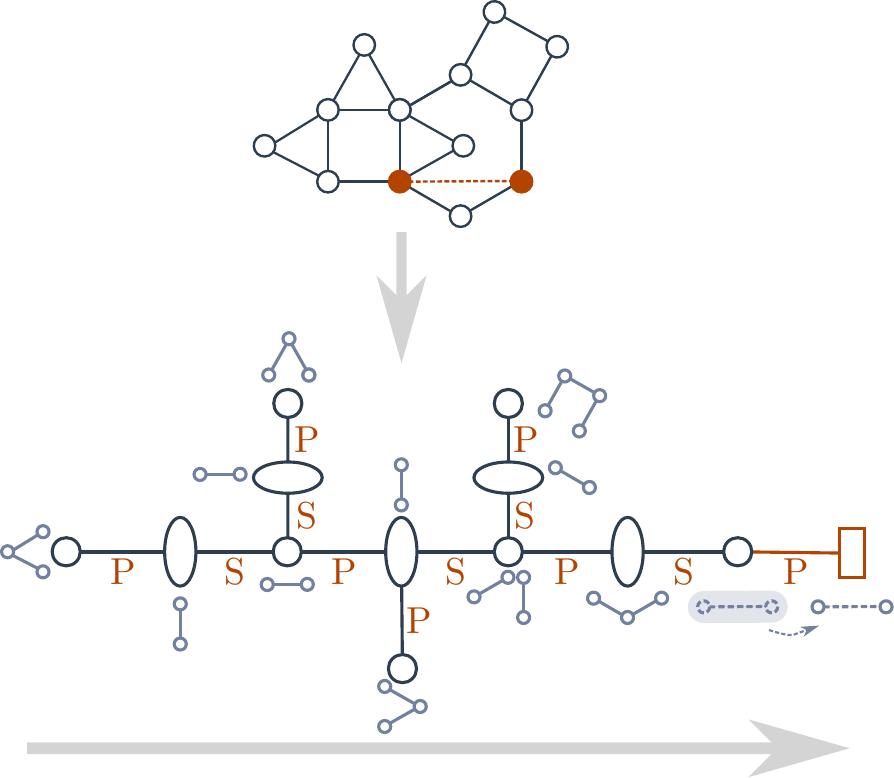}
    \caption{Determination and traversal of a tree-construction graph. (Top) Illustration of the original graph with terminals colored in orange. As the terminals are not connected, we add an edge between the terminals, depicted as dashed orange line. (Bottom) Construction graph where circular nodes are cycle nodes and oval nodes are segment nodes. The hopping node (orange rectangle) is chosen as the root for the bottom-up traversal. Each node is first calculated individually by serial composition of the represented edges, drawn in gray. The tree is traversed from leaves to root (as indicated by the gray arrow). At each node the result is updated by the operation indicated in orange at the edge (S for serial composition and P for parallel composition). The hopping node represents the added edge between the terminals with multiplicity $m_{e_\text{add}}=0$ that evaluates to identity, depicted as dashed gray edge. }
    \label{fig:construction_graph}
\end{figure}

We show an example for the traversal of a construction graph in Fig.~\ref{fig:construction_graph}. As indicated by the arrows, the hopping node is chosen as the root that all edges are oriented towards. The tree is traversed from the leaves and the result is updated at each node according to the rules, as indicated by the labels S and P at the edges.

\subsubsection{Evaluation of construction graphs with loops for highly connected blocks}\label{sec:loop_construction_graphs}
Graphs with treewidth $\mathrm{tw}\geq 3$ in categories (D) and (E) have construction graphs containing one or more loops. 
In contrast to SP blocks with tree-construction graphs, such graphs do not admit a rooted traversal and consequently cannot be evaluated using the procedure described in Sec.~\ref{sec:tree_construction_graphs}. Further, we find that graphs with treewidth $\mathrm{tw}$ require $\mathrm{tw}-1$ momentum channels such that graphs with $\mathrm{tw}\geq3$ require two or more momentum channels as demonstrated by the example below and are evaluated by the tensor network, see Sec.~\ref{sec:tn_blocks}.

We therefore limit the following discussion to illustrating how the construction graph evaluation formalism extends to one specific higher-treewidth example, demonstrating the need for more momentum channels in the process. 
A general algorithm and semi-analytic implementation remains as future work.
The central idea is to temporarily open the loops in the construction graph by removing cycle nodes, reducing the remaining structure to a tree graph whose evaluation mirrors the one discussed in Sec.~\ref{sec:tree_construction_graphs}. 
The loop structure is restored by convolution with the removed cycle nodes, which requires additional momentum channels. 
For 1qp graphs, the choice of the removed cycle node determines where the external momentum channel is closed in the final convolution.

\begin{figure}[t]
    \centering
    \includegraphics[width=\linewidth]{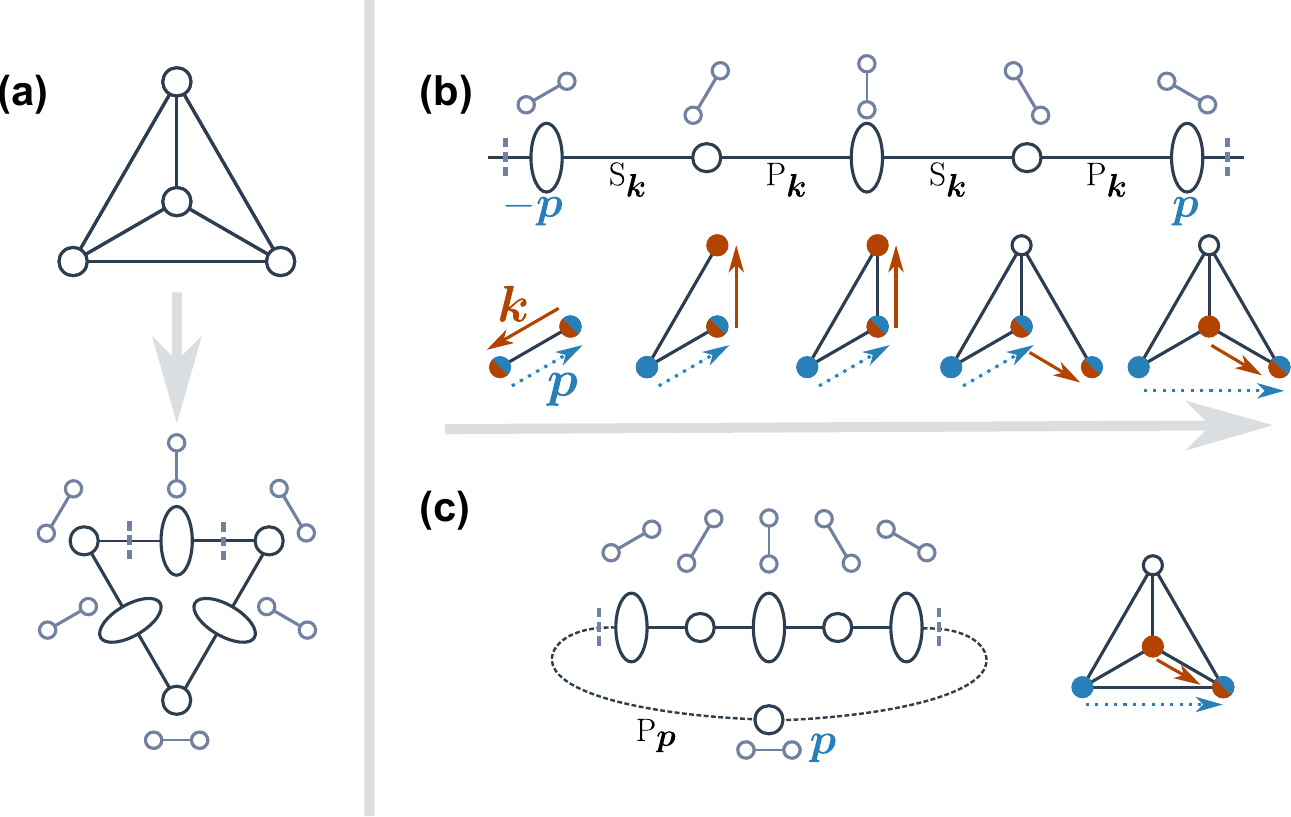}
    \caption{Evaluation of the $K_4$ graph with $\mathrm{tw}=3$. (a) The construction graph is a simple loop. One cycle node is cut out of the construction graph to open the loop, indicated by gray dashed lines.
    (b) This results in a chain graph which is constructed from serial (S$_{\bm k}$) and parallel (P$_{\bm k}$) composition of bridges with momentum $\bm k $ associated between the terminals (see orange colored vertices and respective orange arrows). The two endpoint segment nodes additionally carry momentum $\bm p$ with opposite directions associated between terminals (see blue colored terminals and dashed blue arrows) which is required for the closing of the construction graph loop in step 2. 
    (c) The loop of the construction graph is restored by parallel composition (P$_{\bm p}$) of the intermediate result with the removed cycle node in $\bm p$. This yields the $K_4$ graph with momenta $\bm k$ and $\bm p$ assigned as depicted.}
    \label{fig:K4}
\end{figure}

As an example we consider the simplest $\mathrm{tw}=3$ graph, the fully connected graph with four nodes ($K_4$) depicted in Fig.~\ref{fig:K4} and also refer to this figure for an illustration of the procedure. For simplicity all multiplicities are set to $m_e=1$.
The $K_4$ construction graph is given by a simple loop. 
The first step is the removal of one cycle node from the construction graph. 
The removed cycle node introduces an auxiliary momentum channel $\bm p$.
This momentum enters and exits through the two endpoints of the resulting chain graph, while all internal nodes of the chain graph depend only on the momentum channel $\bm k$ that constructs the chain following the SP procedure. 
To ensure momentum conservation when the loop is closed again, the two endpoint segment nodes carry opposite directions of the momentum $\bm p$.
Starting at one endpoint of the chain, evaluate it according to the SP tree procedure.
For each intermediate node $n$ in the chain, we calculate $Z_n(\bm k)$ by serial composition of the represented edges, which for the $K_4$ reduces to $Z_{\Lambda,\alpha}(\bm k)$ for each node. 
We perform a multiplication at an intermediate cycle node and a convolution at a segment node. 
The loop is restored by convolution with the removed cycle node, $Z_{\Lambda,\alpha}(\bm p)$. The final result is thus
\begin{align}
    \zeta_{K_4}(\bm k, \bm p) = \Big(&Z_{\alpha}(\bm k, -\bm p) \cdot Z_{\alpha}(\bm k) *_{\bm k} Z_{\alpha}(\bm k) \cdot \\ &\cdot Z_{\alpha}(\bm k)  *_{\bm k}Z_{\alpha}(\bm k, \bm p)\Big)\nonumber *_{\bm p} Z_{\alpha} (\bm p)\,,
\end{align}
dropping the specification of the $\Lambda$ index for shorter notation. The construction described here is the Fourier space representation of the tensor network representation in the previous section. Note, that two momentum channels are required, leading to a quadratic scaling in work and memory, consistent with the scaling exponent $\mathrm{tw}-1=2$.

For a 0qp graph the resulting expression is evaluated at $\bm k = \bm p = 0$. For a 1qp graph, one of the momenta corresponds to the external momentum between the terminals while the other one is set to zero. The corresponding momentum assignment is depicted in Fig.~\ref{fig:K4}.

\section{Application to the long-range transverse-field Ising model with power-law interactions}\label{sec:benchmark}

In this section we apply the graph zeta method to the long-range transverse-field Ising model with power-law interactions (LRTFIM). We start by demonstrating that the graph zeta method yields deterministic series coefficients, in contrast to previously employed Monte-Carlo (MC) approaches which are inherently stochastic \cite{Fey2019,Langheld2022,Adelhardt2024}. 
We then discuss the central physical quantity, the 1qp gap, which provides access to continuous phase transitions and their universality class
and demonstrate that the graph zeta method provides access to the 1qp dispersion across the full Brillouin zone in a single calculation.
Finally, we obtain the quantum phase diagram of the LRTFIM for four lattice geometries in one to three dimensions with a high resolution in the long-range decay exponent not previously achievable. 
This dense sampling is enabled by the significant decreased computational effort from cluster-scale MC runs to minutes on a personal computer for the parameters studied in this work.

As introduced in Sec.~\ref{sec:setting}, the Hamiltonian of the LRTFIM reads 
\begin{align}
    \label{eq:lrtfimHamiltonian}
    H = \frac{1}{2}\sum_{\bm{i}} \sigma_{\bm{i}}^z - \frac{\lambda}{2} \sum_{\bm{i} \neq \bm{j}} \frac{1}{\vert \bm{j} - \bm{i}\vert^{\sigma + d}}\sigma_{\bm{i}}^x \sigma_{\bm{j}}^x\,,
\end{align}
where we replaced the generic kernel from Eq.~\eqref{eq:H_TFIM} by the power-law kernel $\mathcal{K}_{\sigma+d}(\bm x)$
with $\sigma + d = \alpha$.
The perturbative expansion in the parameter $\lambda$ starts from the $z$-polarized high-field phase at $\lambda=0$ and
the perturbation parameter $\lambda=J/2h$ controls the strength of Ising interactions $J$ relative to the transverse field $h>0$. 
We give a short outline on the established quantum phase diagram. For a more comprehensive review see Ref.~\cite{Adelhardt2024} and references therein.

For ferromagnetic interactions ($\lambda>0$), the system undergoes a continuous quantum phase transition at $\lambda_c$ towards a ferromagnetically ordered low-field phase with spontaneously broken $\mathbb{Z}_2$ symmetry, regardless of lattice geometry and dimension. 
In the nearest-neighbor limit, the transition belongs to the $(d+1)$d Ising universality class \cite{Suzuki1976,SachdevBook}. 
Long-range interactions stabilize the ordered phase such that $\lambda_c$ decreases monotonously with decreasing decay exponent $\sigma$ compared to the nearest-neighbor model and approaches zero for $\sigma\to0$ \cite{Fey2016,Zhu2018,Puebla2019,Koziol2021,Winter2026}.
Long-range interactions enrich this picture by introducing three universality regimes as a function of the decay exponent $\sigma$ \cite{Dutta2001,Defenu2017,Defenu2023}:
nearest-neighbor (NN) universality for $\sigma\geq2-\eta_\text{SR}$, long-range mean-field (LRMF) universality for $\sigma\leq\sigma_\text{uc}=2d/3$, and a non-trivial intermediate regime where the critical exponents change continuously between the two limiting cases \cite{Adelhardt2024}.
Here $\eta_\text{SR}$ denotes the anomalous dimension of the short-range model and $\sigma_{\text{uc}}$ denotes the so-called upper critical exponent.

For antiferromagnetic nearest-neighbor interactions ($\lambda<0$) on bipartite lattices, the system exhibits a quantum phase transition between the high-field phase and a $\mathbb{Z}_2$-symmetry-broken antiferromagnetically ordered low-field phase.
In the NN model there is an exact duality mapping via a sublattice rotation between ferro- and antiferromagnetic models such that the universality is identical, \ie, ($d$+1)-dimensional Ising \cite{SachdevBook}.
Long-range interactions introduce a hierarchy of competing interactions which destabilize the ordered phase and shift the transition to larger values of $|\lambda_c|$ \cite{Koffel2012,Vodola2016,Sun2017,Fey2016,Fey2019,Puebla2019,Koziol2021}. 
Existing numerical studies indicate that long-range interactions do not lead to the emergence of distinct universality regimes and the system remains in the NN regime for all $(\sigma+d) > 0$ \cite{Sun2017}. 

For antiferromagnetic interactions on non-bipartite lattices, geometric frustration is present already in the nearest-neighbor case and no bipartite sublattice-rotation mapping exists. 
The prime example is the TFIM on the triangular lattice which exhibits 3d XY universality in the nearest-neighbor limit towards a clock-ordered low-field phase \cite{Powalski2013,Moessner2003}. 
For $\sigma<\infty$ the quantum phase diagram hosts a sixfold-degenerate plain stripe phase, an intermediate clock-ordered phase, and the high-field polarized phase \cite{Koziol2024,Koziol2023,Koziol2019,Fey2019,Adelhardt2024,Humeniuk2016}.
The system remains in that universality class down to $\sigma \approx 1$ \cite{Fey2019}. 
For smaller $\sigma$ there is evidence for a potential first-order transition towards a stripe-ordered low-field phase \cite{Humeniuk2016,Koziol2019,Fey2019,Saadatmand2018}.

In the following we use the LRTFIM as a benchmark demonstration for our deterministic series expansion approach.
As illustrated by the short summary above, the reasoning behind this choice is the rich low-energy physics of the model, paired with the extensive background literature. 
Furthermore, in recent years, the model became a go-to benchmark for developing methods in the field of long-range interacting models \cite{RocaJerat2024,Mcnaughton2025,Winter2026}.
We start with a brief summary of the problem size in Sec.~\ref{sec:CompositionOfBlocks}, before comparing the exact series with the MC-based evaluations of the hardcore sums in Sec~\ref{sec:Coefficients}.
Physical results can be found in Sec.~\ref{sec:PhysicalQuantities} and Sec.~\ref{sec:CriticalExponents}.

\subsection{Block statistics}
\label{sec:CompositionOfBlocks}
We first examine the statistics of the blocks contributing to the series coefficients of the LRTFIM, considering orders up to 13 in the 0qp sector and up to 11 in the 1qp sector.
Before block decomposition, the corresponding sets contain 8\,403 and 22\,677 graphs, respectively. 
We recall that here, edge multiplicities are regarded as graph attributes.
Topologically identical graphs with different edge multiplicities are therefore treated as distinct graphs, since their edge exponents differ and their contributions must be evaluated separately.
Block decomposing the graphs hosting the perturbative processes into minimal blocks yields 19\,207 blocks in the 0qp sector and 70\,000 blocks in the 1qp sector.

In the 0qp sector, 70.4\,\% of blocks admit an analytic evaluation, separated into 7\,902 bridges and 5\,611 zero-momentum circles. Furthermore, 23.3\,\% (4\,483) are non-analytic SP blocks. The remaining 6.3\,\% (1\,211) are $\mathrm{tw}\ge 3$ blocks.
Among those, we encounter 1\,180 cases of $\mathrm{tw}=3$ and 31 blocks with $\mathrm{tw}=4$.
In the 1qp sector, 78.6\,\% of the blocks evaluate analytically (44\,439 bridges and 10\,606 zero-momentum circles), 19.3\,\% (13\,530) are SP blocks and 2\,\% are non-SP (1\,418 with $\mathrm{tw}=3$ and 7 with $\mathrm{tw}=4$). 

As discussed in Sec.~\ref{sec:block_caching}, many of these blocks occur repeatedly, allowing the number of distinct evaluations to be reduced substantially through block caching. 
In the 0qp sector, caching reduces the number of blocks by 87\,\% from 19\,207 to 2\,602 unique blocks. Among these, 4.7\,\% admit analytic evaluation, 62.3\,\% are SP blocks, 32.0\,\% have $\mathrm{tw}=3$ and 1.0\,\% (26 blocks) have $\mathrm{tw}=4$.
In the 1qp sector, caching yields an even stronger reduction of 95\,\% from 70\,000 to 3\,423 unique blocks. Of these, 1.9\,\% are analytic, 76.8\,\% are SP blocks, 21.2\,\% have $\mathrm{tw}=3$ and only 0.1\,\% (4 blocks) have $\mathrm{tw}=4$.

\subsection{Series coefficients from the graph zeta method}
\label{sec:Coefficients}

A central advantage of the graph zeta method is that it yields deterministic perturbative series coefficients. 
In contrast, previously employed Monte Carlo (MC) summation approaches determine estimates for the coefficients which are inherently affected by statistical errors. 
For graph contributions which can be evaluated analytically within the introduced framework (see the discussion in Sec.~\ref{Sec:evaluation}), the results are obtained to machine precision. For the remaining graph contributions, the numerical error is controlled by the discretization parameter $n_p$ and decreases algebraically with increasing resolution. 
Throughout this section, we refer to the graph-zeta coefficients as numerically exact, meaning they are free of sampling errors and remaining uncertainty can be reduced by tuning the discretization.

These numerically exact coefficients provide a natural benchmark for previously published MC series.
To this end, we calculate the relative deviation between the MC coefficient $c_\text{MC}$ and the graph-zeta coefficient $c_\text{zeta}$ per order,
\begin{align}\label{eq:rel_deviation}
    \frac{|c_\text{MC}-c_\text{zeta}|}{|c_\text{zeta}|}\,,
\end{align}
with $c_\text{zeta}$ serving as the reference value. 

\begin{figure}[t]
    \centering
    \includegraphics[width=\linewidth]{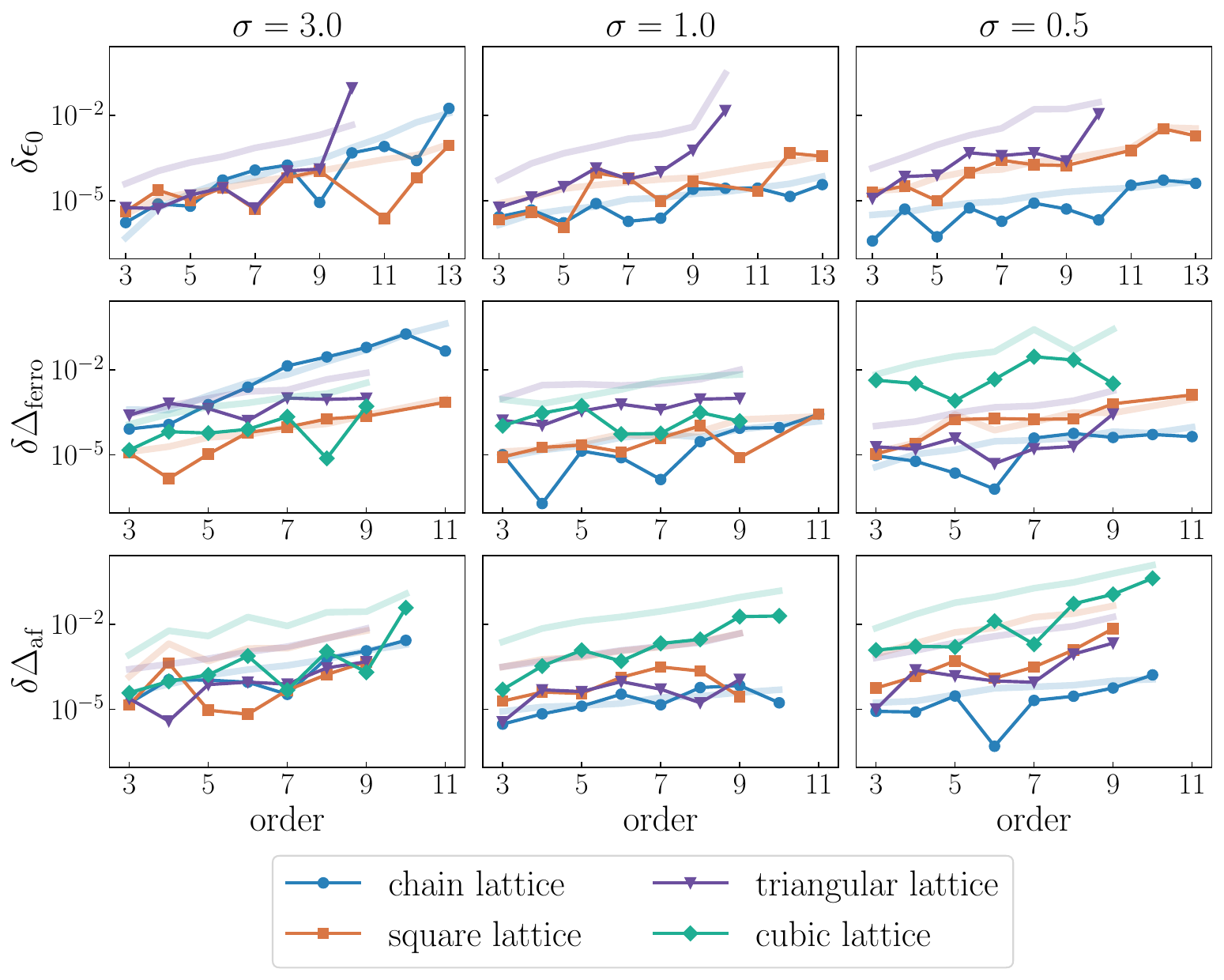}
    \caption{Relative deviation [Eq.~\eqref{eq:rel_deviation}] of the MC coefficients from the graph zeta coefficients in the LRTFIM, shown for the ground-state energy density $\epsilon_0$ (top) and the 1qp gap $\Delta$ in the ferromagnetic (middle) and antiferromagnetic (bottom) case. Deviations are shown as dots, one per expansion order, for long-range decay exponents ${\sigma\in\{3.0,1.0,0.5\}}$ for the chain (1d), square and triangular (2d), and cubic (3d) lattices. For comparison, the uncertainty of the MC data is indicated by thick, semi-transparent lines in the color of the corresponding lattice. MC data are taken from Ref.~\cite{Langheld2022} for the chain (0qp and ferromagnetic 1qp), Refs.~\cite{Adelhardt2020,Adelhardt2024} for the antiferromagnetic 1qp chain, Refs.~\cite{Fey2019,Adelhardt2024} for the square lattice 0qp and ferromagnetic 1qp series, Ref.~\cite{Fey2019} for the square lattice antiferromagnetic 1qp and all triangular-lattice 1qp series, and Ref.~\cite{Fey2020Diss} for the triangular lattice 0qp series as well as all cubic-lattice series.}
    \label{fig:MC_zeta_rel_dev}
\end{figure}

We show this relative deviation for the two quantities introduced in Sec.~\ref{sec:setting}, the ground-state energy density $\epsilon_0$ and the 1qp excitation gap $\Delta$, in the chain, square, triangular and cubic lattices in Fig.~\ref{fig:MC_zeta_rel_dev}. For each lattice, we consider both ferro- and antiferromagnetic interactions and three representative values of the decay exponent $\sigma$. 
We note that the ferromagnetic 1qp gap is obtained by evaluating the dispersion $\omega(\bm k)$ at $\bm k_c=\bm 0$ for all lattices, while the antiferromagnetic gap is located at $\bm k_c = \bm \pi$ for chain, square, and cubic lattices and at $\bm k_c= \left(2\pi/3, -2\pi/3\right)$ for the triangular lattice. The zeta-coefficients are calculated with $n_p=512$ for $d=1$, $n_p=66$ for antiferromagnetic interactions on the two-dimensional triangular lattice and $n_p=64$ for all other $d=2$ cases, and $n_p=10$ for $d=3$ unless stated otherwise.

Across all lattices, the relative deviation generally increases with perturbative order. This trend is consistent with the increasing difficulty of accurately estimating higher-order coefficients by stochastic sampling. 

We further compare the observed deviations with the statistical uncertainties reported for the MC coefficients, which were estimated from the sample standard deviation over independent MC runs with different random seeds. For the majority of coefficients, the deviation from the graph zeta result is within the quoted uncertainty. 

\begin{figure}
    \centering
    \includegraphics[width=\linewidth]{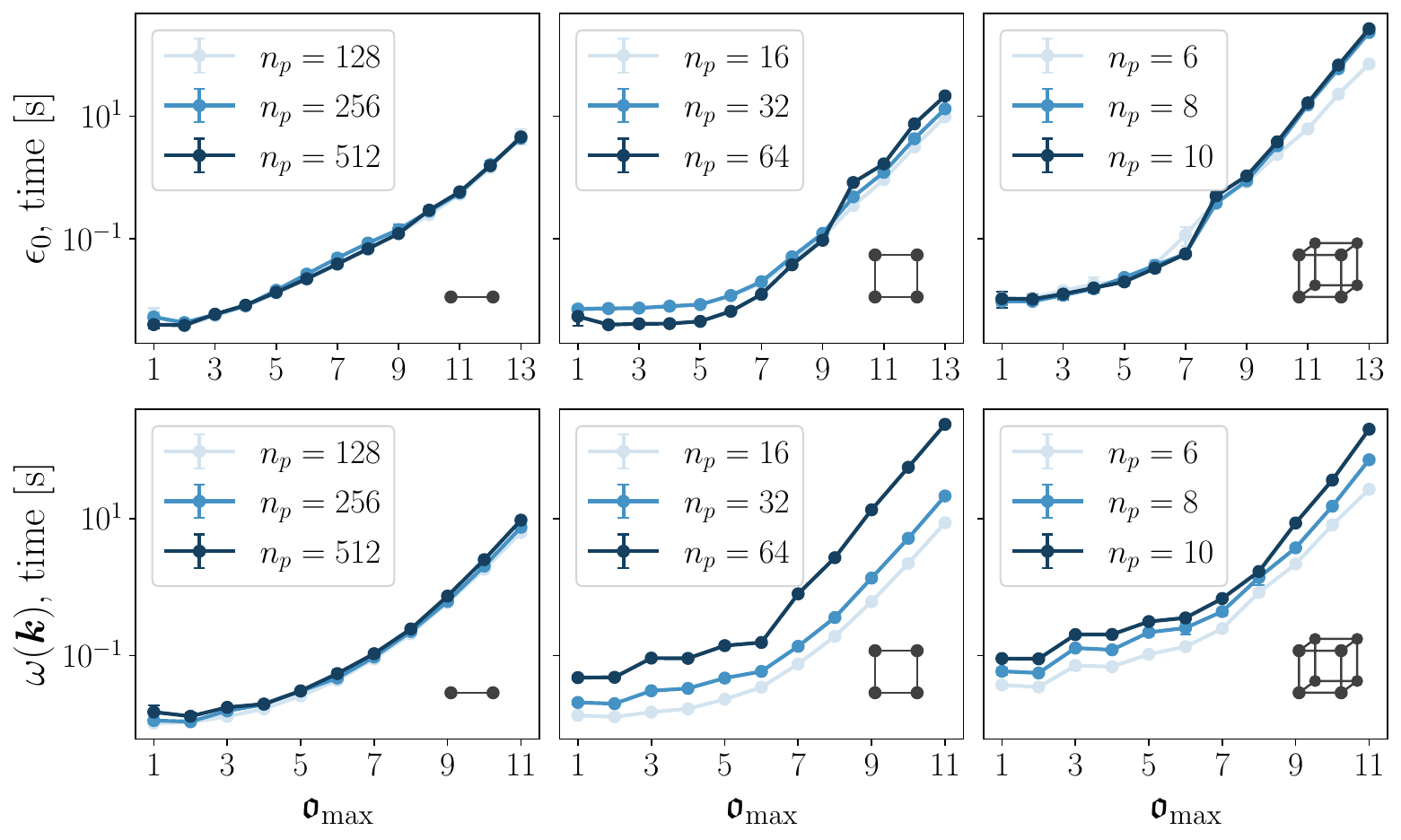}
    \caption{
    Runtime of the graph zeta method for the ground-state energy density $\epsilon_0$ (top) and the 1qp dispersion $\omega(\bm k)$ of the LRTFIM as a function of the maximal perturbative order $\mathfrak{o}_\text{max}$ for the chain (1d, left), square (2d, center) and cubic (3d, right) lattice and several discretization values $n_p$ at $\sigma=0.5$. Each data point is averaged over 10 runs on a personal computer on a single core (Apple M4 chip) and the error bars give the standard deviation.}
    \label{fig:graph-zeta-timing}
\end{figure}

We also demonstrate the runtime of the graph zeta method for the calculation of the 0qp and 1qp series coefficients in Fig.~\ref{fig:graph-zeta-timing}. For both quantities, the ground-state energy density $\epsilon_0$ and the 1qp dispersion $\omega(\bm k)$, we show the computation time as a function of the maximal perturbative order $\mathfrak{o}_\text{max}$ for the one, two and three dimensional LRTFIM for various values of the discretization $n_p$. Notably, we obtain the 0qp and 1qp series coefficients for the three-dimensional system in around 200 seconds. 

\subsection{Physical quantities}
\label{sec:PhysicalQuantities}
Having established the accuracy of the graph zeta coefficients, we now turn to the physical quantities described by the series.
The 1qp gap provides access to quantum critical properties of the model. A second-order quantum phase transition is indicated by a closing of the gap $\Delta(\lambda_c)=0$. Around the critical point, the 1qp gap follows a power-law behavior 
\begin{align}
    \Delta (\lambda) \sim |\lambda-\lambda_c|^{z\nu}\,,
\end{align}
from which we can extract both the critical point $\lambda_c$ and $z\nu$ via extrapolations of the series expansion. Here, $z$ and $\nu$ are the dynamical and correlation length exponents, respectively.

Throughout this section, we focus on the 1qp gap rather than the 0qp (ground-state) energy density and the associated susceptibility $\chi=\partial_\lambda^2 \epsilon_0$. This is because the critical behavior of the ground-state energy is more difficult to resolve. While the gap is zero at the critical point, the susceptibility diverges. The associated critical exponent $\alpha_c$, with $\chi \sim |\lambda-\lambda_c|^{-\alpha_c}$, is zero or close to zero in the limiting cases $\sigma=0$ and $\sigma=\infty$ and is heavily influenced by corrections to the dominant power-law behavior. 
The 1qp gap which vanishes with a finite exponent $z\nu$ thus provides a much more reliable extraction of the critical behavior.

\begin{figure}[t]
    \centering
    \includegraphics[width=\linewidth]{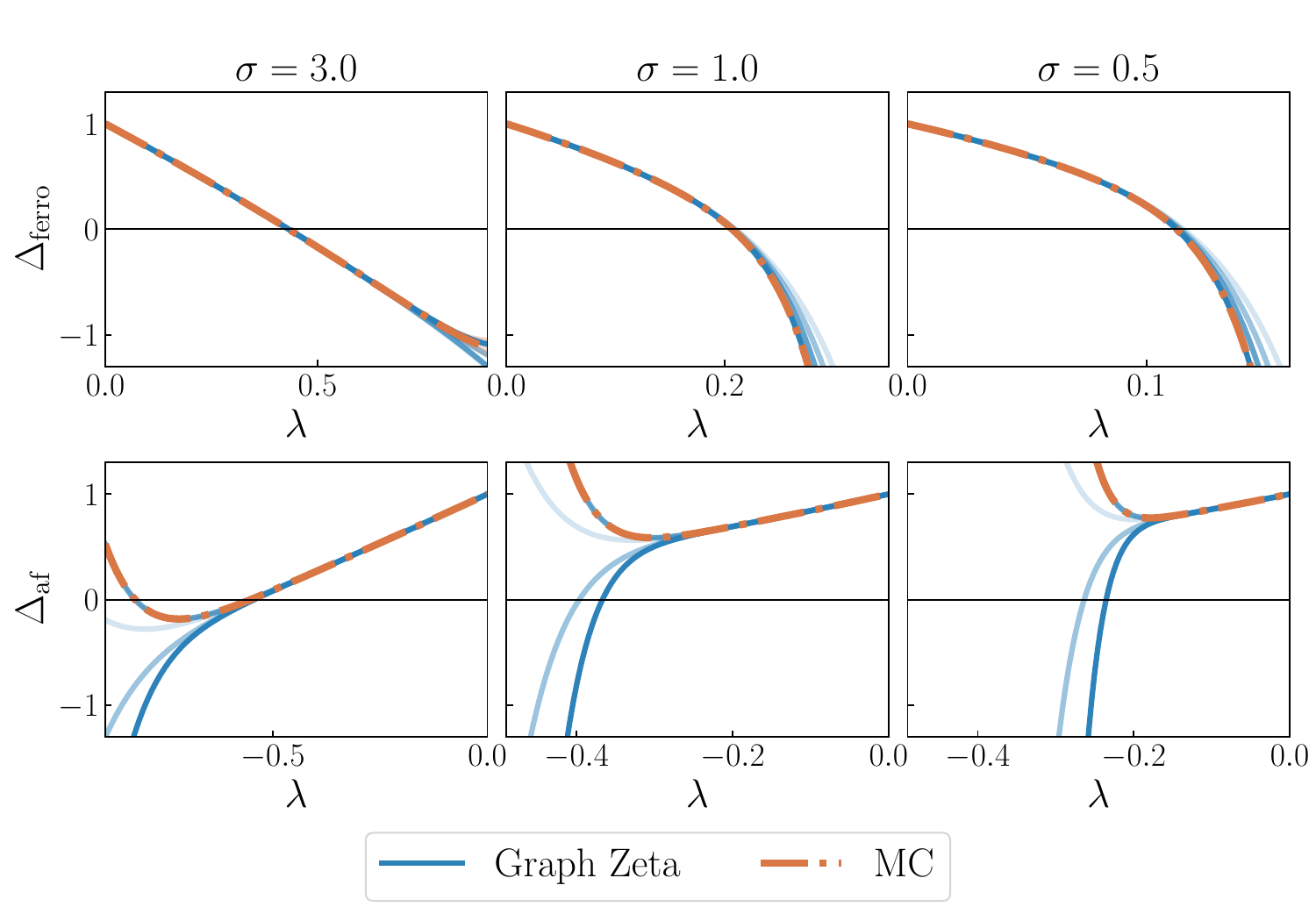}
    \caption{1qp excitation gap $\Delta$ as a function of the perturbation parameter $\lambda$ in the one-dimensional LRTFIM for $\sigma \in \{3.0,1.0,0.5\}$ in order 11 for ferromagnetic (top) and antiferromagnetic (bottom) Ising interactions in blue. Lower orders are shown in lighter blues to illustrate series convergence. MC series from Ref.~\cite{Langheld2022} (ferromagnetic, order 11) and Refs.~\cite{Adelhardt2020,Adelhardt2024} (antiferromagnetic, order 10) are shown as reference as orange dash-dotted lines. 
    }
    \label{fig:1qp_gap}
\end{figure}

Fig.~\ref{fig:1qp_gap} shows the 1qp gap as a function of the perturbation parameter $\lambda=J/2h$ for the 1d chain, computed to order 11, alongside the series in lower orders 8 to 10.
The series converges well in the order for ferromagnetic interactions across the whole range of $\sigma$. 
The critical point shifts to smaller $\lambda_c$ with decreasing $\sigma$, reflecting the stabilization of the ordered phase by long-range interactions.
For antiferromagnetic interactions, the spread between orders increases for smaller $\sigma$, due to the decreasing radius of convergence.
The qualitative behavior is consistent across all lattices considered.
For reference, previously published MC series from Refs.~\cite{Langheld2022,Adelhardt2020,Adelhardt2024} are shown and found to agree with the graph-zeta results on the scale of the figure. 

A qualitative advantage of the graph zeta method over the MC approach is that the full $\bm k$-dependence of the dispersion $\omega(\bm k)$ across the entire Brillouin zone is obtained from a single calculation. In the MC approach, each $\bm k$ point requires an independent simulation with several runs to obtain an average, restricting practical calculations to a few selected high-symmetry points. By contrast, the graph zeta method evaluates $\omega(\bm k)$ via the Fourier representation of Eq.~\eqref{eq:graph_zeta_fourier}, which gives the full dispersion on a $\bm k$ grid of $n_p$ points per dimension at no additional cost beyond computing the series coefficients themselves. 

This capability is particularly important when the critical momentum $\bm k_c$ deviates from its nearest-neighbor value at small decay exponents $\sigma$, potentially due to an as-yet-unknown physical mechanism. Without the full $\bm k$ dependence a change in $\bm k_c$ could remain undetected.

\begin{figure}[t]
    \centering
    \includegraphics[width=\linewidth]{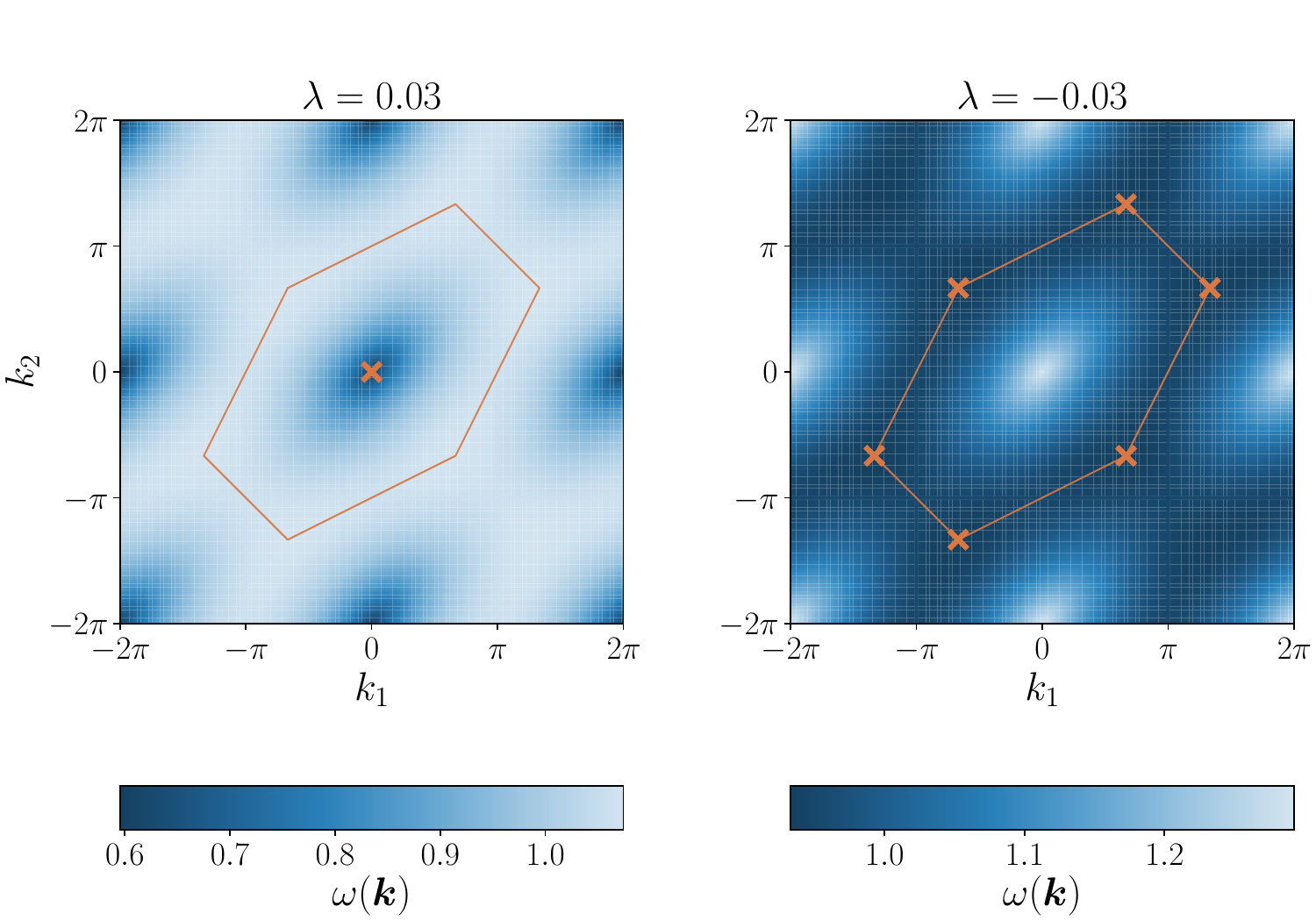}\\
    \vspace{.5cm}
    \includegraphics[width=\linewidth]{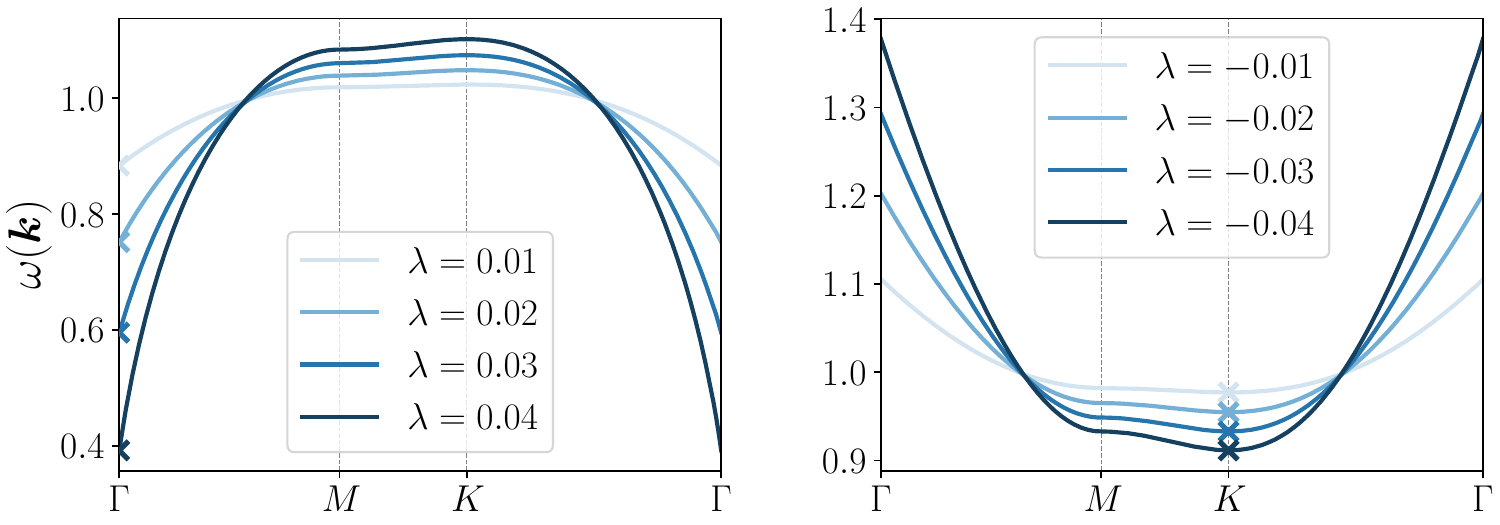}
    \caption{1qp dispersion $\omega(\bm k)$ from the graph zeta method in the LRTFIM on the triangular lattice for $\sigma=1.0$, evaluated for ferromagnetic (left) and antiferromagnetic (right) interactions in the high-field phase. 
    The dispersion is obtained in a single calculation on a $\bm k$-grid with $n_p=66$ points per dimension. The series is calculated up to order $\mathfrak{o}=11$. The momentum components $k_1$ and $k_2$ are defined with respect to the lattice vectors $\bm a_1=(1/2,\sqrt{3}/2)$ and $\bm a_2=(-1/2,\sqrt{3}/2)$. 
    The minima of the dispersion lie at $\bm k_c = (0,0)$ ($\Gamma$) for ferromagnetic and at $\bm k_c = \pm (2\pi/3,-2\pi/3)$ ($K$) and symmetry-related points for antiferromagnetic interactions and are marked by orange crosses. 
    (Top) Full dispersion over the Brillouin zone, evaluated at $\lambda=0.03$ for ferromagnetic and $\lambda=-0.03$ for antiferromagnetic interactions. The Brillouin-zone boundary is shown as orange outline.
    (Bottom) Dispersion along the high-symmetry path for various values of $\lambda$. }
    \label{fig:dispersion}
\end{figure}

We illustrate the access to the full Brillouin zone for the triangular lattice in Fig.~\ref{fig:dispersion}, showing the full dispersion at $\sigma=1.0$ for both ferromagnetic ($\lambda=0.03$) and antiferromagnetic ($\lambda=-0.03$) interactions at $\lambda$ values well within the polarized phase. Across all decay exponents considered, we find that $\bm k_c$ remains at its NN value of $\bm k_c=\bm 0 $ for ferromagnetic and $\bm k_c = \pm (2\pi/3,-2\pi/3)$ for antiferromagnetic interactions. For chain, square and cubic lattices, we find analogous behavior.

\subsection{Critical properties from gap extrapolation}
\label{sec:CriticalExponents}
We now turn to the extraction of quantum critical properties from the graph zeta results. We obtain estimates of the critical point $\lambda_c$ and the critical exponent $z\nu$ via DlogPadé extrapolations of the 1qp gap series. 
This is a standard procedure for series expansions and is typically employed for the extraction of critical properties from \eg MC series or exact NN series. For a detailed description, see Refs.~\cite{Baker1975,Guttmann1989}. 
The general procedure is to determine Padé approximants $P(L,M)$ of the logarithmic derivative of the gap, $\frac{{\rm d}}{{\rm d}\lambda} \ln(\Delta)$ with $L+M\leq \mathfrak{o}_{\rm max}-1$. 
Here, $L$ is the order of the numerator polynomial and $M$ the order of the denominator polynomial.
Sorting out defective extrapolants one can then determine the quantum-critical point $\lambda_{\rm c}$ from poles of the extrapolant and $z\nu$ from its residuum. DlogPadé extrapolants are grouped into families with $L-M=\text{const.}$ and we estimate $\lambda_c$ by averaging over the highest-order extrapolants of families with $|L-M|\leq 1$, taking the standard deviation as measure for the uncertainty of the extrapolation.

\begin{figure*}[t]
    \centering
    \includegraphics[width=0.48\linewidth]{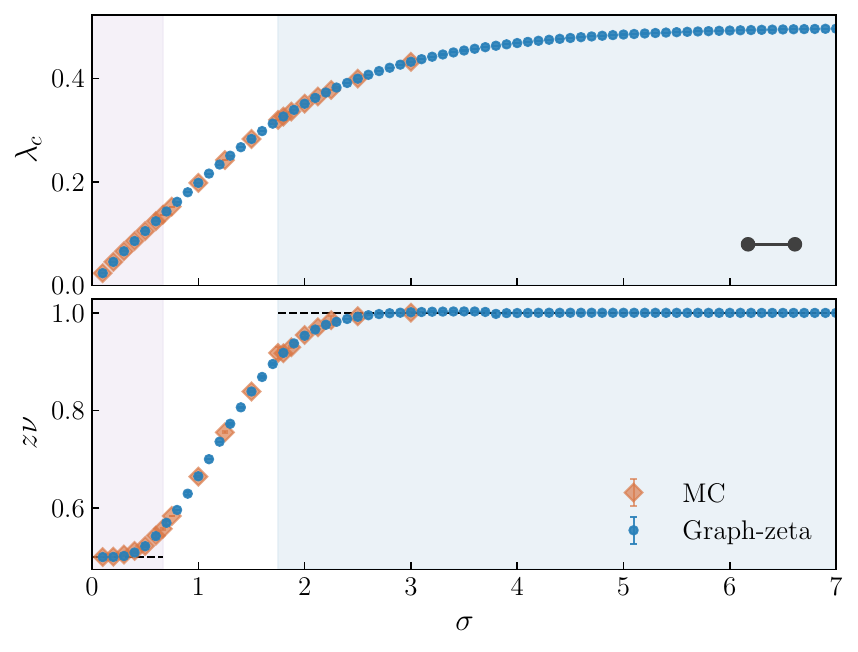}
    \includegraphics[width=0.48\linewidth]{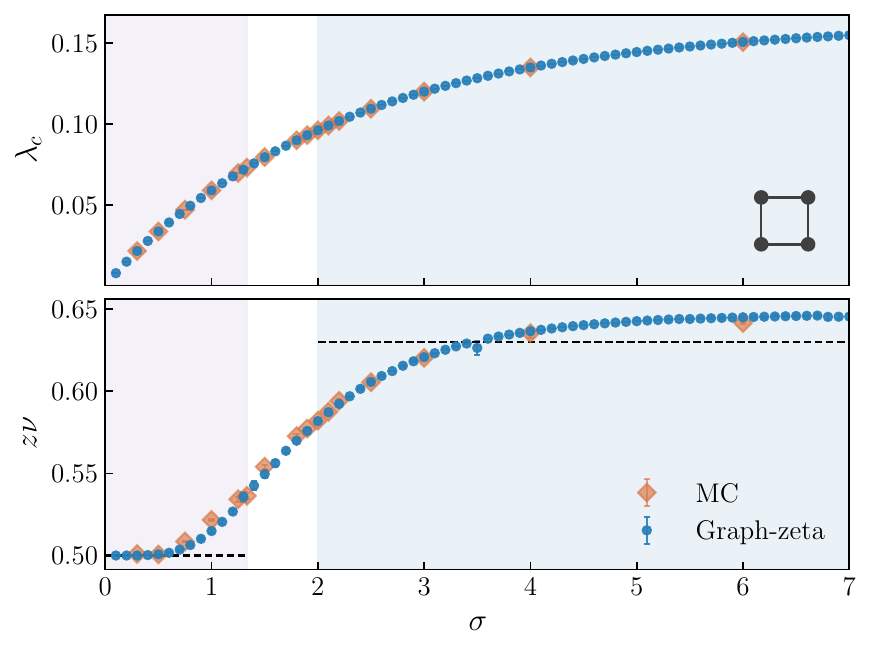}
    \includegraphics[width=0.48\linewidth]{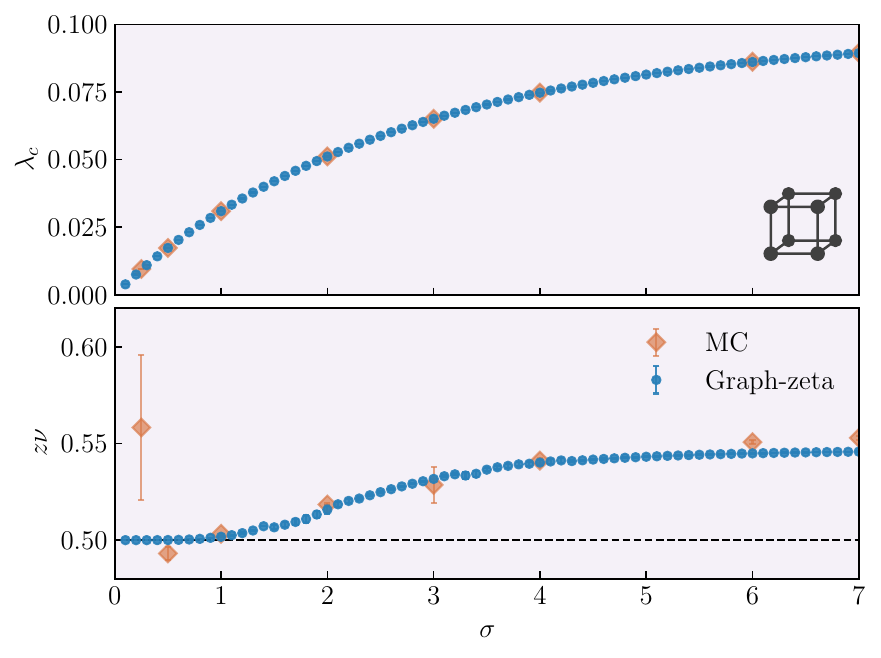}
    \includegraphics[width=0.48\linewidth]{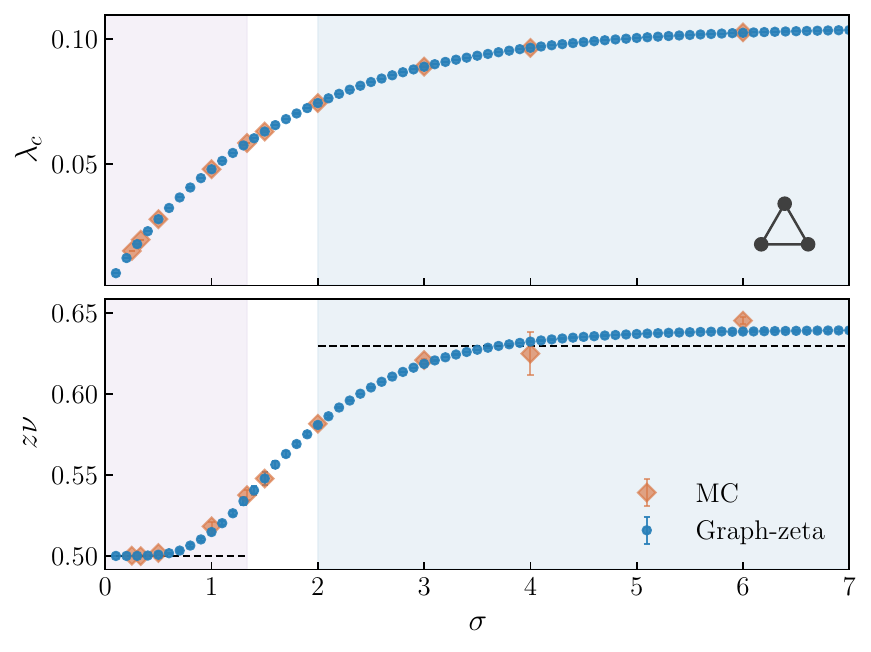}
    \caption{Criticality of the ferromagnetic LRTFIM on various lattices in one, two and three dimensions. We show the critical point $\lambda_c$ and the critical exponent $z\nu$ as a function of the decay exponent $\sigma$.
    For $\lambda>\lambda_c$ the system is in a ferromagnetically ordered phase. Both $\lambda_c$ and $z\nu$ are obtained from DlogPadé extrapolations of the 1qp gap series in the maximal perturbative order $\mathfrak{o}_\text{max}=11$. Results from the graph zeta method are shown in blue.
    The horizontal black dashed lines depict the expected values for $z\nu$. For one and two dimensions, the shaded regions indicate the different universality regimes whose borders depend on the dimension of the system. 
    In 1d (2d), we expect LRMF criticality with $z\nu = 1/2$ for $\sigma < 2/3\,(4/3)$ (purple shaded region), NN criticality in the $(d+1)$d Ising universality class with $z\nu = 1 \,(0.629971)$ for $\sigma > 1.75\,(2)$ (blue shaded region) and a non-trivial intermediate regime with continuously varying critical exponents interpolating between the two limiting cases. In 3d, we expect the mean-field regime to extend over the whole range of $\sigma$. Rounding effects at the regime boundaries are typically attributed to finite perturbative orders \cite{Adelhardt2024}. Additionally, at the upper critical dimension $\sigma_\text{uc}=2d/3$ there are multiplicative logarithmic corrections to the dominant power-law behavior at the critical point \cite{Fey2019}. For the NN limit $\sigma=\infty$, an overestimation of the critical exponent is a well-known limitation of the approach which also influences the NN regime for large $\sigma$. 
    Monte Carlo results shown for reference in orange are taken from Ref.~\cite{Langheld2022} for the chain, \cite{Fey2019,Adelhardt2024} for the square lattice, \cite{Fey2019} for the triangular lattice and \cite{Fey2020Diss} for the cubic lattice.}
    \label{fig:criticality_ferro}
\end{figure*}

\begin{figure*}[t]
    \centering
    \includegraphics[width=0.48\linewidth]{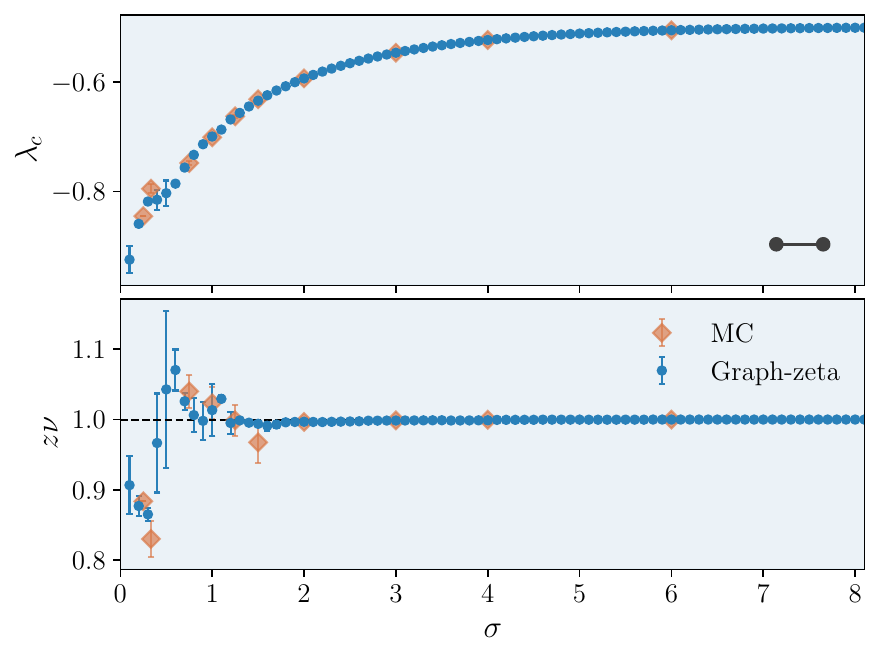}
    \includegraphics[width=0.48\linewidth]{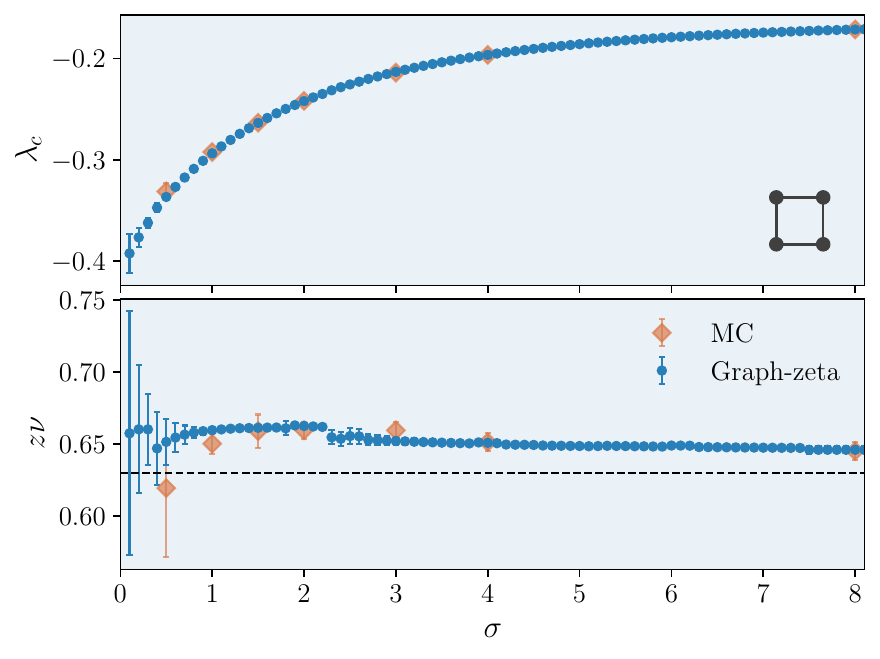}
    \includegraphics[width=0.48\linewidth]{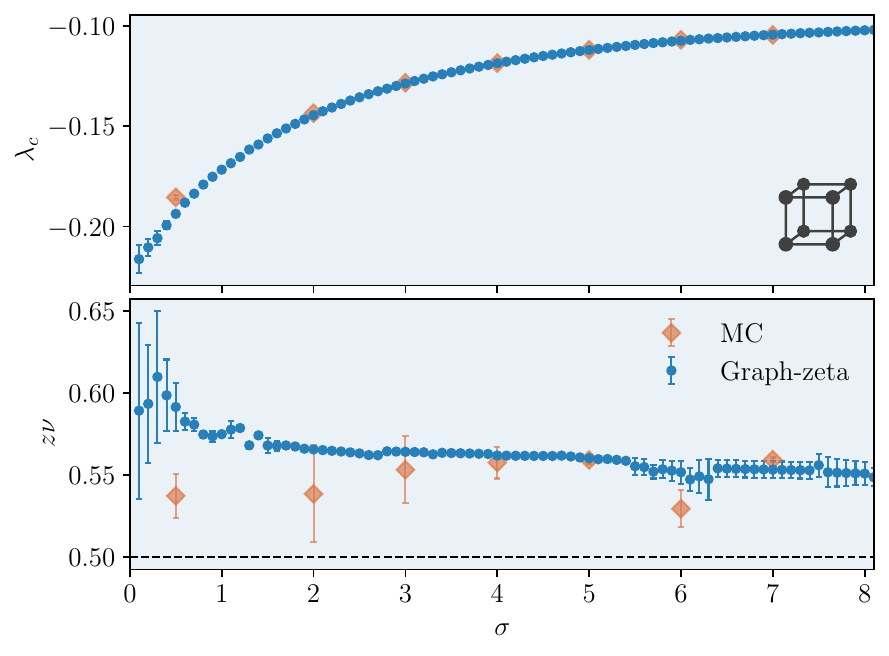}
    \includegraphics[width=0.48\linewidth]{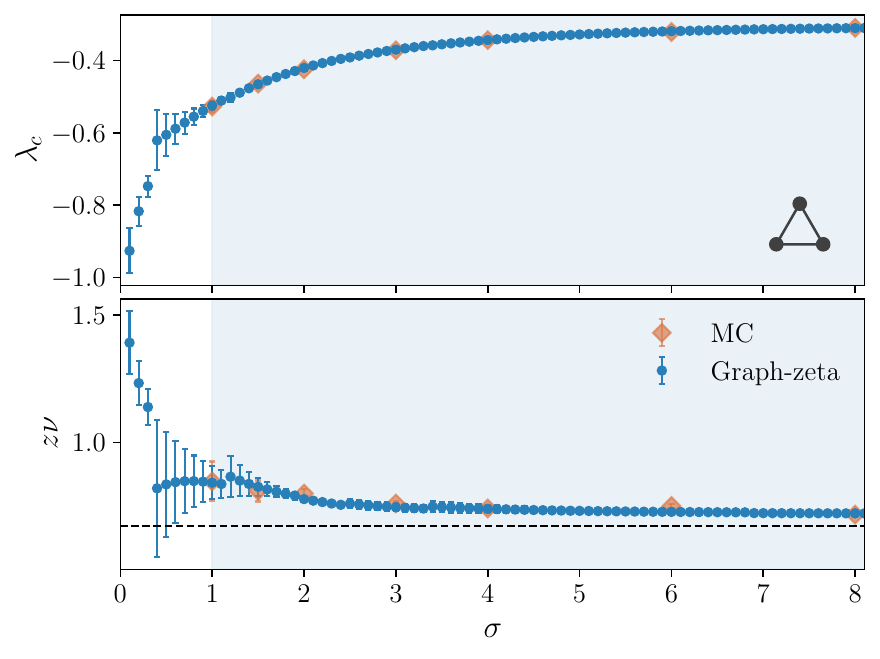}
    \caption{
    Criticality of the antiferromagnetic LRTFIM on various lattices in one, two and three dimensions. We show the critical point $\lambda_c$ and the critical exponent $z\nu$ as a function of the decay exponent $\sigma$.
    For $\lambda>\lambda_c$ the system is in an antiferromagnetically ordered phase. Both $\lambda_c$ and $z\nu$ are obtained from DlogPadé extrapolations of the 1qp gap series in the maximal perturbative order $\mathfrak{o}_\text{max}=11$. Results from the graph zeta method are shown in blue.
    The horizontal black dashed lines depict the expected values for $z\nu$. 
    For bipartite lattices (chain, square, cubic) we expect the NN regime (blue shaded region) with ($d+1$)d Ising universality to extend over the full range of $\sigma$ with $z\nu = 1$ for the chain, $z\nu = 0.629971$ for the square and $z\nu = 1/2$ for the cubic lattice. For the triangular lattice, the situation is not yet resolved conclusively, current literature suggest that the NN regime with 3d XY universality and $z\nu=0.67175(10)$ \cite{smerald2018,hasenbusch2019} extends over $\sigma\gtrsim 1$ \cite{Saadatmand2018,Koziol2019,Fey2019}. Similar to Ref.~\cite{Fey2019}, a breakdown of the graph-zeta extrapolation around $\sigma \approx 0.6$ is observed.
    Monte Carlo results are shown as reference in blue and are taken from Refs.~\cite{Adelhardt2020,Adelhardt2024} for the chain, \cite{Fey2019} for square and triangular lattice and \cite{Fey2020Diss} for the cubic lattice.
    }
    \label{fig:criticality_af}
\end{figure*}

The critical points and exponents, $\lambda_c$ and $z\nu$, extracted from the graph zeta series using DlogPadé extrapolations, are shown in Fig.~\ref{fig:criticality_ferro} for ferromagnetic and Fig.~\ref{fig:criticality_af} for antiferromagnetic interactions.
Each figure contains results for the four lattice geometries considered in this work (chain, square, triangular, and cubic), allowing for a systematic comparison across spatial dimension and coordination number. 
The graph zeta method enables a dense sampling of the parameter space such that the dependence of $\lambda_c$ and $z\nu$ can be resolved with high resolution. This provides a detailed characterization of the quantum phase diagram and resolves the crossover between universality regimes not previously achievable. 
Previously published Monte Carlo data from Refs.~\cite{Fey2019,Fey2020Diss,Adelhardt2020,Langheld2022,Adelhardt2024} are shown as reference and confirm consistency with the graph zeta results across all lattices and interaction signs. 
At the same time, the substantially denser sampling in $\sigma$ provided by the graph zeta method reveals smooth trends that were previously only accessible in a coarser form within the MC data. In particular, this allows the identification of outliers among the MC data points. The small jumps in the extrapolated $z\nu$ for some combinations of lattice and interaction sign can be attributed to changes in which extrapolants are non-defective and contribute to the averaged result, rather than to physical discontinuities.

For antiferromagnetic interactions, the quality of the extrapolation of $z\nu$, and, to a lesser degree, of $\lambda_c$ decreases significantly for $\sigma \lesssim 1$. While the point of the antiferromagnetic quantum phase transition moves further away from the unperturbed limit $\lambda=0$, the radius of convergence---given by the ferromagnetic phase transition point moving closer to $\lambda=0$---decreases at the same time. Therefore, the ability of the series to accurately determine the phase transition point decreases significantly and both high perturbative orders and exact coefficients are crucial as the extrapolation depends heavily on small variations in the series. 
This difficulty was recognized in prior MC work as well and the computational effort of the MC simulations was increased significantly compared to the case of ferromagnetic interactions, both in terms of runtime and the number of random seeds per data point, to obtain more reliable data.

\section{Application to the modeling of the quantum Ising magnet $\text{KTmSe}_2$}
\label{sec:KTmSe2}
Having benchmarked the graph zeta method on the LRTFIM with isotropic power-law interactions, we now use the framework as a tool for the modeling of a real material.
The method's efficiency and applicability to short- and long-range interacting systems make it possible to evaluate different microscopic descriptions at high perturbative order across the entire Brillouin zone and to systematically discriminate between them.
This capability is directly accessible through the open-source Graph Zeta Library \cite{gzl2026}, which allows rapidly testing microscopic models and evaluating them across many parameter sets. 
Here, we apply it to the quantum magnet $\mathrm{KTmSe}_2$ and compare one-quasiparticle dispersions with inelastic neutron-scattering (INS) data \cite{Zheng2023KTmSe2}, demonstrating the graph zeta method on a model that combines short-range exchange with a long-range dipolar tail and the treatment of the angular dependence of dipolar interactions.

Triangular-lattice antiferromagnets based on alkali-rare-earth-chalcogen compounds form an important class of quantum materials \cite{Liu2018REC,Bordelon2019NaYbO2,Dai2021NaYbSe2,Zhang2021NaYbSe2CEF,Pocs2021CsYbSe2CEF,Scheie2020ErSe2CEF,Zheng2023KTmSe2}.
$\mathrm{KTmSe}_2$ consists of triangular sheets of non-Kramers $\mathrm{Tm}^{3+}$ ions in an ABC stacking.
Its low-energy magnetism has been described by an effectively two-dimensional antiferromagnetic transverse-field Ising model in the quantum-disordered, transverse-field-polarized regime \cite{Zheng2023KTmSe2}.
The separation between sheets is roughly twice the nearest-neighbor distance within a sheet.
The intrinsic transverse field originates from the crystal-electric-field splitting of the two lowest singlet states, which together form an effective pseudospin-$1/2$ quasidoublet \cite{Zheng2023KTmSe2}. 

Thermodynamic and INS measurements were interpreted in terms of an antiferromagnetic $J_1$-$J_2$ TFIM on the triangular lattice \cite{Zheng2023KTmSe2}, where the excitation spectrum was modeled using linear spin-wave theory for the effective TFIM and mean-field RPA for the full crystal-field Hamiltonian.
Here, we revisit this effective model using high-field series expansions beyond these approximations, allowing us to reassess the fitted couplings and gather evidence whether long-range dipolar interactions are relevant for the measured dispersion.

We first consider a two-dimensional model with nearest-neighbor exchange and isotropic dipolar interactions within each sheet (Sec.~\ref{sec:2DModeling}).
We then quantify the effect of interlayer dipolar interactions in Sec.~\ref{sec:3DModeling}.

\subsection{Two-dimensional modeling}
\label{sec:2DModeling}

For the two-dimensional description of $\mathrm{KTmSe}_2$, we consider the Hamiltonian
\begin{equation}
\label{eq:J1J2AndDipolar}
\begin{aligned}
H =& \frac{J_{\mathrm{dip}}}{2}\sum_{\bm i\neq \bm j}\frac{1}{|\bm j-\bm i|^3}\sigma_{\bm i}^z\sigma_{\bm j}^z
+J_{1}\sum_{\langle \bm i,\bm j\rangle}\sigma_{\bm i}^z\sigma_{\bm j}^z \\
&+J_{2}\sum_{\langle\langle \bm i,\bm j\rangle\rangle}\sigma_{\bm i}^z\sigma_{\bm j}^z
+h\sum_{\bm i}\sigma_{\bm i}^x \,,
\end{aligned}
\end{equation}
where $J_1$ and $J_2$ are the nearest- and next-nearest-neighbor Ising exchange couplings, and $J_{\mathrm{dip}}$ is the dipolar coupling at the nearest-neighbor distance. The positions $\bm i$ and $\bm j$ are expressed in units of the lattice spacing. 
Following the convention used for magnetic compounds, we choose the Ising interaction along the $\sigma_i^z$-quantization axis.
This convention differs from Eq.~\eqref{eq:H_TFIM} only by a global rotation of spin axes and therefore leaves the perturbative construction unchanged.
Because the easy Ising axis of $\mathrm{KTmSe}_2$ is perpendicular to the triangular-lattice planes \cite{Zheng2023KTmSe2}, the interaction is isotropic in this two-dimensional approximation.

We compare three microscopic descriptions.
First, we study the $J_1$-$J_2$ model with the parameters of Ref.~\cite{Zheng2023KTmSe2}, converted to our Pauli-operator convention.
Second, we optimize $J_1$ and $h$ in the same model at fixed $J_2$ by fitting the dispersion to the ridge of maximum INS intensity \cite{Zheng2023KTmSe2}.
Third, we replace the phenomenological $J_2$ interaction by the complete antiferromagnetic dipolar tail and calculate the resulting $J_1$-dipolar dispersion.
For all three cases, we evaluate the high-field expansion to perturbative order $11$ using the Graph Zeta Library and compare the resulting dispersions directly with the experimental INS spectrum \cite{Zheng2023KTmSe2}. 
For the experimentally relevant parameters, the series are already converged in perturbative order, and extrapolations \cite{Guttmann1989} do not deviate significantly from the bare series. 
We note that the precise extraction of model parameters is currently limited by the extraction of the experimental data from the INS colorplot in article Ref.~\cite{Zheng2023KTmSe2}.
We therefore expect the qualitative statements about the model to be reasonable, while the precise coupling values may still depend on extraction error, as well as not considered experimental uncertainties.

\begin{figure}[h]
\centering
\includegraphics[width=0.9\linewidth]{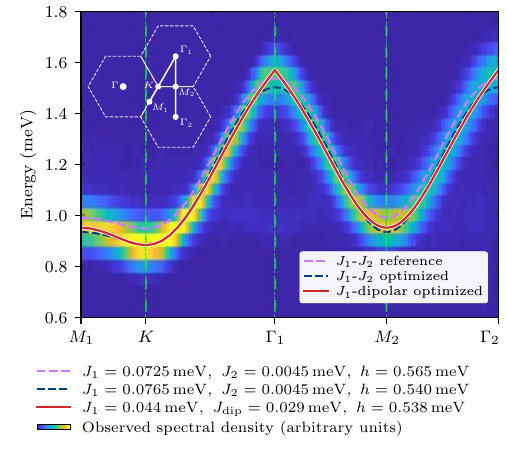}
\caption{One-quasiparticle dispersion obtained for the three two-dimensional descriptions along the high-symmetry path of the hexagonal Brillouin zone, calculated to order $11$ in the high-field expansion using the graph zeta method.
The background shows the measured spectral density from Ref.~\cite{Zheng2023KTmSe2}.
The $J_1$-$J_2$ dispersions use either the parameters of Ref.~\cite{Zheng2023KTmSe2} (purple dashed line) or optimized $J_1$ and $h$ at fixed $J_2$ (blue dashed line).
For the $J_1$-dipolar model (red solid line), $J_1$ and $h$ are fitted to the ridge of maximum INS intensity \cite{Zheng2023KTmSe2}, and $J_{\mathrm{dip}}$ is estimated from the measured saturation moment.
The legend gives the interaction parameters and transverse field $h$ in $\mathrm{meV}$.
}
\label{fig:material}
\end{figure}

Among the three models, the $J_1$-dipolar model provides the best overall description of the experimental dispersion, see Fig.~\ref{fig:material}.
Its 1qp dispersion shows the smallest deviation from the ridge of maximum intensity in the INS data and simultaneously reproduces the energies at the $\Gamma_{1,2}$, $K$, and $M$ points.
In contrast, neither of the $J_1$-$J_2$ models captures the experimental dispersion at all three high-symmetry points with comparable accuracy.

The microscopic scales of the $J_1$-dipolar description are physically reasonable.
From the crystal-field analysis and saturation measurements reported in Ref.~\cite{Zheng2023KTmSe2}, we estimate a dipolar coupling $J_{\mathrm{dip}}$ in the range of $0.02$ to $0.03\,\mathrm{meV}$.
We fix $J_{\mathrm{dip}}$ at the saturation-based estimate, assuming that the saturation moment represents the low-energy quasidoublet moment.
The fitted transverse field corresponds to an unperturbed quasidoublet splitting consistent with the scale obtained from the point-charge crystal-field calculation of Ref.~\cite{Zheng2023KTmSe2}.

Three further independent observations support the dipolar interpretation.
First, in the Pauli convention the fitted next-nearest-neighbor coupling of the short-range model from Ref.~\cite{Zheng2023KTmSe2} is comparable to the dipolar coupling between next-nearest neighbors
\begin{align}
    J_2=&\,0.0045(5)\,\mathrm{meV}\,,\\
    \frac{J_{\text{dip}}}{\big(\sqrt{3}\big)^3}=&\,\frac{0.0215\,\mathrm{meV}}{\big(\sqrt{3}\big)^3}\simeq0.0041\,\mathrm{meV}\, .
\end{align}
The phenomenological $J_2$ may therefore partly encode a neglected dipolar tail. 

Second, the $J_1$-dipolar model gives a linear dispersion near the $\Gamma_{1,2}$ points.
In two dimensions, a $1/r^3$ interaction generates a leading nonanalytic contribution proportional to $|\bm k|$ near $\bm k=\bm 0$, whereas finite-range $J_1$ and $J_1$-$J_2$ models have analytic, generically quadratic leading behavior. 
We observe that the measured dispersion tends to follow this linear behavior more closely than the quadratic behavior of the purely short-range models (see Fig.~\ref{fig:material}).
The measured maxima are slightly rounded, which could arise from three-dimensional effects or instrumental broadening. 
We investigate the three-dimensional contributions quantitatively in Sec.~\ref{sec:3DModeling}.

Third, a free fit of the $J_1$-$J_2$-dipolar model (not shown) in Eq.~\eqref{eq:J1J2AndDipolar} gives a fitted $J_2$ much smaller than the dipolar contribution at the same distance and improves the fit only marginally compared to the $J_1$-dipolar model.
This indicates that $J_2$ in the $J_1$-$J_2$ model originates potentially from a truncated dipolar tail rather than a distinct exchange interaction.

\subsection{Three-dimensional modeling}
\label{sec:3DModeling}

Beyond the intralayer Ising interactions, dipolar interactions also couple the Thulium moments in different sheets.
We therefore extend the calculations for the $J_1$-dipolar model to the full ABC stack to quantify the interlayer contribution and assess the accuracy of the two-dimensional description along the measured momentum path.

Starting from Eq.~\eqref{eq:J1J2AndDipolar} with $J_2=0$, we extend the dipolar sum to the entire three-dimensional lattice and replace the isotropic interaction kernel by
\begin{equation}
\label{eq:KTmSe2_dipolar_3D}
K_{\mathrm{3D}}(\bm r)
=\frac{1-3r_z^2/|\bm r|^2}{|\bm r|^3}\,,
\end{equation}
where $\bm r=\bm j-\bm i$ and the component $r_z$ is the separation along the crystallographic $c$ axis.
We take the lateral displacement between neighboring sheets and their separation from the crystal structure published in Ref.~\cite{Zheng2023KTmSe2}.
The short-range exchange $J_1$ remains restricted to nearest neighbors within each sheet.

We evaluate the angular lattice sums using anisotropic Epstein zeta functions~\cite{buchheit2026zeta,buchheit2026anisotropic,epsteinlib}, whose meromorphic continuation from the absolutely convergent regime defines the conditionally convergent three-dimensional dipolar sums.
The potential pole at the marginal decay exponent cancels, since the dipolar kernel has a vanishing angular average. 
We compare the three-dimensional 1qp dispersions at perturbative order $5$ \footnote{For this proof-of-principle calculation we used an experimental code extension to anisotropic dipolar sums.} with the two-dimensional results at order $11$.
Higher orders are in principle also accessible for the angle-dependent dipolar kernel, but a stable and efficient implementation in the Graph Zeta Library is still under development.

Keeping $J_{\mathrm{dip}}$ fixed at the saturation-based estimate used in the previous section, we fit $J_1$ and $h$ to the digitized ridge of maximum INS intensity.
For the three-dimensional model, this gives $J_1\simeq0.040\,\mathrm{meV}$ and $h\simeq0.534\,\mathrm{meV}$ which are reasonable values.
The root-mean-square deviations from the experimental data extracted from Ref.~\cite{Zheng2023KTmSe2} are approximately the same between the two- and three-dimensional dipolar modeling.
For comparison, the root-mean-square deviation of the optimized two-dimensional $J_1$-$J_2$ model is approximately twice that value.
Again, it is important to put these values into context.
The experimental data is extracted from an INS color plot in Ref.~\cite{Zheng2023KTmSe2} and therefore carries an extraction error.
Accordingly, we interpret the root-mean-square deviation as a qualitative measure.
Once experimental data is available a full quantitative analysis of couplings is directly possible. 

\begin{figure}
\centering
\includegraphics[width=0.9\linewidth]{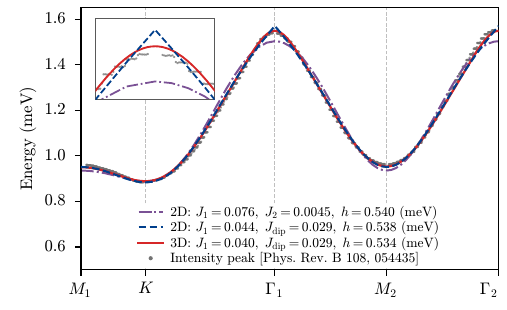}
\caption{Comparison of the two- and three-dimensional $J_1$-dipolar models for $\mathrm{KTmSe}_2$ with the experimental data from Ref.~\cite{Zheng2023KTmSe2}, using the optimized two-dimensional $J_1$-$J_2$ model as a reference.
Gray points show peak centers extracted from the neutron-scattering intensity in Fig.~3(a) of Ref.~\cite{Zheng2023KTmSe2}.
The dispersion of the two-dimensional models (blue and purple dashed lines) are calculated to order $11$ and are the same as in Fig.~\ref{fig:material}.
The dispersion of the three-dimensional model (red solid lines) is calculated to order $5$.
}
\label{fig:3DMaterial}
\end{figure}

As shown in Fig.~\ref{fig:3DMaterial}, the interlayer interaction mainly modifies the dispersion near the band maximum.
The interlayer contribution smooths the linear cusp of the isolated-sheet dispersion, producing a slightly rounded maximum.
The dispersion away from the band maximum changes only little.

The small effect on the maxima at $\Gamma_{1,2}$ may seem surprising, since bulk dipolar interactions can strongly lower the excitation energy obtained from an two-dimensional in-plane modeling at a true three-dimensional Brillouin zone center.
However, the measured $\Gamma_{1,2}$ points are centers of repeated two-dimensional zones at finite momentum transfer, and are not equivalent to the $\Gamma$ point at $\bm k=\bm 0$ of the full ABC stack \footnote{Finite-energy inelastic neutron scattering requires nonzero momentum transfer. 
A true three-dimensional zone center can nevertheless be probed near a nonzero reciprocal-lattice vector of the full crystal.}.
At these $\Gamma_{1,2}$ points, the excitation amplitude is uniform within each triangular layer, but its phase changes by $\pm 2\pi/3$ between successive layers.
This phase relation suppresses the collective interlayer contribution that is responsible for the strong softening of the dispersion at a three-dimensional Brillouin zone center.

Overall, these observations suggest that dipolar interactions contribute substantially to the low-energy magnetism of $\mathrm{KTmSe}_2$.
A definitive assessment will require a systematic quantitative analysis of the INS spectral intensity across the entire Brillouin zone.
Higher-resolution measurements near the measured $\Gamma_{1,2}$ points and a true three-dimensional $\Gamma$ point would help distinguish the momentum dependence of dipolar and finite-range contributions, while
measurements near a nonzero reciprocal-lattice vector of the full crystal would test the predicted softening of a two-dimensional in-plane modeling at a three-dimensional zone center.
Measurements of the field dependence of the excitation spectrum would provide an independent test of the proposed $J_1$-dipolar description.

Along the measured momentum path, the two-dimensional model with isotropic dipolar interactions already captures the low-energy dispersion reasonably well.
Its comparatively simple description also makes $\mathrm{KTmSe}_2$ a promising candidate for a one-to-one quantum-simulation-to-material experiment on platforms capable of simulating dipolar interactions \cite{Britton2012,Bohnet2016,Chomaz2022,Christakis2023,Li2023,Miller2024,Laupretre2026,Leclerc2026,Davis2023,Chen2025XY}.

\section{Conclusions and Outlook}
\label{sec:OutlookAndConclusion}

In this work, we introduced the graph zeta method as a deterministic and efficient approach for evaluating the high-dimensional lattice sums arising in the embedding step of linked-cluster expansions for quantum models with long-range interactions.
An essential preparatory step enabling the application of this method is an exact softcore mapping that removes explicit summation constraints of the original hardcore embedding problem and expresses the perturbative series coefficients in terms of unconstrained graph lattice sums.
A central step of the graph zeta method is the factorization of these graph zeta functions into minimal computable blocks which are categorized by their treewidth $\mathrm{tw}$.
For arbitrary combinations of power-law and short-range interaction kernels,
bridges ($\mathrm{tw}=1$) and zero-momentum cycles ($\mathrm{tw}=2$) can be evaluated analytically, remaining series-parallel blocks with treewidth $\mathrm{tw} = 2$ are treated through an algebra of products and convolutions of Epstein zeta functions, and highly connected blocks $(\mathrm{tw} > 2)$ are evaluated using tensor-network bucket elimination.
Together with block caching, this decomposition shifts the dominant exponential complexity from the total number of graph vertices to the treewidth of the constituent blocks and permits extensive reuse of recurring contributions.
Beyond power-law and short-range interactions, the core technical elements apply without modification to any absolutely summable, translationally invariant and even interaction kernel and only particular block evaluations must be adapted accordingly.

We applied the graph zeta method to the long-range transverse-field Ising model with isotropic power-law interactions on chain, square, triangular, and cubic lattices, obtaining high-order series for the ground-state energy and one-quasiparticle excitation energies in one, two, and three dimensions.
The resulting coefficients agree with previously published Monte Carlo estimates within their statistical uncertainties while providing a more accurate reference that is free of stochastic sampling errors.
Such deterministic, numerically exact evaluations were previously computationally inaccessible.
Further, we obtain the momentum-dependence of calculated quantities over the full Brillouin zone in a single calculation, enabling access to the full dispersion relation.
These evaluations are possible in minutes on a personal computer, compared to previous cluster-scale computation times for Monte Carlo results. 
This strongly reduced computational cost permits a dense scan of the interaction decay exponent and consequently a higher-resolution determination of critical points, critical exponents, and crossovers between universality regimes. 

The graph zeta method is implemented in the open-source Graph Zeta Library \cite{gzl2026}, which evaluates arbitrary graph lattice sums for combinations of power-law and short-range kernels. 
We foresee the Graph Zeta Library as a practical tool for rapidly testing microscopic models starting from a perturbative limit and evaluating them across large parameter sets. 
We demonstrated this capability by comparing competing transverse-field Ising descriptions of $\mathrm{KTmSe}_2$ directly with its measured INS spectrum \cite{Zheng2023KTmSe2}. 
Among the models considered, the $J_1$-dipolar model provides the best overall description of the dispersion, giving suggestive evidence that dipolar interactions contribute substantially to the low-energy magnetism of $\mathrm{KTmSe}_2$.

Beyond this demonstrated example, 
the method is directly relevant to several quantum-simulation platforms. 
For example, Rydberg atom arrays realize long-range transverse-field Ising models in which the chosen Rydberg states interact through a van-der-Waals potential.
As dynamical structure factors are now experimentally accessible in these platforms \cite{Sun2026}, the momentum-resolved excitation spectra and, in future extensions, spectral weights produced by the graph zeta method can be compared directly with measurements.
Three-dimensional dipolar transverse-field Ising materials provide a further application.
Representative systems include $\mathrm{LiHoF}_4$ \cite{Hansen1975,Bitko1996,Chakraborty2004,Tabei2008,Gingras2011}, the molecular nanomagnet $\mathrm{Fe}_8$ \cite{MartnezHidalgo2001,Buruzi2011}, $\mathrm{Mn}_{12}$ acetates \cite{Li2010,Millis2010,Subedi2012}, and the rare-earth materials $\mathrm{RE(OH)}_3$ \cite{Gingras2011REOH}. 
Their full dipolar coupling is angle dependent. 
The spatial dimension itself poses no obstacle for the graph zeta method and series for three-dimensional systems are obtained in minutes, on a single desktop core. 
Also, the block decomposition is kernel agnostic, so these materials can be addressed by supplying the appropriate anisotropic lattice Fourier transform while adapting the semi-analytic zeta algebra accordingly. For a recent treatment of anisotropic kernels, see \cite{buchheit2026zeta}. The associated lattice Fourier transforms are already implemented in EpsteinLib \cite{epsteinlib}. 

Future developments include the extension to multi-atomic lattices, broader classes of interaction kernels, convergence improvements based on Epstein zeta derivatives, and refinements of the tensor-network architecture, including a singularity treatment comparable to the zeta algebra for series-parallel graphs. 
We also aim to apply the method to additional observables, such as spectral weights and higher quasiparticle sectors, to other quantum models and lattice architectures, and finite temperatures \cite{Burkard2026A,Burkard2026B}. 
Together with the $\mathrm{KTmSe}_2$ application, these directions establish the graph zeta method as a fast, deterministic route from microscopic long-range Hamiltonians to quantitative, momentum-resolved predictions for quantum matter.

\section*{Author contributions}
A.D.: Conceptualization; Methodology; Software (preliminary development and Graph Zeta Library testing); Data curation, formal analysis, investigation (LRTFIM); Writing--- original draft (graph zeta method overview and algorithms, LRTFIM application); Visualization. 

P.A.: Conceptualization; Methodology; Writing---original draft (Linked-cluster expansion to hardcore problem).

J.A.K.: Conceptualization; Methodology; Software (softcore mapping); Data curation, formal analysis, investigation, visualization (experimental application); Writing---original draft (softcore mapping, experimental application).

A.A.B.: Conceptualization; Methodology (mathematical foundations); Software (Graph Zeta Library); Data analysis; Investigation; Writing---original draft (graph zeta method overview).

K.P.S.: Conceptualization; Funding acquisition; Supervision.

All authors contributed to Writing---review \& editing.

\section*{Data availability}

The open-source high-performance Graph Zeta Library \cite{gzl2026}, including its main dependency, EpsteinLib \cite{epsteinlib}, is publicly available.
The raw data and the code to generate the results of this paper can be made available from A.D. (\href{antonia.duft@fau.de}{antonia.duft@fau.de}) upon reasonable request. 
All results presented in this work are easily reproducible using the Graph Zeta Library.

\section*{Acknowledgments}

A.A.B. and J.A.K. remember the great spirit of the ``SiQuMa24 Conference'' with focus on the simulation of Quantum Matter with long-range interactions held at Schloss Dagstuhl in summer 2024, where first discussions on the subject of this work have started.
We are grateful to Jonathan Kaspar Busse whose implementation of the angular dependence in EpsteinLib was essential for assessing the full three-dimensional dipolar model of $\text{KTmSe}_2$.
We gratefully acknowledge the scientific support and HPC resources provided by the Erlangen National High Performance Computing Center (NHR@FAU) of the Friedrich-Alexander-Universität Erlangen-Nürnberg (FAU) under the NHR projects b177dc (``SELRIQS'') and n101af (``SuperSEM'').
The hardware of NHR@FAU is funded by the German Research Foundation (DFG).
A.D., P.A., J.A.K., and K.P.S. acknowledge support by the Deutsche Forschungsgemeinschaft (DFG, German Research Foundation), Project-ID 429529648–TRR 306 QuCoLiMa (Quantum Cooperativity of Light and Matter) and by the Munich Quantum Valley, which is supported by the Bavarian state government with funds from the Hightech Agenda Bayern Plus.
J.A.K. is funded by the Austrian Science Fund (FWF) [10.55776/COE1, 10.55776/F101200] and the European Union (NextGenerationEU). 
A.A.B. acknowledges support of this work by the Klaus-Tschira Stiftung under Grant No. 00.025.2025.

\bibliographystyle{apsrev4-1-etal}
\bibliography{bibliography.bib}

@article{Adelhardt2020,
  author = {Adelhardt, Patrick and Koziol, Jan Alexander and Schellenberger, Andreas and Schmidt, Kai Phillip},
  title = {Quantum Criticality and Excitations of a Long-Range Anisotropic {{XY}} Chain in a Transverse Field},
  journal = {Physical Review B},
  volume = {102},
  number = {17},
  pages = {174424},
  year = {2020},
  month = nov,
  publisher = {American Physical Society},
  doi = {10.1103/PhysRevB.102.174424},
  url = {https://doi.org/10.1103/PhysRevB.102.174424},
  urldate = {2025-02-03}
}

@article{Adelhardt2023,
  author = {Adelhardt, Patrick and Schmidt, Kai P.},
  title = {Continuously varying critical exponents in long-range quantum spin ladders},
  journal = {SciPost Physics},
  volume = {15},
  number = {3},
  pages = {087},
  year = {2023},
  month = sep,
  publisher = {Stichting SciPost},
  doi = {10.21468/scipostphys.15.3.087},
  url = {https://doi.org/10.21468/scipostphys.15.3.087},
  issn = {2542-4653}
}

@article{Adelhardt2024,
  author = {Adelhardt, Patrick and Koziol, Jan A. and Langheld, Anja and Schmidt, Kai P.},
  title = {Monte {{Carlo Based Techniques}} for {{Quantum Magnets}} with {{Long-Range Interactions}}},
  journal = {Entropy},
  volume = {26},
  number = {5},
  pages = {401},
  year = {2024},
  month = may,
  publisher = {Multidisciplinary Digital Publishing Institute},
  doi = {10.3390/e26050401},
  url = {https://doi.org/10.3390/e26050401},
  issn = {1099-4300},
  urldate = {2025-02-03}
}

@article{Adelhardt2025,
  author = {Adelhardt, Patrick and Duft, Antonia and Schmidt, Kai Phillip},
  title = {Quantum-critical and dynamical properties of the {XXZ} bilayer with long-range interactions},
  journal = {Physical Review B},
  volume = {111},
  number = {2},
  pages = {024409},
  year = {2025},
  month = jan,
  publisher = {American Physical Society (APS)},
  doi = {10.1103/physrevb.111.024409},
  url = {https://doi.org/10.1103/physrevb.111.024409},
  issn = {2469-9969}
}

@article{Adelhardt2026,
  author = {Adelhardt, Patrick and Muleady, Sean R. and Schmidt, Kai P. and Gorshkov, Alexey V.},
  title = {Unconventional entanglement scaling and quantum criticality in the long-range spin-one {Heisenberg} chain with single-ion anisotropy},
  journal = {Physical Review Research},
  year = {2026},
  publisher = {American Physical Society},
  doi = {10.1103/3cyg-8wsw},
  url = {https://doi.org/10.1103/3cyg-8wsw},
  eprint = {2604.12754},
  archivePrefix = {arXiv},
  primaryClass = {cond-mat.str-el},
  note = {Accepted for publication, 5 September 2026}
}

@article{Baier2016,
  author = {Baier, S. and Mark, M. J. and Petter, D. and Aikawa, K. and Chomaz, L. and Cai, Z. and Baranov, M. and Zoller, P. and Ferlaino, F.},
  title = {Extended {Bose-Hubbard} models with ultracold magnetic atoms},
  journal = {Science},
  volume = {352},
  number = {6282},
  pages = {201--205},
  year = {2016},
  month = apr,
  publisher = {American Association for the Advancement of Science (AAAS)},
  doi = {10.1126/science.aac9812},
  url = {https://doi.org/10.1126/science.aac9812},
  issn = {1095-9203}
}

@book{Baker1975,
  author = {Baker, Jr., George A.},
  title = {Essentials of {Pad{\'e}} Approximants},
  year = {1975},
  publisher = {Academic Press},
  isbn = {0-12-074855-X}
}

@article{Barredo2015,
  author = {Barredo, Daniel and Labuhn, Henning and Ravets, Sylvain and Lahaye, Thierry and Browaeys, Antoine and Adams, Charles S.},
  title = {Coherent Excitation Transfer in a Spin Chain of Three {Rydberg} Atoms},
  journal = {Physical Review Letters},
  volume = {114},
  number = {11},
  pages = {113002},
  year = {2015},
  month = mar,
  publisher = {American Physical Society},
  doi = {10.1103/PhysRevLett.114.113002},
  url = {https://doi.org/10.1103/PhysRevLett.114.113002}
}

@article{Bauls2023,
  author = {Ba{\~n}uls, Mari Carmen},
  title = {Tensor Network Algorithms: A Route Map},
  journal = {Annual Review of Condensed Matter Physics},
  volume = {14},
  number = {1},
  pages = {173--191},
  year = {2023},
  month = mar,
  publisher = {Annual Reviews},
  doi = {10.1146/annurev-conmatphys-040721-022705},
  url = {https://doi.org/10.1146/annurev-conmatphys-040721-022705},
  issn = {1947-5462}
}

@book{Becca2017,
  author = {Becca, Federico and Sorella, Sandro},
  title = {Quantum {Monte Carlo} Approaches for Correlated Systems},
  year = {2017},
  month = nov,
  publisher = {Cambridge University Press},
  doi = {10.1017/9781316417041},
  url = {https://doi.org/10.1017/9781316417041},
  isbn = {9781316417041}
}

@article{Bernien2017,
  author = {Bernien, Hannes and Schwartz, Sylvain and Keesling, Alexander and Levine, Harry and Omran, Ahmed and Pichler, Hannes and Choi, Soonwon and Zibrov, Alexander S. and Endres, Manuel and Greiner, Markus and Vuleti{\'c}, Vladan and Lukin, Mikhail D.},
  title = {Probing many-body dynamics on a 51-atom quantum simulator},
  journal = {Nature},
  volume = {551},
  number = {7682},
  pages = {579--584},
  year = {2017},
  month = nov,
  publisher = {Springer Science and Business Media LLC},
  doi = {10.1038/nature24622},
  url = {https://doi.org/10.1038/nature24622},
  issn = {1476-4687}
}

@article{Bitko1996,
  author = {Bitko, D. and Rosenbaum, T. F. and Aeppli, G.},
  title = {Quantum Critical Behavior for a Model Magnet},
  journal = {Physical Review Letters},
  volume = {77},
  number = {5},
  pages = {940--943},
  year = {1996},
  month = jul,
  publisher = {American Physical Society},
  doi = {10.1103/PhysRevLett.77.940},
  url = {https://doi.org/10.1103/PhysRevLett.77.940}
}

@article{Bloch2008,
  author = {Bloch, Immanuel and Dalibard, Jean and Zwerger, Wilhelm},
  title = {Many-body physics with ultracold gases},
  journal = {Reviews of Modern Physics},
  volume = {80},
  number = {3},
  pages = {885--964},
  year = {2008},
  month = jul,
  publisher = {American Physical Society},
  doi = {10.1103/RevModPhys.80.885},
  url = {https://doi.org/10.1103/RevModPhys.80.885}
}

@article{bodlaender1998,
  author = {Bodlaender, Hans L.},
  title = {A Partial {$k$}-Arboretum of Graphs with Bounded Treewidth},
  journal = {Theoretical Computer Science},
  volume = {209},
  number = {1--2},
  pages = {1--45},
  year = {1998},
  doi = {10.1016/S0304-3975(97)00228-4},
  url = {https://doi.org/10.1016/S0304-3975(97)00228-4}
}

@article{Bohn2017,
  author = {Bohn, John L. and Rey, Ana Maria and Ye, Jun},
  title = {Cold molecules: Progress in quantum engineering of chemistry and quantum matter},
  journal = {Science},
  volume = {357},
  number = {6355},
  pages = {1002--1010},
  year = {2017},
  month = sep,
  publisher = {American Association for the Advancement of Science (AAAS)},
  doi = {10.1126/science.aam6299},
  url = {https://doi.org/10.1126/science.aam6299},
  issn = {1095-9203}
}

@article{Bohnet2016,
  author = {Bohnet, Justin G. and Sawyer, Brian C. and Britton, Joseph W. and Wall, Michael L. and Rey, Ana Maria and Foss-Feig, Michael and Bollinger, John J.},
  title = {Quantum spin dynamics and entanglement generation with hundreds of trapped ions},
  journal = {Science},
  volume = {352},
  number = {6291},
  pages = {1297--1301},
  year = {2016},
  month = jun,
  publisher = {American Association for the Advancement of Science (AAAS)},
  doi = {10.1126/science.aad9958},
  url = {https://doi.org/10.1126/science.aad9958},
  issn = {1095-9203}
}

@book{bondy2008graph,
  author = {Bondy, John Adrian and Murty, Uppaluri Siva Ramachandra},
  title = {Graph Theory},
  series = {Graduate Texts in Mathematics},
  volume = {244},
  year = {2008},
  publisher = {Springer London},
  doi = {10.1007/978-1-84628-970-5},
  url = {https://doi.org/10.1007/978-1-84628-970-5}
}

@article{Bordelon2019NaYbO2,
  author = {Bordelon, Mitchell M. and Kenney, Eric and Liu, Chunxiao and Hogan, Tom and Posthuma, Lorenzo and Kavand, Marzieh and Lyu, Yuanqi and Sherwin, Mark and Butch, N. P. and Brown, Craig and Graf, M. J. and Balents, Leon and Wilson, Stephen D.},
  title = {Field-tunable quantum disordered ground state in the triangular-lattice antiferromagnet ${\mathrm{NaYbO}}_{2}$},
  journal = {Nature Physics},
  volume = {15},
  number = {10},
  pages = {1058--1064},
  year = {2019},
  month = oct,
  doi = {10.1038/s41567-019-0594-5},
  url = {https://doi.org/10.1038/s41567-019-0594-5}
}

@article{Britton2012,
  author = {Britton, Joseph W. and Sawyer, Brian C. and Keith, Adam C. and Wang, C.-C. Joseph and Freericks, James K. and Uys, Hermann and Biercuk, Michael J. and Bollinger, John J.},
  title = {Engineered two-dimensional {Ising} interactions in a trapped-ion quantum simulator with hundreds of spins},
  journal = {Nature},
  volume = {484},
  number = {7395},
  pages = {489--492},
  year = {2012},
  month = apr,
  publisher = {Springer Science and Business Media LLC},
  doi = {10.1038/nature10981},
  url = {https://doi.org/10.1038/nature10981},
  issn = {1476-4687}
}

@article{Browaeys2020,
  author = {Browaeys, Antoine and Lahaye, Thierry},
  title = {Many-body physics with individually controlled {Rydberg} atoms},
  journal = {Nature Physics},
  volume = {16},
  number = {2},
  pages = {132--142},
  year = {2020},
  month = jan,
  publisher = {Springer Science and Business Media LLC},
  doi = {10.1038/s41567-019-0733-z},
  url = {https://doi.org/10.1038/s41567-019-0733-z},
  issn = {1745-2481}
}

@article{buchheit2024epstein,
  author = {Buchheit, Andreas A and Busse, Jonathan K and Gutendorf, Ruben},
  title = {Computation and properties of the {Epstein} zeta function with applications to quantum systems},
  journal = {IMA Journal of Numerical Analysis},
  pages = {drag057},
  year = {2026},
  month = jul,
  publisher = {Oxford University Press},
  doi = {10.1093/imanum/drag057},
  url = {https://doi.org/10.1093/imanum/drag057}
}

@article{buchheit2025,
  author = {Buchheit, Andreas A. and Busse, Jonathan K.},
  title = {{Epstein} zeta method for many-body lattice sums},
  journal = {Numerische Mathematik},
  year = {2026},
  month = jul,
  doi = {10.1007/s00211-026-01558-y},
  url = {https://doi.org/10.1007/s00211-026-01558-y}
}

@article{buchheit2026,
  author        = {Buchheit, Andreas A. and Rupp, Andreas},
  title         = {Graph lattice sums and graph zeta functions for long-range interacting quantum lattice models},
  year          = {2026},
  eprint        = {2609.18918},
  archivePrefix = {arXiv},
  primaryClass  = {math.NA}
}

@article{buchheit2026zeta,
  author = {Buchheit, Andreas Alexander and Busse, Jonathan Kaspar and Ke{\ss}ler, Torsten and Rybakov, Filipp N},
  title = {Zeta expansion for long-range interactions under periodic boundary conditions with applications to micromagnetics},
  journal = {Journal of Computational Physics},
  volume = {559},
  pages = {114885},
  year = {2026},
  publisher = {Elsevier},
  doi = {10.1016/j.jcp.2026.114885},
  url = {https://doi.org/10.1016/j.jcp.2026.114885}
}

@article{Burkard2026A,
  author = {Burkard, Ruben and Schneider, Benedikt and Sbierski, Bj\"orn},
  title = {Dynamic Correlations of Frustrated Quantum Spins from High-Temperature Expansion},
  journal = {Physical Review Letters},
  volume = {136},
  number = {5},
  pages = {056501},
  year = {2026},
  month = feb,
  publisher = {American Physical Society},
  doi = {10.1103/jtjk-x2lw},
  url = {https://doi.org/10.1103/jtjk-x2lw}
}

@article{Burkard2026B,
  author = {Burkard, Ruben and Schneider, Benedikt and Sbierski, Bj\"orn},
  title = {High-temperature series expansion of the dynamic {Matsubara} spin correlator},
  journal = {Physical Review B},
  volume = {113},
  number = {7},
  pages = {075102},
  year = {2026},
  month = feb,
  publisher = {American Physical Society},
  doi = {10.1103/1l92-z6qd},
  url = {https://doi.org/10.1103/1l92-z6qd}
}

@article{Buruzi2011,
  author = {Burzur\'{\i}, E. and Luis, F. and Barbara, B. and Ballou, R. and Ressouche, E. and Montero, O. and Campo, J. and Maegawa, S.},
  title = {Magnetic Dipolar Ordering and Quantum Phase Transition in an ${\mathrm{Fe}}_{8}$ Molecular Magnet},
  journal = {Physical Review Letters},
  volume = {107},
  number = {9},
  pages = {097203},
  year = {2011},
  month = aug,
  publisher = {American Physical Society},
  doi = {10.1103/PhysRevLett.107.097203},
  url = {https://doi.org/10.1103/PhysRevLett.107.097203}
}

@article{Carr2009,
  author = {Carr, Lincoln D and DeMille, David and Krems, Roman V and Ye, Jun},
  title = {Cold and ultracold molecules: science, technology and applications},
  journal = {New Journal of Physics},
  volume = {11},
  number = {5},
  pages = {055049},
  year = {2009},
  month = may,
  publisher = {IOP Publishing},
  doi = {10.1088/1367-2630/11/5/055049},
  url = {https://doi.org/10.1088/1367-2630/11/5/055049},
  issn = {1367-2630}
}

@book{Carr2010,
  editor = {Carr, Lincoln D.},
  title = {Understanding Quantum Phase Transitions},
  year = {2010},
  publisher = {CRC Press},
  doi = {10.1201/b10273},
  url = {https://doi.org/10.1201/b10273},
  isbn = {9781439802618}
}

@article{Cederbaum1989,
  author = {Cederbaum, L S and Schirmer, J and Meyer, H -D},
  title = {Block diagonalisation of {Hermitian} matrices},
  journal = {Journal of Physics A: Mathematical and General},
  volume = {22},
  number = {13},
  pages = {2427--2439},
  year = {1989},
  month = jul,
  publisher = {IOP Publishing},
  doi = {10.1088/0305-4470/22/13/035},
  url = {https://doi.org/10.1088/0305-4470/22/13/035},
  issn = {1361-6447}
}

@article{Chakraborty2004,
  author = {Chakraborty, P. B. and Henelius, P. and Kj\o{}nsberg, H. and Sandvik, A. W. and Girvin, S. M.},
  title = {Theory of the magnetic phase diagram of {LiHoF}$_{4}$},
  journal = {Physical Review B},
  volume = {70},
  number = {14},
  pages = {144411},
  year = {2004},
  month = oct,
  publisher = {American Physical Society},
  doi = {10.1103/PhysRevB.70.144411},
  url = {https://doi.org/10.1103/PhysRevB.70.144411}
}

@article{Chen2023,
  author = {Chen, Cheng and Bornet, Guillaume and Bintz, Marcus and Emperauger, Gabriel and Leclerc, Lucas and Liu, Vincent S. and Scholl, Pascal and Barredo, Daniel and Hauschild, Johannes and Chatterjee, Shubhayu and Schuler, Michael and L\"{a}uchli, Andreas M. and Zaletel, Michael P. and Lahaye, Thierry and Yao, Norman Y. and Browaeys, Antoine},
  title = {Continuous symmetry breaking in a two-dimensional {Rydberg} array},
  journal = {Nature},
  volume = {616},
  number = {7958},
  pages = {691--695},
  year = {2023},
  month = feb,
  publisher = {Springer Science and Business Media LLC},
  doi = {10.1038/s41586-023-05859-2},
  url = {https://doi.org/10.1038/s41586-023-05859-2},
  issn = {1476-4687}
}

@article{Chen2025XY,
  author = {Cheng Chen and Gabriel Emperauger and Guillaume Bornet and Filippo Caleca and Bastien G{\'e}ly and Marcus Bintz and Shubhayu Chatterjee and Vincent Liu and Daniel Barredo and Norman Y. Yao and Thierry Lahaye and Fabio Mezzacapo and Tommaso Roscilde and Antoine Browaeys},
  title = {Spectroscopy of elementary excitations from quench dynamics in a dipolar {XY} {Rydberg} simulator},
  journal = {Science},
  volume = {389},
  number = {6759},
  pages = {483--487},
  year = {2025},
  doi = {10.1126/science.adn0618},
  url = {https://doi.org/10.1126/science.adn0618}
}

@article{Chomaz2022,
  author = {Chomaz, Lauriane and Ferrier-Barbut, Igor and Ferlaino, Francesca and Laburthe-Tolra, Bruno and Lev, Benjamin L and Pfau, Tilman},
  title = {Dipolar physics: a review of experiments with magnetic quantum gases},
  journal = {Reports on Progress in Physics},
  volume = {86},
  number = {2},
  pages = {026401},
  year = {2023},
  month = feb,
  publisher = {IOP Publishing},
  doi = {10.1088/1361-6633/aca814},
  url = {https://doi.org/10.1088/1361-6633/aca814},
  issn = {1361-6633}
}

@article{Christakis2023,
  author = {Christakis, Lysander and Rosenberg, Jason S. and Raj, Ravin and Chi, Sungjae and Morningstar, Alan and Huse, David A. and Yan, Zoe Z. and Bakr, Waseem S.},
  title = {Probing site-resolved correlations in a spin system of ultracold molecules},
  journal = {Nature},
  volume = {614},
  number = {7946},
  pages = {64--69},
  year = {2023},
  month = feb,
  publisher = {Springer Science and Business Media LLC},
  doi = {10.1038/s41586-022-05558-4},
  url = {https://doi.org/10.1038/s41586-022-05558-4},
  issn = {1476-4687}
}

@article{Cirac2021,
  author = {Cirac, J. Ignacio and P\'erez-Garc\'{\i}a, David and Schuch, Norbert and Verstraete, Frank},
  title = {Matrix product states and projected entangled pair states: Concepts, symmetries, theorems},
  journal = {Reviews of Modern Physics},
  volume = {93},
  number = {4},
  pages = {045003},
  year = {2021},
  month = dec,
  publisher = {American Physical Society},
  doi = {10.1103/RevModPhys.93.045003},
  url = {https://doi.org/10.1103/RevModPhys.93.045003}
}

@article{Cloizeaux1960,
  author = {des Cloizeaux, Jacques},
  title = {Extension d'une formule de {Lagrange} {\`a} des probl{\`e}mes de valeurs propres},
  journal = {Nuclear Physics},
  volume = {20},
  pages = {321--346},
  year = {1960},
  month = oct,
  publisher = {Elsevier BV},
  doi = {10.1016/0029-5582(60)90177-2},
  url = {https://doi.org/10.1016/0029-5582(60)90177-2},
  issn = {0029-5582}
}

@article{Coester2015,
  author = {Coester, K. and Schmidt, K. P.},
  title = {Optimizing linked-cluster expansions by white graphs},
  journal = {Physical Review E},
  volume = {92},
  number = {2},
  pages = {022118},
  year = {2015},
  month = aug,
  publisher = {American Physical Society (APS)},
  doi = {10.1103/physreve.92.022118},
  url = {https://doi.org/10.1103/physreve.92.022118},
  issn = {1550-2376}
}

@article{Cordella2004VF2,
  author = {Cordella, Luigi P. and Foggia, Pasquale and Sansone, Carlo and Vento, Mario},
  title = {A (Sub)Graph Isomorphism Algorithm for Matching Large Graphs},
  journal = {IEEE Transactions on Pattern Analysis and Machine Intelligence},
  volume = {26},
  number = {10},
  pages = {1367--1372},
  year = {2004},
  doi = {10.1109/TPAMI.2004.75},
  url = {https://doi.org/10.1109/TPAMI.2004.75}
}

@article{Cornish2024,
  author = {Cornish, Simon L. and Tarbutt, Michael R. and Hazzard, Kaden R. A.},
  title = {Quantum computation and quantum simulation with ultracold molecules},
  journal = {Nature Physics},
  volume = {20},
  number = {5},
  pages = {730--740},
  year = {2024},
  month = may,
  publisher = {Springer Science and Business Media LLC},
  doi = {10.1038/s41567-024-02453-9},
  url = {https://doi.org/10.1038/s41567-024-02453-9},
  issn = {1745-2481}
}

@article{Dai2021NaYbSe2,
  author = {Dai, Peng-Ling and Zhang, Gaoning and Xie, Yaofeng and Duan, Chunruo and Gao, Yonghao and Zhu, Zihao and Feng, Erxi and Tao, Zhen and Huang, Chien-Lung and Cao, Huibo and Podlesnyak, Andrey and Granroth, Garrett E. and Everett, Michelle S. and Neuefeind, Joerg C. and Voneshen, David and Wang, Shun and Tan, Guotai and Morosan, Emilia and Wang, Xia and Lin, Hai-Qing and Shu, Lei and Chen, Gang and Guo, Yanfeng and Lu, Xingye and Dai, Pengcheng},
  title = {Spinon {Fermi} Surface Spin Liquid in a Triangular Lattice Antiferromagnet ${\mathrm{NaYbSe}}_{2}$},
  journal = {Physical Review X},
  volume = {11},
  number = {2},
  pages = {021044},
  year = {2021},
  month = may,
  doi = {10.1103/PhysRevX.11.021044},
  url = {https://doi.org/10.1103/PhysRevX.11.021044}
}

@article{Davis2023,
  author = {Davis, E. J. and Ye, B. and Machado, F. and Meynell, S. A. and Wu, W. and Mittiga, T. and Schenken, W. and Joos, M. and Kobrin, B. and Lyu, Y. and Wang, Z. and Bluvstein, D. and Choi, S. and Zu, C. and Bleszynski Jayich, A. C. and Yao, N. Y.},
  title = {Probing Many-body Dynamics in a Two-dimensional Dipolar Spin Ensemble},
  journal = {Nature Physics},
  volume = {19},
  number = {6},
  pages = {836--844},
  year = {2023},
  doi = {10.1038/s41567-023-01944-5},
  url = {https://doi.org/10.1038/s41567-023-01944-5}
}

@article{Dechter1999,
  author = {Dechter, Rina},
  title = {Bucket elimination: A unifying framework for reasoning},
  journal = {Artificial Intelligence},
  volume = {113},
  number = {1-2},
  pages = {41--85},
  year = {1999},
  doi = {10.1016/S0004-3702(99)00059-4},
  url = {https://doi.org/10.1016/S0004-3702(99)00059-4}
}

@article{Defenu2017,
  author = {Defenu, Nicol{\`o} and Trombettoni, Andrea and Ruffo, Stefano},
  title = {Criticality and phase diagram of quantum long-range {$O(N)$} models},
  journal = {Physical Review B},
  volume = {96},
  number = {10},
  pages = {104432},
  year = {2017},
  month = sep,
  publisher = {American Physical Society},
  doi = {10.1103/PhysRevB.96.104432},
  url = {https://doi.org/10.1103/PhysRevB.96.104432}
}

@article{Defenu2023,
  author = {Defenu, Nicol\`o and Donner, Tobias and Macr\`{\i}, Tommaso and Pagano, Guido and Ruffo, Stefano and Trombettoni, Andrea},
  title = {Long-range interacting quantum systems},
  journal = {Reviews of Modern Physics},
  volume = {95},
  number = {3},
  pages = {035002},
  year = {2023},
  month = aug,
  publisher = {American Physical Society},
  doi = {10.1103/RevModPhys.95.035002},
  url = {https://doi.org/10.1103/RevModPhys.95.035002}
}

@article{dibattista1996,
  author = {Di Battista, Giuseppe and Tamassia, Roberto},
  title = {On-line planarity testing},
  journal = {SIAM Journal on Computing},
  volume = {25},
  number = {5},
  pages = {956--997},
  year = {1996},
  doi = {10.1137/S0097539794280736},
  url = {https://doi.org/10.1137/S0097539794280736}
}

@book{diestel2025graph,
  author = {Diestel, Reinhard},
  title = {Graph Theory},
  series = {Graduate Texts in Mathematics},
  edition = {6th},
  volume = {173},
  year = {2025},
  publisher = {Springer},
  address = {Heidelberg},
  doi = {10.1007/978-3-662-70107-2},
  url = {https://doi.org/10.1007/978-3-662-70107-2},
  isbn = {978-3-662-70106-5}
}

@article{Duffin1965,
  author = {Duffin, R. J.},
  title = {Topology of series-parallel networks},
  journal = {Journal of Mathematical Analysis and Applications},
  volume = {10},
  number = {2},
  pages = {303--318},
  year = {1965},
  doi = {10.1016/0022-247X(65)90125-3},
  url = {https://doi.org/10.1016/0022-247X(65)90125-3}
}

@article{Dutta2001,
  author = {Dutta, Amit and Bhattacharjee, J. K.},
  title = {Phase Transitions in the Quantum {{Ising}} and Rotor Models with a Long-Range Interaction},
  journal = {Physical Review B},
  volume = {64},
  number = {18},
  pages = {184106},
  year = {2001},
  month = oct,
  publisher = {American Physical Society},
  doi = {10.1103/PhysRevB.64.184106},
  url = {https://doi.org/10.1103/PhysRevB.64.184106}
}

@article{Ebadi2021,
  author = {Ebadi, Sepehr and Wang, Tout T. and Levine, Harry and Keesling, Alexander and Semeghini, Giulia and Omran, Ahmed and Bluvstein, Dolev and Samajdar, Rhine and Pichler, Hannes and Ho, Wen Wei and Choi, Soonwon and Sachdev, Subir and Greiner, Markus and Vuleti{\'c}, Vladan and Lukin, Mikhail D.},
  title = {Quantum phases of matter on a 256-atom programmable quantum simulator},
  journal = {Nature},
  volume = {595},
  number = {7866},
  pages = {227--232},
  year = {2021},
  month = jul,
  publisher = {Springer Science and Business Media LLC},
  doi = {10.1038/s41586-021-03582-4},
  url = {https://doi.org/10.1038/s41586-021-03582-4},
  issn = {1476-4687}
}

@article{epstein1903theorieI,
  author = {Epstein, P.},
  title = {{Zur Theorie allgemeiner Zetafunctionen}},
  journal = {Mathematische Annalen},
  volume = {56},
  number = {4},
  pages = {615--644},
  year = {1903},
  publisher = {Springer},
  doi = {10.1007/bf01444309},
  url = {https://doi.org/10.1007/bf01444309}
}

@article{epstein1903theorieII,
  author = {Epstein, P.},
  title = {{Zur Theorie allgemeiner Zetafunktionen. II}},
  journal = {Mathematische Annalen},
  volume = {63},
  number = {2},
  pages = {205--216},
  year = {1906},
  publisher = {Springer},
  doi = {10.1007/bf01449900},
  url = {https://doi.org/10.1007/bf01449900}
}

@misc{epsteinlib,
  author = {Buchheit, Andreas A. and Busse, Jonathan K. and Gutendorf, Ruben},
  title = {{EpsteinLib v0.6.2}},
  year = {2026},
  url = {https://github.com/epsteinlib/epsteinlib},
  version = {0.6.2},
  license = {AGPL-3.0-only},
  note = {Software, version 0.6.2},
  urldate = {2026-09-11}
}

@article{Fey2016,
  author = {Fey, Sebastian and Schmidt, Kai Phillip},
  title = {Critical behavior of quantum magnets with long-range interactions in the thermodynamic limit},
  journal = {Physical Review B},
  volume = {94},
  number = {7},
  pages = {075156},
  year = {2016},
  month = aug,
  publisher = {American Physical Society},
  doi = {10.1103/PhysRevB.94.075156},
  url = {https://doi.org/10.1103/PhysRevB.94.075156}
}

@article{Fey2019,
  author = {Fey, Sebastian and Kapfer, Sebastian C and Schmidt, Kai Phillip},
  title = {Quantum criticality of two-dimensional quantum magnets with long-range interactions},
  journal = {Physical Review Letters},
  volume = {122},
  number = {1},
  pages = {017203},
  year = {2019},
  publisher = {APS},
  doi = {10.1103/physrevlett.122.017203},
  url = {https://doi.org/10.1103/physrevlett.122.017203}
}

@phdthesis{Fey2020Diss,
  author = {Fey, Sebastian},
  title = {Investigation of zero-temperature transverse-field {Ising} models with long-range interactions},
  school = {Friedrich-Alexander-Universit{\"a}t Erlangen-N{\"u}rnberg},
  year = {2020},
  url = {https://www.emergent.physics.nat.fau.de/files/2025/02/2020_Dissertation_Fey.pdf}
}

@article{Friedenauer2008,
  author = {Friedenauer, A. and Schmitz, H. and Glueckert, J. T. and Porras, D. and Schaetz, T.},
  title = {Simulating a quantum magnet with trapped ions},
  journal = {Nature Physics},
  volume = {4},
  number = {10},
  pages = {757--761},
  year = {2008},
  month = jul,
  publisher = {Springer Science and Business Media LLC},
  doi = {10.1038/nphys1032},
  url = {https://doi.org/10.1038/nphys1032},
  issn = {1745-2481}
}

@article{Gelfand2000,
  author = {Gelfand, Martin P. and Singh, Rajiv R. P.},
  title = {High-order convergent expansions for quantum many particle systems},
  journal = {Advances in Physics},
  volume = {49},
  number = {1},
  pages = {93--140},
  year = {2000},
  month = jan,
  publisher = {Informa UK Limited},
  doi = {10.1080/000187300243390},
  url = {https://doi.org/10.1080/000187300243390},
  issn = {1460-6976}
}

@article{Gingras2011,
  author = {Gingras, Michel J P and Henelius, Patrik},
  title = {Collective Phenomena in the {$\mathrm{LiHo}_{x}\mathrm{Y}_{1-x}\mathrm{F}_{4}$} Quantum {Ising} Magnet: Recent Progress and Open Questions},
  journal = {Journal of Physics: Conference Series},
  volume = {320},
  pages = {012001},
  year = {2011},
  month = sep,
  publisher = {IOP Publishing},
  doi = {10.1088/1742-6596/320/1/012001},
  url = {https://doi.org/10.1088/1742-6596/320/1/012001},
  issn = {1742-6596}
}

@article{Gingras2011REOH,
  author = {Stasiak, Pawel and Gingras, Michel J. P.},
  title = {Assessment of the $\mathrm{RE}{(\text{OH})}_{3}$ {Ising} magnetic materials as possible candidates for the study of transverse-field-induced quantum phase transitions},
  journal = {Physical Review B},
  volume = {78},
  number = {22},
  pages = {224412},
  year = {2008},
  month = dec,
  publisher = {American Physical Society},
  doi = {10.1103/PhysRevB.78.224412},
  url = {https://doi.org/10.1103/PhysRevB.78.224412}
}

@article{Guo2024,
  author = {Guo, S.-A. and Wu, Y.-K. and Ye, J. and Zhang, L. and Lian, W.-Q. and Yao, R. and Wang, Y. and Yan, R.-Y. and Yi, Y.-J. and Xu, Y.-L. and Li, B.-W. and Hou, Y.-H. and Xu, Y.-Z. and Guo, W.-X. and Zhang, C. and Qi, B.-X. and Zhou, Z.-C. and He, L. and Duan, L.-M.},
  title = {A site-resolved two-dimensional quantum simulator with hundreds of trapped ions},
  journal = {Nature},
  volume = {630},
  number = {8017},
  pages = {613--618},
  year = {2024},
  month = may,
  publisher = {Springer Science and Business Media LLC},
  doi = {10.1038/s41586-024-07459-0},
  url = {https://doi.org/10.1038/s41586-024-07459-0},
  issn = {1476-4687}
}

@incollection{Guttmann1989,
  author = {Guttmann, A. J.},
  editor = {Domb, C. and Lebowitz, J. L.},
  title = {{Asymptotic Analysis of Power-Series Expansions}},
  booktitle = {Phase Transitions and Critical Phenomena},
  volume = {13},
  year = {1989},
  publisher = {Academic Press},
  isbn = {0122203135}
}

@misc{gzl2026,
  author    = {Buchheit, Andreas A.},
  title     = {Graph {Z}eta {L}ibrary ({GZL})},
  year      = {2026},
  version   = {1.0.0},
  publisher = {Zenodo},
  doi       = {10.5281/zenodo.22961920},
  url       = {https://github.com/graph-zeta/gzl},
  license   = {AGPL-3.0-or-later},
  note      = {v1.0.0, Zenodo}
}

@article{Hansen1975,
  author = {Hansen, P. E. and Johansson, T. and Nevald, R.},
  title = {Magnetic properties of lithium rare-earth fluorides: Ferromagnetism in LiErF$_{4}$ and {LiHoF}$_{4}$ and crystal-field parameters at the rare-earth and Li sites},
  journal = {Physical Review B},
  volume = {12},
  number = {11},
  pages = {5315--5324},
  year = {1975},
  month = dec,
  publisher = {American Physical Society},
  doi = {10.1103/PhysRevB.12.5315},
  url = {https://doi.org/10.1103/PhysRevB.12.5315}
}

@article{hasenbusch2019,
  author = {Hasenbusch, Martin},
  title = {Monte {Carlo} study of an improved clock model in three dimensions},
  journal = {Physical Review B},
  volume = {100},
  number = {22},
  pages = {224517},
  year = {2019},
  publisher = {American Physical Society},
  doi = {10.1103/PhysRevB.100.224517},
  url = {https://doi.org/10.1103/PhysRevB.100.224517},
  urldate = {2026-08-11}
}

@article{Hormann2023,
  author = {H\"{o}rmann, Max and Schmidt, Kai P.},
  title = {Projective cluster-additive transformation for quantum lattice models},
  journal = {SciPost Physics},
  volume = {15},
  number = {3},
  pages = {097},
  year = {2023},
  month = sep,
  publisher = {Stichting SciPost},
  doi = {10.21468/scipostphys.15.3.097},
  url = {https://doi.org/10.21468/scipostphys.15.3.097},
  issn = {2542-4653}
}

@article{Humeniuk2016,
  author = {Humeniuk, Stephan},
  title = {Quantum {Monte} {Carlo} study of long-range transverse-field {Ising} models on the triangular lattice},
  journal = {Physical Review B},
  volume = {93},
  number = {10},
  pages = {104412},
  year = {2016},
  month = mar,
  publisher = {American Physical Society},
  doi = {10.1103/PhysRevB.93.104412},
  url = {https://doi.org/10.1103/PhysRevB.93.104412},
  urldate = {2026-07-06}
}

@article{Islam2011,
  author = {Islam, R. and Edwards, E.E. and Kim, K. and Korenblit, S. and Noh, C. and Carmichael, H. and Lin, G.-D. and Duan, L.-M. and Joseph Wang, C.-C. and Freericks, J.K. and Monroe, C.},
  title = {Onset of a quantum phase transition with a trapped ion quantum simulator},
  journal = {Nature Communications},
  volume = {2},
  number = {1},
  pages = {377},
  year = {2011},
  month = jul,
  publisher = {Springer Science and Business Media LLC},
  doi = {10.1038/ncomms1374},
  url = {https://doi.org/10.1038/ncomms1374},
  issn = {2041-1723}
}

@article{Islam2013,
  author = {Islam, R. and Senko, C. and Campbell, W. C. and Korenblit, S. and Smith, J. and Lee, A. and Edwards, E. E. and Wang, C.-C. J. and Freericks, J. K. and Monroe, C.},
  title = {Emergence and Frustration of Magnetism with Variable-Range Interactions in a Quantum Simulator},
  journal = {Science},
  volume = {340},
  number = {6132},
  pages = {583--587},
  year = {2013},
  month = may,
  publisher = {American Association for the Advancement of Science (AAAS)},
  doi = {10.1126/science.1232296},
  url = {https://doi.org/10.1126/science.1232296},
  issn = {1095-9203}
}

@article{Jurcevic2014,
  author = {Jurcevic, P. and Lanyon, B. P. and Hauke, P. and Hempel, C. and Zoller, P. and Blatt, R. and Roos, C. F.},
  title = {Quasiparticle engineering and entanglement propagation in a quantum many-body system},
  journal = {Nature},
  volume = {511},
  number = {7508},
  pages = {202--205},
  year = {2014},
  month = jul,
  publisher = {Springer Science and Business Media LLC},
  doi = {10.1038/nature13461},
  url = {https://doi.org/10.1038/nature13461},
  issn = {1476-4687}
}

@article{Kiesenhofer2023,
  author = {Kiesenhofer, Dominik and Hainzer, Helene and Zhdanov, Artem and Holz, Philip C. and Bock, Matthias and Ollikainen, Tuomas and Roos, Christian F.},
  title = {Controlling Two-Dimensional {Coulomb} Crystals of More Than 100 Ions in a Monolithic Radio-Frequency Trap},
  journal = {PRX Quantum},
  volume = {4},
  number = {2},
  pages = {020317},
  year = {2023},
  month = apr,
  publisher = {American Physical Society},
  doi = {10.1103/PRXQuantum.4.020317},
  url = {https://doi.org/10.1103/PRXQuantum.4.020317}
}

@article{Kim2010,
  author = {Kim, K. and Chang, M.-S. and Korenblit, S. and Islam, R. and Edwards, E. E. and Freericks, J. K. and Lin, G.-D. and Duan, L.-M. and Monroe, C.},
  title = {Quantum simulation of frustrated {Ising} spins with trapped ions},
  journal = {Nature},
  volume = {465},
  number = {7298},
  pages = {590--593},
  year = {2010},
  month = jun,
  publisher = {Springer Science and Business Media LLC},
  doi = {10.1038/nature09071},
  url = {https://doi.org/10.1038/nature09071},
  issn = {1476-4687}
}

@article{Knetter2000,
  author = {C. Knetter and G.S. Uhrig},
  title = {Perturbation theory by flow equations: dimerized and frustrated {S} = 1/2 chain},
  journal = {The European Physical Journal B},
  volume = {13},
  number = {2},
  pages = {209--225},
  year = {2000},
  doi = {10.1007/s100510050026},
  url = {https://doi.org/10.1007/s100510050026}
}

@article{Knetter2003,
  author = {C. Knetter and K. P. Schmidt and G. S. Uhrig},
  title = {{The} structure of operators in effective particle-conserving models},
  journal = {Journal of Physics A: Mathematical and General},
  volume = {36},
  number = {29},
  pages = {7889--7907},
  year = {2003},
  doi = {10.1088/0305-4470/36/29/302},
  url = {https://doi.org/10.1088/0305-4470/36/29/302}
}

@article{Koffel2012,
  author = {Koffel, Thomas and Lewenstein, M. and Tagliacozzo, Luca},
  title = {Entanglement Entropy for the Long-Range {Ising} Chain in a Transverse Field},
  journal = {Physical Review Letters},
  volume = {109},
  number = {26},
  pages = {267203},
  year = {2012},
  month = dec,
  publisher = {American Physical Society},
  doi = {10.1103/PhysRevLett.109.267203},
  url = {https://doi.org/10.1103/PhysRevLett.109.267203}
}

@article{Koziol2019,
  author = {Koziol, Jan and Fey, Sebastian and Kapfer, Sebastian C. and Schmidt, Kai Phillip},
  title = {Quantum criticality of the transverse-field {Ising} model with long-range interactions on triangular-lattice cylinders},
  journal = {Physical Review B},
  volume = {100},
  number = {14},
  pages = {144411},
  year = {2019},
  month = oct,
  publisher = {American Physical Society (APS)},
  doi = {10.1103/physrevb.100.144411},
  url = {https://doi.org/10.1103/physrevb.100.144411},
  issn = {2469-9969}
}

@article{Koziol2021,
  author = {Koziol, Jan Alexander and Langheld, Anja and Kapfer, Sebastian C. and Schmidt, Kai Phillip},
  title = {Quantum-critical properties of the long-range transverse-field {Ising} model from quantum {Monte Carlo} simulations},
  journal = {Physical Review B},
  volume = {103},
  number = {24},
  pages = {245135},
  year = {2021},
  month = jun,
  publisher = {American Physical Society},
  doi = {10.1103/PhysRevB.103.245135},
  url = {https://doi.org/10.1103/PhysRevB.103.245135}
}

@article{Koziol2023,
  author = {Koziol, Jan Alexander and Duft, Antonia and Morigi, Giovanna and Schmidt, Kai Phillip},
  title = {Systematic analysis of crystalline phases in bosonic lattice models with algebraically decaying density-density interactions},
  journal = {SciPost Physics},
  volume = {14},
  number = {5},
  pages = {136},
  year = {2023},
  month = may,
  doi = {10.21468/SciPostPhys.14.5.136},
  url = {https://doi.org/10.21468/SciPostPhys.14.5.136},
  issn = {2542-4653},
  urldate = {2025-02-03}
}

@article{Koziol2024,
  author = {Koziol, Jan Alexander and M{\"u}hlhauser, Matthias and Schmidt, Kai Phillip},
  title = {Order-by-disorder and long-range interactions in the antiferromagnetic transverse-field {Ising} model on the triangular lattice---{A} perturbative point of view},
  journal = {Results in Physics},
  volume = {61},
  pages = {107794},
  year = {2024},
  month = jun,
  doi = {10.1016/j.rinp.2024.107794},
  url = {https://doi.org/10.1016/j.rinp.2024.107794},
  issn = {2211-3797},
  urldate = {2026-07-03}
}

@incollection{KrauthBook,
  author = {Krauth, Werner},
  editor = {Kert{\'e}sz, J{\'a}nos and Kondor, Imre},
  title = {Introduction to {Monte Carlo} algorithms},
  booktitle = {Advances in Computer Simulation},
  series = {Lecture Notes in Physics},
  volume = {501},
  pages = {1--35},
  year = {1998},
  publisher = {Springer Berlin Heidelberg},
  doi = {10.1007/bfb0105457},
  url = {https://doi.org/10.1007/bfb0105457},
  isbn = {9783540696759}
}

@article{Labuhn2016,
  author = {Labuhn, Henning and Barredo, Daniel and Ravets, Sylvain and de L{\'e}s{\'e}leuc, Sylvain and Macr{\`i}, Tommaso and Lahaye, Thierry and Browaeys, Antoine},
  title = {Tunable two-dimensional arrays of single {Rydberg} atoms for realizing quantum {Ising} models},
  journal = {Nature},
  volume = {534},
  number = {7609},
  pages = {667--670},
  year = {2016},
  month = jun,
  publisher = {Springer Science and Business Media LLC},
  doi = {10.1038/nature18274},
  url = {https://doi.org/10.1038/nature18274},
  issn = {1476-4687}
}

@article{Lange2024,
  author = {Lange, Hannah and Van de Walle, Anka and Abedinnia, Atiye and Bohrdt, Annabelle},
  title = {From architectures to applications: a review of neural quantum states},
  journal = {Quantum Science and Technology},
  volume = {9},
  number = {4},
  pages = {040501},
  year = {2024},
  month = sep,
  publisher = {IOP Publishing},
  doi = {10.1088/2058-9565/ad7168},
  url = {https://doi.org/10.1088/2058-9565/ad7168},
  issn = {2058-9565}
}

@article{Langheld2022,
  author = {Langheld, Anja and Koziol, Jan Alexander and Adelhardt, Patrick and Kapfer, Sebastian and Schmidt, Kai Phillip},
  title = {Scaling at Quantum Phase Transitions above the Upper Critical Dimension},
  journal = {SciPost Physics},
  volume = {13},
  number = {4},
  pages = {088},
  year = {2022},
  month = oct,
  doi = {10.21468/SciPostPhys.13.4.088},
  url = {https://doi.org/10.21468/SciPostPhys.13.4.088},
  issn = {2542-4653},
  urldate = {2026-02-16}
}

@article{Laupretre2026,
  author = {Laupr{\^e}tre, Thomas and Bernal, Jose Daniel and Baamara, Youcef and Rey, Ana Maria and Vernac, Laurent and Laburthe-Tolra, Bruno},
  title = {Probing Coherences and Itinerant Magnetism in a Dipolar Lattice Gas},
  journal = {Physical Review Letters},
  volume = {136},
  number = {10},
  pages = {103401},
  year = {2026},
  doi = {10.1103/h3fr-chgk},
  url = {https://doi.org/10.1103/h3fr-chgk}
}

@misc{Leclerc2026,
  author = {Lucas Leclerc and Sergi Juli{\`a}-Farr{\'e} and Gabriel Silva Freitas and Guillaume Villaret and Boris Albrecht and Lucas B{\'e}guin and Lilian Bourachot and Cl{\'e}mence Briosne-Frejaville and Dorian Claveau and Antoine Cornillot and Julius de Hond and Djibril Diallo and Cl{\'e}ment Dupays and Robin Dupont and Thomas Eritzpokhoff and Emmanuel Gottlob and Lo{\"i}c Henriet and Michael Kaicher and Lucas Lassabli{\`e}re and Arvid Lindberg and Yohann Machu and Hadriel Mamann and Thomas Pansiot and Julien Ripoll and Eun Sang Choi and Adrien Signoles and Joseph Vovrosh and Bruno Ximenez and Vivien Zapf and Shengzhi Zhang and Haidong Zhou and Minseong Lee and Tiagos Mendes-Santos and Constantin Dalyac and Antoine Browaeys and Alexandre Dauphin},
  title = {One-to-one quantum simulation of a frustrated magnet with 256 qubits},
  year = {2026},
  doi = {10.48550/arXiv.2603.20372},
  url = {https://doi.org/10.48550/arXiv.2603.20372},
  eprint = {2603.20372},
  archivePrefix = {arXiv},
  primaryClass = {quant-ph}
}

@article{Leseleuc2019,
  author = {de L{\'e}s{\'e}leuc, Sylvain and Lienhard, Vincent and Scholl, Pascal and Barredo, Daniel and Weber, Sebastian and Lang, Nicolai and B\"{u}chler, Hans Peter and Lahaye, Thierry and Browaeys, Antoine},
  title = {Observation of a symmetry-protected topological phase of interacting bosons with {Rydberg} atoms},
  journal = {Science},
  volume = {365},
  number = {6455},
  pages = {775--780},
  year = {2019},
  month = aug,
  publisher = {American Association for the Advancement of Science (AAAS)},
  doi = {10.1126/science.aav9105},
  url = {https://doi.org/10.1126/science.aav9105},
  issn = {1095-9203}
}

@article{Li2010,
  author = {Li, Shiqi and Bo, Lin and Wen, Bo and Sarachik, M. P. and Subedi, P. and Kent, A. D. and Yeshurun, Y. and Millis, A. J. and Lampropoulos, C. and Mukherjee, S. and Christou, G.},
  title = {Experimental determination of the {Weiss} temperature of ${\text{Mn}}_{12}\text{-ac}$ and ${\text{Mn}}_{12}\text{-ac-MeOH}$},
  journal = {Physical Review B},
  volume = {82},
  number = {17},
  pages = {174405},
  year = {2010},
  month = nov,
  publisher = {American Physical Society},
  doi = {10.1103/PhysRevB.82.174405},
  url = {https://doi.org/10.1103/PhysRevB.82.174405}
}

@article{Li2023,
  author = {Li, Jun-Ru and Matsuda, Kyle and Miller, Calder and Carroll, Annette N. and Tobias, William G. and Higgins, Jacob S. and Ye, Jun},
  title = {Tunable Itinerant Spin Dynamics with Polar Molecules},
  journal = {Nature},
  volume = {614},
  number = {7946},
  pages = {70--74},
  year = {2023},
  doi = {10.1038/s41586-022-05479-2},
  url = {https://doi.org/10.1038/s41586-022-05479-2}
}

@article{Liu2018REC,
  author = {Liu, Weiwei and Zhang, Zheng and Ji, Jianting and Liu, Yixuan and Li, Jianshu and Wang, Xiaoqun and Lei, Hechang and Chen, Gang and Zhang, Qingming},
  title = {Rare-Earth Chalcogenides: A Large Family of Triangular Lattice Spin Liquid Candidates},
  journal = {Chinese Physics Letters},
  volume = {35},
  number = {11},
  pages = {117501},
  year = {2018},
  month = oct,
  publisher = {IOP Publishing},
  doi = {10.1088/0256-307x/35/11/117501},
  url = {https://doi.org/10.1088/0256-307x/35/11/117501},
  issn = {1741-3540}
}

@article{Lwdin1962,
  author = {L\"{o}wdin, Per-Olov},
  title = {Studies in Perturbation Theory. IV. Solution of Eigenvalue Problem by Projection Operator Formalism},
  journal = {Journal of Mathematical Physics},
  volume = {3},
  number = {5},
  pages = {969--982},
  year = {1962},
  month = sep,
  publisher = {AIP Publishing},
  doi = {10.1063/1.1724312},
  url = {https://doi.org/10.1063/1.1724312},
  issn = {1089-7658}
}

@article{MarkovShi2008,
  author = {Markov, Igor L. and Shi, Yaoyun},
  title = {Simulating Quantum Computation by Contracting Tensor Networks},
  journal = {SIAM Journal on Computing},
  volume = {38},
  number = {3},
  pages = {963--981},
  year = {2008},
  doi = {10.1137/050644756},
  url = {https://doi.org/10.1137/050644756}
}

@article{MartnezHidalgo2001,
  author = {Mart{\'i}nez-Hidalgo, X and Chudnovsky, E. M and Aharony, A},
  title = {Dipolar ordering in Fe$_{8}$},
  journal = {Europhysics Letters (EPL)},
  volume = {55},
  number = {2},
  pages = {273--279},
  year = {2001},
  month = jul,
  publisher = {IOP Publishing},
  doi = {10.1209/epl/i2001-00331-2},
  url = {https://doi.org/10.1209/epl/i2001-00331-2},
  issn = {1286-4854}
}

@article{Matsubara1956,
  author = {Matsubara, Takeo and Matsuda, Hirotsugu},
  title = {A Lattice Model of Liquid Helium, I},
  journal = {Progress of Theoretical Physics},
  volume = {16},
  number = {6},
  pages = {569--582},
  year = {1956},
  month = dec,
  publisher = {Oxford University Press (OUP)},
  doi = {10.1143/ptp.16.569},
  url = {https://doi.org/10.1143/ptp.16.569},
  issn = {0033-068X}
}

@article{Mcnaughton2025,
  author = {Jake McNaughton and Mohamed Hibat-Allah},
  title = {Adaptive neural quantum states: a recurrent neural network perspective},
  journal = {Machine Learning: Science and Technology},
  volume = {7},
  number = {5},
  pages = {055012},
  year = {2026},
  month = sep,
  publisher = {IOP Publishing},
  doi = {10.1088/2632-2153/aea1da},
  url = {https://doi.org/10.1088/2632-2153/aea1da},
  eprint = {2507.18700},
  archivePrefix = {arXiv},
  primaryClass = {cond-mat.dis-nn}
}

@article{Medvidovi2024,
  author = {Medvidovi{\'c}, Matija and Moreno, Javier Robledo},
  title = {Neural-network quantum states for many-body physics},
  journal = {The European Physical Journal Plus},
  volume = {139},
  number = {7},
  pages = {631},
  year = {2024},
  month = jul,
  publisher = {Springer Science and Business Media LLC},
  doi = {10.1140/epjp/s13360-024-05311-y},
  url = {https://doi.org/10.1140/epjp/s13360-024-05311-y},
  issn = {2190-5444}
}

@book{Mila2011,
  editor = {Lacroix, Claudine and Mendels, Philippe and Mila, Fr{\'e}d{\'e}ric},
  title = {Introduction to Frustrated Magnetism: Materials, Experiments, Theory},
  series = {Springer Series in Solid-State Sciences},
  volume = {164},
  year = {2011},
  publisher = {Springer Berlin Heidelberg},
  doi = {10.1007/978-3-642-10589-0},
  url = {https://doi.org/10.1007/978-3-642-10589-0},
  isbn = {9783642105890},
  issn = {0171-1873}
}

@article{Miller2024,
  author = {Miller, Calder and Carroll, Annette N. and Lin, Junyu and Hirzler, Henrik and Gao, Haoyang and Zhou, Hengyun and Lukin, Mikhail D. and Ye, Jun},
  title = {Two-Axis Twisting Using Floquet-Engineered {XYZ} Spin Models with Polar Molecules},
  journal = {Nature},
  volume = {633},
  number = {8029},
  pages = {332--337},
  year = {2024},
  doi = {10.1038/s41586-024-07883-2},
  url = {https://doi.org/10.1038/s41586-024-07883-2}
}

@article{Millis2010,
  author = {Millis, A. J. and Kent, A. D. and Sarachik, M. P. and Yeshurun, Y.},
  title = {Pure and random-field quantum criticality in the dipolar {Ising} model: Theory of ${\text{Mn}}_{12}$ acetates},
  journal = {Physical Review B},
  volume = {81},
  number = {2},
  pages = {024423},
  year = {2010},
  month = jan,
  publisher = {American Physical Society},
  doi = {10.1103/PhysRevB.81.024423},
  url = {https://doi.org/10.1103/PhysRevB.81.024423}
}

@article{Moessner2003,
  author = {Isakov, S. V. and Moessner, R.},
  title = {Interplay of quantum and thermal fluctuations in a frustrated magnet},
  journal = {Physical Review B},
  volume = {68},
  number = {10},
  pages = {104409},
  year = {2003},
  month = sep,
  publisher = {American Physical Society},
  doi = {10.1103/PhysRevB.68.104409},
  url = {https://doi.org/10.1103/PhysRevB.68.104409},
  urldate = {2025-02-04}
}

@article{Monroe2021,
  author = {Monroe, C. and Campbell, W. C. and Duan, L.-M. and Gong, Z.-X. and Gorshkov, A. V. and Hess, P. W. and Islam, R. and Kim, K. and Linke, N. M. and Pagano, G. and Richerme, P. and Senko, C. and Yao, N. Y.},
  title = {Programmable quantum simulations of spin systems with trapped ions},
  journal = {Reviews of Modern Physics},
  volume = {93},
  number = {2},
  pages = {025001},
  year = {2021},
  month = apr,
  publisher = {American Physical Society},
  doi = {10.1103/RevModPhys.93.025001},
  url = {https://doi.org/10.1103/RevModPhys.93.025001}
}

@article{Moses2015,
  author = {Moses, Steven A. and Covey, Jacob P. and Miecnikowski, Matthew T. and Yan, Bo and Gadway, Bryce and Ye, Jun and Jin, Deborah S.},
  title = {Creation of a low-entropy quantum gas of polar molecules in an optical lattice},
  journal = {Science},
  volume = {350},
  number = {6261},
  pages = {659--662},
  year = {2015},
  month = nov,
  publisher = {American Association for the Advancement of Science (AAAS)},
  doi = {10.1126/science.aac6400},
  url = {https://doi.org/10.1126/science.aac6400},
  issn = {1095-9203}
}

@article{Moses2016,
  author = {Moses, Steven A. and Covey, Jacob P. and Miecnikowski, Matthew T. and Jin, Deborah S. and Ye, Jun},
  title = {New frontiers for quantum gases of polar molecules},
  journal = {Nature Physics},
  volume = {13},
  number = {1},
  pages = {13--20},
  year = {2017},
  month = jan,
  publisher = {Springer Science and Business Media LLC},
  doi = {10.1038/nphys3985},
  url = {https://doi.org/10.1038/nphys3985},
  issn = {1745-2481}
}

@phdthesis{Muehlhauser2024PhD,
  author = {M{\"u}hlhauser, Matthias},
  title = {{Graph Decomposition Techniques for Quantum Spin Systems With Multi-Spin Interactions}},
  school = {Friedrich-Alexander-Universit{\"a}t Erlangen-N{\"u}rnberg},
  year = {2024},
  doi = {10.25593/open-fau-1450},
  url = {https://doi.org/10.25593/open-fau-1450}
}

@book{Oitmaa2006,
  author = {Oitmaa, Jaan and Hamer, Chris and Zheng, Weihong},
  title = {Series Expansion Methods for Strongly Interacting Lattice Models},
  year = {2006},
  month = apr,
  publisher = {Cambridge University Press},
  doi = {10.1017/cbo9780511584398},
  url = {https://doi.org/10.1017/cbo9780511584398},
  isbn = {9780521143592}
}

@article{Orus2014,
  author = {Or{\'u}s, Rom{\'a}n},
  title = {A practical introduction to tensor networks: Matrix product states and projected entangled pair states},
  journal = {Annals of Physics},
  volume = {349},
  pages = {117--158},
  year = {2014},
  doi = {10.1016/j.aop.2014.06.013},
  url = {https://doi.org/10.1016/j.aop.2014.06.013}
}

@article{Paeckel2019,
  author = {Paeckel, Sebastian and K\"{o}hler, Thomas and Swoboda, Andreas and Manmana, Salvatore R. and Schollw\"{o}ck, Ulrich and Hubig, Claudius},
  title = {Time-evolution methods for matrix-product states},
  journal = {Annals of Physics},
  volume = {411},
  pages = {167998},
  year = {2019},
  month = dec,
  publisher = {Elsevier BV},
  doi = {10.1016/j.aop.2019.167998},
  url = {https://doi.org/10.1016/j.aop.2019.167998},
  issn = {0003-4916}
}

@article{Pocs2021CsYbSe2CEF,
  author = {Pocs, Christopher A. and Siegfried, Peter E. and Xing, Jie and Sefat, Athena S. and Hermele, Michael and Normand, B. and Lee, Minhyea},
  title = {Systematic extraction of crystal electric-field effects and quantum magnetic model parameters in triangular rare-earth magnets},
  journal = {Physical Review Research},
  volume = {3},
  number = {4},
  pages = {043202},
  year = {2021},
  month = dec,
  doi = {10.1103/PhysRevResearch.3.043202},
  url = {https://doi.org/10.1103/PhysRevResearch.3.043202}
}

@article{Powalski2013,
  author = {Powalski, M. and Coester, K. and Moessner, R. and Schmidt, K. P.},
  title = {Disorder by disorder and flat bands in the kagome transverse field {Ising} model},
  journal = {Physical Review B},
  volume = {87},
  number = {5},
  pages = {054404},
  year = {2013},
  month = feb,
  publisher = {American Physical Society},
  doi = {10.1103/PhysRevB.87.054404},
  url = {https://doi.org/10.1103/PhysRevB.87.054404},
  urldate = {2025-02-03}
}

@article{Prokofev1998,
  author = {Prokof'ev, N. V. and Svistunov, B. V. and Tupitsyn, I. S.},
  title = {Exact, complete, and universal continuous-time worldline {Monte Carlo} approach to the statistics of discrete quantum systems},
  journal = {Journal of Experimental and Theoretical Physics},
  volume = {87},
  number = {2},
  pages = {310--321},
  year = {1998},
  month = aug,
  publisher = {Pleiades Publishing Ltd},
  doi = {10.1134/1.558661},
  url = {https://doi.org/10.1134/1.558661},
  issn = {1090-6509}
}

@article{Puebla2019,
  author = {Puebla, Ricardo and Marty, Oliver and Plenio, Martin B.},
  title = {Quantum {Kibble-Zurek} physics in long-range transverse-field {Ising} models},
  journal = {Physical Review A},
  volume = {100},
  number = {3},
  pages = {032115},
  year = {2019},
  month = sep,
  publisher = {American Physical Society},
  doi = {10.1103/PhysRevA.100.032115},
  url = {https://doi.org/10.1103/PhysRevA.100.032115}
}

@article{Richerme2014,
  author = {Richerme, Philip and Gong, Zhe-Xuan and Lee, Aaron and Senko, Crystal and Smith, Jacob and Foss-Feig, Michael and Michalakis, Spyridon and Gorshkov, Alexey V. and Monroe, Christopher},
  title = {Non-local propagation of correlations in quantum systems with long-range interactions},
  journal = {Nature},
  volume = {511},
  number = {7508},
  pages = {198--201},
  year = {2014},
  month = jul,
  publisher = {Springer Science and Business Media LLC},
  doi = {10.1038/nature13450},
  url = {https://doi.org/10.1038/nature13450},
  issn = {1476-4687}
}

@article{robles2025exact,
  author = {Robles-Navarro, Andres and Cooper, Shaun and Buchheit, Andreas A and Busse, Jonathan K and Burrows, Antony and Smits, Odile and Schwerdtfeger, Peter},
  title = {Exact lattice summations for {Lennard-Jones} potentials coupled to a three-body {Axilrod--Teller--Muto} term applied to cuboidal phase transitions},
  journal = {The Journal of Chemical Physics},
  volume = {163},
  number = {9},
  pages = {094104},
  year = {2025},
  publisher = {AIP Publishing},
  doi = {10.1063/5.0276677},
  url = {https://doi.org/10.1063/5.0276677}
}

@article{RocaJerat2024,
  author = {Roca-Jerat, Sebasti{\'a}n and Gallego, Manuel and Luis, Fernando and Carrete, Jes{\'u}s and Zueco, David},
  title = {Transformer wave function for quantum long-range models},
  journal = {Physical Review B},
  volume = {110},
  number = {20},
  pages = {205147},
  year = {2024},
  month = nov,
  publisher = {American Physical Society (APS)},
  doi = {10.1103/physrevb.110.205147},
  url = {https://doi.org/10.1103/physrevb.110.205147},
  issn = {2469-9969}
}

@article{rota1964,
  author = {Rota, Gian-Carlo},
  title = {On the foundations of combinatorial theory {I}. {T}heory of {M}\"obius functions},
  journal = {Zeitschrift f{\"u}r Wahrscheinlichkeitstheorie und Verwandte Gebiete},
  volume = {2},
  number = {4},
  pages = {340--368},
  year = {1964},
  doi = {10.1007/BF00531932},
  url = {https://doi.org/10.1007/BF00531932}
}

@article{Saadatmand2018,
  author = {Saadatmand, S.N. and Bartlett, S.D. and McCulloch, I.P.},
  title = {Phase diagram of the quantum {Ising} model with long-range interactions on an infinite-cylinder triangular lattice},
  journal = {Physical Review B},
  volume = {97},
  number = {15},
  pages = {155116},
  year = {2018},
  publisher = {American Physical Society},
  doi = {10.1103/PhysRevB.97.155116},
  url = {https://doi.org/10.1103/PhysRevB.97.155116},
  issn = {2469-9950}
}

@book{SachdevBook,
  author = {Sachdev, Subir},
  title = {Quantum Phase Transitions},
  edition = {2},
  year = {2011},
  publisher = {Cambridge University Press},
  address = {Cambridge},
  doi = {10.1017/CBO9780511973765},
  url = {https://doi.org/10.1017/CBO9780511973765}
}

@article{Sandvik2002,
  author = {Sylju{\r{a}}sen, Olav F. and Sandvik, Anders W.},
  title = {Quantum {Monte Carlo} with directed loops},
  journal = {Physical Review E},
  volume = {66},
  number = {4},
  pages = {046701},
  year = {2002},
  month = oct,
  publisher = {American Physical Society (APS)},
  doi = {10.1103/physreve.66.046701},
  url = {https://doi.org/10.1103/physreve.66.046701},
  issn = {1095-3787}
}

@article{Sandvik2003,
  author = {Sandvik, Anders W.},
  title = {Stochastic series expansion method for quantum {Ising} models with arbitrary interactions},
  journal = {Physical Review E},
  volume = {68},
  number = {5},
  pages = {056701},
  year = {2003},
  month = nov,
  publisher = {American Physical Society},
  doi = {10.1103/PhysRevE.68.056701},
  url = {https://doi.org/10.1103/PhysRevE.68.056701}
}

@inproceedings{Sandvik2010,
  author = {Sandvik, Anders W.},
  title = {Computational Studies of Quantum Spin Systems},
  booktitle = {AIP Conference Proceedings},
  volume = {1297},
  number = {1},
  pages = {135--338},
  year = {2010},
  publisher = {AIP},
  doi = {10.1063/1.3518900},
  url = {https://doi.org/10.1063/1.3518900},
  issn = {0094-243X}
}

@article{Schauss2018,
  author = {Schauss, Peter},
  title = {Quantum simulation of transverse {Ising} models with {Rydberg} atoms},
  journal = {Quantum Science and Technology},
  volume = {3},
  number = {2},
  pages = {023001},
  year = {2018},
  month = jan,
  publisher = {IOP Publishing},
  doi = {10.1088/2058-9565/aa9c59},
  url = {https://doi.org/10.1088/2058-9565/aa9c59},
  issn = {2058-9565}
}

@article{Scheie2020ErSe2CEF,
  author = {Scheie, A. and Garlea, V. O. and Sanjeewa, L. D. and Xing, J. and Sefat, A. S.},
  title = {Crystal-field Hamiltonian and anisotropy in ${\mathrm{KErSe}}_{2}$ and ${\mathrm{CsErSe}}_{2}$},
  journal = {Physical Review B},
  volume = {101},
  number = {14},
  pages = {144432},
  year = {2020},
  month = apr,
  doi = {10.1103/PhysRevB.101.144432},
  url = {https://doi.org/10.1103/PhysRevB.101.144432}
}

@article{Schneider2012,
  author = {Schneider, Ch and Porras, Diego and Schaetz, Tobias},
  title = {Experimental quantum simulations of many-body physics with trapped ions},
  journal = {Reports on Progress in Physics},
  volume = {75},
  number = {2},
  pages = {024401},
  year = {2012},
  month = jan,
  publisher = {IOP Publishing},
  doi = {10.1088/0034-4885/75/2/024401},
  url = {https://doi.org/10.1088/0034-4885/75/2/024401},
  issn = {1361-6633}
}

@article{Scholl2021,
  author = {Scholl, Pascal and Schuler, Michael and Williams, Hannah J. and Eberharter, Alexander A. and Barredo, Daniel and Schymik, Kai-Niklas and Lienhard, Vincent and Henry, Louis-Paul and Lang, Thomas C. and Lahaye, Thierry and L\"{a}uchli, Andreas M. and Browaeys, Antoine},
  title = {Quantum simulation of {2D} antiferromagnets with hundreds of {Rydberg} atoms},
  journal = {Nature},
  volume = {595},
  number = {7866},
  pages = {233--238},
  year = {2021},
  month = jul,
  publisher = {Springer Science and Business Media LLC},
  doi = {10.1038/s41586-021-03585-1},
  url = {https://doi.org/10.1038/s41586-021-03585-1},
  issn = {1476-4687}
}

@article{Scholl2022,
  author = {Scholl, P. and Williams, H. J. and Bornet, G. and Wallner, F. and Barredo, D. and Henriet, L. and Signoles, A. and Hainaut, C. and Franz, T. and Geier, S. and Tebben, A. and Salzinger, A. and Z\"urn, G. and Lahaye, T. and Weidem\"uller, M. and Browaeys, A.},
  title = {Microwave Engineering of Programmable $XXZ$ Hamiltonians in Arrays of {Rydberg} Atoms},
  journal = {PRX Quantum},
  volume = {3},
  number = {2},
  pages = {020303},
  year = {2022},
  month = apr,
  publisher = {American Physical Society},
  doi = {10.1103/PRXQuantum.3.020303},
  url = {https://doi.org/10.1103/PRXQuantum.3.020303}
}

@article{Schollwck2011,
  author = {Schollw\"{o}ck, Ulrich},
  title = {The density-matrix renormalization group in the age of matrix product states},
  journal = {Annals of Physics},
  volume = {326},
  number = {1},
  pages = {96--192},
  year = {2011},
  month = jan,
  publisher = {Elsevier BV},
  doi = {10.1016/j.aop.2010.09.012},
  url = {https://doi.org/10.1016/j.aop.2010.09.012},
  issn = {0003-4916}
}

@article{Semeghini2021,
  author = {Semeghini, G. and Levine, H. and Keesling, A. and Ebadi, S. and Wang, T. T. and Bluvstein, D. and Verresen, R. and Pichler, H. and Kalinowski, M. and Samajdar, R. and Omran, A. and Sachdev, S. and Vishwanath, A. and Greiner, M. and Vuleti{\'c}, V. and Lukin, M. D.},
  title = {Probing topological spin liquids on a programmable quantum simulator},
  journal = {Science},
  volume = {374},
  number = {6572},
  pages = {1242--1247},
  year = {2021},
  month = dec,
  publisher = {American Association for the Advancement of Science (AAAS)},
  doi = {10.1126/science.abi8794},
  url = {https://doi.org/10.1126/science.abi8794},
  issn = {1095-9203}
}

@article{Shavitt1980,
  author = {Shavitt, Isaiah and Redmon, Lynn T.},
  title = {Quasidegenerate perturbation theories. A canonical {van Vleck} formalism and its relationship to other approaches},
  journal = {The Journal of Chemical Physics},
  volume = {73},
  number = {11},
  pages = {5711--5717},
  year = {1980},
  month = dec,
  publisher = {AIP Publishing},
  doi = {10.1063/1.440050},
  url = {https://doi.org/10.1063/1.440050},
  issn = {1089-7690}
}

@article{Shervashidze2011WL,
  author = {Shervashidze, Nino and Schweitzer, Pascal and van Leeuwen, Erik Jan and Mehlhorn, Kurt and Borgwardt, Karsten M.},
  title = {{Weisfeiler-Lehman} Graph Kernels},
  journal = {Journal of Machine Learning Research},
  volume = {12},
  number = {77},
  pages = {2539--2561},
  year = {2011},
  url = {https://jmlr.org/papers/v12/shervashidze11a.html}
}

@book{sidi2003practical,
  author = {Sidi, Avram},
  title = {Practical Extrapolation Methods: Theory and Applications},
  year = {2003},
  publisher = {Cambridge University Press},
  address = {Cambridge},
  doi = {10.1017/cbo9780511546815},
  url = {https://doi.org/10.1017/cbo9780511546815}
}

@article{smerald2018,
  author = {Smerald, Andrew and Mila, Fr{\'e}d{\'e}ric},
  title = {Spin-liquid behaviour and the interplay between {Pokrovsky}-{Talapov} and {Ising} criticality in the distorted, triangular-lattice, dipolar {Ising} antiferromagnet},
  journal = {SciPost Physics},
  volume = {5},
  number = {3},
  pages = {030},
  year = {2018},
  doi = {10.21468/SciPostPhys.5.3.030},
  url = {https://doi.org/10.21468/SciPostPhys.5.3.030},
  issn = {2542-4653},
  urldate = {2026-08-11}
}

@article{Su2023,
  author = {Su, Lin and Douglas, Alexander and Szurek, Michal and Groth, Robin and Ozturk, S. Furkan and Krahn, Aaron and H{\'e}bert, Anne H. and Phelps, Gregory A. and Ebadi, Sepehr and Dickerson, Susannah and Ferlaino, Francesca and Markovi{\'c}, Ognjen and Greiner, Markus},
  title = {Dipolar quantum solids emerging in a {Hubbard} quantum simulator},
  journal = {Nature},
  volume = {622},
  number = {7984},
  pages = {724--729},
  year = {2023},
  month = oct,
  publisher = {Springer Science and Business Media LLC},
  doi = {10.1038/s41586-023-06614-3},
  url = {https://doi.org/10.1038/s41586-023-06614-3},
  issn = {1476-4687}
}

@article{Subedi2012,
  author = {Subedi, P. and Kent, A. D. and Wen, Bo and Sarachik, M. P. and Yeshurun, Y. and Millis, A. J. and Mukherjee, S. and Christou, G.},
  title = {Transverse field {Ising} ferromagnetism in Mn${}_{12}$-acetate-MeOH},
  journal = {Physical Review B},
  volume = {85},
  number = {13},
  pages = {134441},
  year = {2012},
  month = apr,
  publisher = {American Physical Society},
  doi = {10.1103/PhysRevB.85.134441},
  url = {https://doi.org/10.1103/PhysRevB.85.134441}
}

@article{Sun2017,
  author = {Sun, Gaoyong},
  title = {Fidelity susceptibility study of quantum long-range antiferromagnetic {Ising} chain},
  journal = {Physical Review A},
  volume = {96},
  number = {4},
  pages = {043621},
  year = {2017},
  month = oct,
  publisher = {American Physical Society},
  doi = {10.1103/PhysRevA.96.043621},
  url = {https://doi.org/10.1103/PhysRevA.96.043621}
}

@article{Sun2026,
  author = {Sun, Xiangkai and Le, Yuan and Naus, Stephen and Tsai, Richard Bing-Shiun and Picard, Lewis R. B. and Murciano, Sara and Knap, Michael and Alicea, Jason and Endres, Manuel},
  title = {Observation of conformal field theory spectra in a quantum simulator},
  journal = {Nature},
  volume = {657},
  number = {8130},
  pages = {98--106},
  year = {2026},
  month = sep,
  publisher = {Springer Science and Business Media LLC},
  doi = {10.1038/s41586-026-10904-x},
  url = {https://doi.org/10.1038/s41586-026-10904-x},
  issn = {1476-4687}
}

@article{Suzuki1976,
  author = {Suzuki, M.},
  title = {Relationship between d-Dimensional Quantal Spin Systems and (d+1)-Dimensional {Ising} Systems: Equivalence, Critical Exponents and Systematic Approximants of the Partition Function and Spin Correlations},
  journal = {Progress of Theoretical Physics},
  volume = {56},
  number = {5},
  pages = {1454--1469},
  year = {1976},
  month = nov,
  publisher = {Oxford University Press (OUP)},
  doi = {10.1143/ptp.56.1454},
  url = {https://doi.org/10.1143/ptp.56.1454},
  issn = {1347-4081}
}

@article{Tabei2008,
  author = {Tabei, S. M. A. and Gingras, M. J. P. and Kao, Y.-J. and Yavors'kii, T.},
  title = {Perturbative quantum {Monte Carlo} study of {LiHoF}$_{4}$ in a transverse magnetic field},
  journal = {Physical Review B},
  volume = {78},
  number = {18},
  pages = {184408},
  year = {2008},
  month = nov,
  publisher = {American Physical Society},
  doi = {10.1103/PhysRevB.78.184408},
  url = {https://doi.org/10.1103/PhysRevB.78.184408}
}

@article{Takahashi1977,
  author = {Takahashi, M},
  title = {Half-filled {Hubbard} model at low temperature},
  journal = {Journal of Physics C: Solid State Physics},
  volume = {10},
  number = {8},
  pages = {1289--7301},
  year = {1977},
  month = apr,
  publisher = {IOP Publishing},
  doi = {10.1088/0022-3719/10/8/031},
  url = {https://doi.org/10.1088/0022-3719/10/8/031},
  issn = {0022-3719}
}

@article{Tarjan1972,
  author = {Tarjan, Robert},
  title = {Depth-First Search and Linear Graph Algorithms},
  journal = {SIAM Journal on Computing},
  volume = {1},
  number = {2},
  pages = {146--160},
  year = {1972},
  doi = {10.1137/0201010},
  url = {https://doi.org/10.1137/0201010}
}

@incollection{Toulouse2016,
  author = {Toulouse, Julien and Assaraf, Roland and Umrigar, Cyrus J.},
  title = {Introduction to the Variational and Diffusion {Monte Carlo} Methods},
  booktitle = {Electron Correlation in Molecules -- ab initio Beyond Gaussian Quantum Chemistry},
  pages = {285--314},
  year = {2016},
  publisher = {Elsevier},
  doi = {10.1016/bs.aiq.2015.07.003},
  url = {https://doi.org/10.1016/bs.aiq.2015.07.003},
  isbn = {9780128030608},
  issn = {0065-3276}
}

@article{ValdesTarjanLawler1982,
  author = {Valdes, Jacobo and Tarjan, Robert E. and Lawler, Eugene L.},
  title = {The recognition of series parallel digraphs},
  journal = {SIAM Journal on Computing},
  volume = {11},
  number = {2},
  pages = {298--313},
  year = {1982},
  doi = {10.1137/0211023},
  url = {https://doi.org/10.1137/0211023}
}

@article{Vodola2016,
  author = {Davide Vodola and Luca Lepori and Elisa Ercolessi and Guido Pupillo},
  title = {Long-range {Ising} and {Kitaev} models: phases, correlations and edge modes},
  journal = {New Journal of Physics},
  volume = {18},
  number = {1},
  pages = {015001},
  year = {2015},
  month = dec,
  publisher = {IOP Publishing},
  doi = {10.1088/1367-2630/18/1/015001},
  url = {https://doi.org/10.1088/1367-2630/18/1/015001}
}

@misc{Winter2026,
  author = {Lucas Winter and Andreas Nunnenkamp},
  title = {{DysonNet}: Constant-Time Local Updates for Neural Quantum States},
  year = {2026},
  doi = {10.48550/arXiv.2603.11189},
  url = {https://doi.org/10.48550/arXiv.2603.11189},
  eprint = {2603.11189},
  archivePrefix = {arXiv},
  primaryClass = {quant-ph}
}

@article{Yao2000,
  author = {Yao, Demin and Shi, Jicong},
  title = {Projection operator approach to time-independent perturbation theory in quantum mechanics},
  journal = {American Journal of Physics},
  volume = {68},
  number = {3},
  pages = {278--281},
  year = {2000},
  month = mar,
  publisher = {American Association of Physics Teachers (AAPT)},
  doi = {10.1119/1.19419},
  url = {https://doi.org/10.1119/1.19419},
  issn = {1943-2909}
}

@article{Zhang2021NaYbSe2CEF,
  author = {Zhang, Zheng and Ma, Xiaoli and Li, Jianshu and Wang, Guohua and Adroja, D. T. and Perring, T. P. and Liu, Weiwei and Jin, Feng and Ji, Jianting and Wang, Yimeng and Kamiya, Yoshitomo and Wang, Xiaoqun and Ma, Jie and Zhang, Qingming},
  title = {Crystalline electric field excitations in the quantum spin liquid candidate ${\mathrm{NaYbSe}}_{2}$},
  journal = {Physical Review B},
  volume = {103},
  number = {3},
  pages = {035144},
  year = {2021},
  month = jan,
  doi = {10.1103/PhysRevB.103.035144},
  url = {https://doi.org/10.1103/PhysRevB.103.035144}
}

@article{Zheng2023KTmSe2,
  author = {Zheng, Shiyi and Wo, Hongliang and Gu, Yiqing and Luo, Rui Leonard and Gu, Yimeng and Zhu, Yinghao and Steffens, Paul and Boehm, Martin and Wang, Qisi and Chen, Gang and Zhao, Jun},
  title = {Exchange-renormalized crystal field excitations in the quantum {Ising} magnet ${\mathrm{KTmSe}}_{2}$},
  journal = {Physical Review B},
  volume = {108},
  number = {5},
  pages = {054435},
  year = {2023},
  month = aug,
  publisher = {American Physical Society},
  doi = {10.1103/PhysRevB.108.054435},
  url = {https://doi.org/10.1103/PhysRevB.108.054435}
}

@article{Zhu2018,
  author = {Zhu, Zhangqi and Sun, Gaoyong and You, Wen-Long and Shi, Da-Ning},
  title = {Fidelity and criticality of a quantum {Ising} chain with long-range interactions},
  journal = {Physical Review A},
  volume = {98},
  number = {2},
  pages = {023607},
  year = {2018},
  month = aug,
  publisher = {American Physical Society},
  doi = {10.1103/PhysRevA.98.023607},
  url = {https://doi.org/10.1103/PhysRevA.98.023607}
}

@article{buchheit2026anisotropic,
  author        = {Buchheit, Andreas A. and Busse, Jonathan K.},
  title         = {Computation of anisotropic singular sums from high-order derivatives of Epstein zeta functions},
  year          = {2026},
  eprint        = {2609.28282},
  archivePrefix = {arXiv},
  primaryClass  = {math.NA}
}

\end{document}